\documentclass[reprint,aps,prb,superscriptaddress,english,nolongbibliography,twocolumn]{revtex4-2}

\usepackage[utf8]{inputenc}
\usepackage{fontenc}
\usepackage[normalem]{ulem}

\usepackage{graphicx}
\usepackage{amsmath,amssymb,bm,mathrsfs,amsfonts}
\usepackage{enumitem}

\usepackage[colorlinks=true,citecolor=blue,urlcolor=blue,linkcolor=blue]{hyperref}

\usepackage[table,xcdraw,dvipsnames]{xcolor}
\usepackage[caption=false,subrefformat=parens,labelformat=empty]{subfig}

\newcommand{\av}[1]{\left \langle #1 \right \rangle}
\newcommand{\bra}[1]{\left\langle #1 \right|}
\newcommand{\ket}[1]{\left| #1 \right\rangle}
\newcommand{\braket}[2]{\left\langle #1 \middle| #2 \right\rangle}

\newcommand{\sav}[1]{\langle #1 \rangle}
\newcommand{\sbra}[1]{\langle #1 |}
\newcommand{\sket}[1]{| #1 \rangle}

\let\Re\relax

\DeclareMathOperator{\Re}{Re}

\DeclareMathOperator{\sech}{sech}

\usepackage{tikz}

\newcommand{\urmove}[0]{%
\begin{tikzpicture}[baseline=0.1ex]%
\draw[thick, red] (0,1.5ex) -- (1.5ex,1.5ex);%
\draw[thick, red] (1.5ex,1.5ex) -- (1.5ex,0);%
\end{tikzpicture}%
}

\newcommand{\ulmove}[0]{%
\begin{tikzpicture}[baseline=0.1ex]%
\draw[thick, red] (0,1.5ex) -- (1.5ex,1.5ex);%
\draw[thick, red] (0,1.5ex) -- (0,0);%
\end{tikzpicture}%
}

\newcommand{\upflip}{\rotatebox[origin=c]{45}{\urmove}}
\newcommand{\downflip}{\rotatebox[origin=c]{225}{\urmove}}

\newcommand{\dlmove}[0]{%
\begin{tikzpicture}[baseline=0.1ex]%
\draw[thick, red] (0,1.5ex) -- (0,0);%
\draw[thick, red] (0,0) -- (1.5ex,0);%
\end{tikzpicture}%
}

\newcommand{\drmove}[0]{%
\begin{tikzpicture}[baseline=0.1ex]%
\draw[thick, red] (0,0) -- (1.5ex,0);%
\draw[thick, red] (1.5ex,1.5ex) -- (1.5ex,0ex);%
\end{tikzpicture}%
}

\newcommand{\vvmove}[0]{%
\begin{tikzpicture}[baseline=0.1ex]%
\draw[thick, red] (0,0) -- (0,1.5ex);%
\draw[thick, red] (1.5ex,0) -- (1.5ex,1.5ex);%
\end{tikzpicture}%
}

\newcommand{\hhmove}[0]{%
\begin{tikzpicture}[baseline=0.1ex]%
\draw[thick, red] (0,0) -- (1.5ex,0);%
\draw[thick, red] (1.5ex,1.5ex) -- (0,1.5ex);%
\end{tikzpicture}%
}

\newcommand{\treb}{%
\begin{tikzpicture}[baseline=0.1ex]%
\draw[thick, red] (0,1.5ex) -- (1.5ex,1.5ex);%
\draw[thick, red] (1.5ex,0) -- (1.5ex,1.5ex);%
\draw[thick, red] (0,0) -- (1.5ex,0);
\end{tikzpicture}%
}

\newcommand{\unb}{%
\begin{tikzpicture}[baseline=0.1ex]%
\draw[thick, red] (0,1.5ex) -- (0,0);%
\end{tikzpicture}%
}

\newcommand{\quatb}{%
\begin{tikzpicture}[baseline=0.1ex]%
\draw[thick, red] (0,1.5ex) -- (1.5ex,1.5ex);%
\draw[thick, red] (0,1.5ex) -- (0,0);%
\draw[thick, red] (0,0) -- (1.5ex,0);
\draw[thick, red] (1.5ex,0) -- (1.5ex,1.5ex);
\end{tikzpicture}%
}

\definecolor{mygray}{RGB}{153, 153, 153}
\newcommand{\gsquare}[0]{\textcolor{mygray}{\blacksquare}}

\begin{document}

\title{Interface dynamics in the two-dimensional quantum Ising model}

\author{Federico Balducci}
\email{fbalducc@sissa.it}
\affiliation{SISSA -- International School for Advanced Studies, via Bonomea 265, 34136, Trieste, Italy}
\affiliation{INFN Sezione di Trieste -- Via Valerio 2, 34127 Trieste, Italy}
\affiliation{The Abdus Salam ICTP -- Strada Costiera 11, 34151, Trieste, Italy}
\author{Andrea Gambassi}
\affiliation{SISSA -- International School for Advanced Studies, via Bonomea 265, 34136, Trieste, Italy}
\affiliation{INFN Sezione di Trieste -- Via Valerio 2, 34127 Trieste, Italy}
\author{Alessio Lerose}
\affiliation{Department of Theoretical Physics, University of Geneva -- Quai Ernest-Ansermet 30, 1205 Geneva, Switzerland}
\author{Antonello Scardicchio}
\affiliation{The Abdus Salam ICTP -- Strada Costiera 11, 34151, Trieste, Italy}
\affiliation{INFN Sezione di Trieste -- Via Valerio 2, 34127 Trieste, Italy}
\author{Carlo Vanoni}
\email{cvanoni@sissa.it}
\affiliation{SISSA -- International School for Advanced Studies, via Bonomea 265, 34136, Trieste, Italy}
\affiliation{INFN Sezione di Trieste -- Via Valerio 2, 34127 Trieste, Italy}
\affiliation{The Abdus Salam ICTP -- Strada Costiera 11, 34151, Trieste, Italy}

\date{\today}

\begin{abstract}
    In a recent letter~[\hyperlink{cite.Balducci2022Localization}{Phys.\ Rev.\ Lett.\ \textbf{129}, 120601 (2022)}] we have shown that the dynamics of interfaces, in the symmetry-broken phase of the two-dimensional ferromagnetic quantum Ising model, displays a robust form of ergodicity breaking. In this paper, we elaborate more on the issue. First, we discuss two classes of initial states on the square lattice, the dynamics of which is driven by complementary terms in the effective Hamiltonian and may be solved exactly: (a) strips of  consecutive neighbouring spins aligned in the opposite direction of the surrounding spins, and (b) a large class of initial states, characterized by the presence of a well-defined ``smooth'' interface separating two infinitely extended regions with oppositely aligned spins. The evolution of the latter states can be mapped onto that of an effective one-dimensional fermionic chain, which is integrable in the infinite-coupling limit. In this case, deep connections with noteworthy results in mathematics emerge, as well as with similar problems in classical statistical physics.  We present a detailed analysis of the evolution of these interfaces both on the lattice and in a suitable continuum limit, including the interface fluctuations and the dynamics of entanglement entropy. Second, we provide analytical and numerical evidence supporting the conclusion 
    that the observed non-ergodicity---arising from Stark localization of the effective fermionic excitations---persists away from the infinite-Ising-coupling limit, and we highlight the presence of a timescale $T\sim e^{c L\ln L}$ for the decay of a region of large linear size $L$. The implications of our work for the classic problem of the decay of a false vacuum are also discussed.
\end{abstract}

\maketitle

\section{Introduction}

The dynamical nucleation of a region of true vacuum in a sea of false vacuum is a classic problem in statistical mechanics~\cite{bray1994theory,onuki2002phase,Karthika2016Review}. Most of the progress, however, has been achieved in the context of stochastic dynamics so far, since the unitary quantum dynamics constitutes a significant challenge. Stochastic dynamics often provides an adequate description of equilibrium condensed matter systems, such as magnets or crystal-liquid mixtures, due to the continuous influence of noisy environmental degrees of freedom, which act like a bath at a well-defined temperature. Nevertheless, there are situations in which one cannot neglect the unitary nature of the quantum dynamical evolution from a pure initial state. This is the case, for instance, in a cosmological setting: the problem was studied long ago by Kobzarev, Okun and Voloshin~\cite{Kobzarev1974Bubble}, and then by Coleman and Callan~\cite{Coleman1977False,Callan1977Fate,coleman1988aspects}, finding also applications in inflationary models of the universe~\cite{guth2000inflation}. In addition, unitary evolution plays a crucial role in recent experiments with ultracold matter, which make it possible to investigate analoguous false-vacuum-decay phenomena in coherent quantum many-body systems, where the nucleation is driven by quantum rather than thermal fluctuations (see, e.g.\  Ref.~\cite{song2022realizing} for a recent experiment in this direction). Finally, there are quantum optimization algorithms~\cite{farhi2001quantum,crosson2014different,laumann2015quantum}, which are designed to find the ground state of a classical Ising model (a computationally NP-hard task), but can incur in several dynamical drawbacks associated to classical or quantum effects~\cite{altshuler2010anderson,bapst2013quantum,bellitti2021entropic}. One can only expect that, in the near future, quantum simulators will allow finely controlled explorations of this physics using table-top experiments, allowing the observation of more counter-intuitive effects of coherent quantum dynamics.

With these motivations in mind, and following Ref.~\cite{Balducci2022Localization}, here we set to study the unitary evolution of nucleated vacuum bubbles in the two-dimensional ($2d$) ferromagnetic quantum Ising model with longitudinal and transverse fields of strengths $h$ and $g$, respectively. These vacuum bubbles correspond, to a first approximation, to regions on the lattice with a certain spin orientation, surrounded by a sea of spins with opposite orientation. We find that the limit of large Ising coupling $J\gg |h|$, $|g|$ is amenable to several simplifications: this is due to the emergence of a constraint on the length of the interface, which separates the regions of opposite spin alignment in the initial state.

In this context, we address the issue of Hilbert space fragmentation, recently investigated in Refs.~\cite{Yoshinaga2021Emergence,Hart2022Hilbert}, and elaborate on the effective Hamiltonian governing the dynamics. Such effective Hamiltonian further simplifies, and becomes amenable of analytical treatment, when restricted to two classes of initial states. The first is defined by the presence of a strip of aligned consecutive spins, running along one of the principal axes of the square lattice; the second, by an infinitely long ``smooth'' interface separating regions with oppositely aligned spins. The dynamics of the latter can be mapped onto a one-dimensional chain of fermions, which becomes integrable for $J\to\infty$. The integrability of this effective model is responsible for ergodicity breaking: we will show, for example, that the corner of a large bubble melts and reconstructs itself periodically in time, with period $\propto 1/|h|$. The same periodic dynamics generically characterizes an initially smooth profile, the evolution of which turns out to take a particularly simple form in a suitable continuum limit, which we discuss in detail. The proposed mapping on the fermionic chain allows us to study also interface fluctuations and the evolution of the entanglement entropy for an infinitely extended right-angled corner. In addition, we will also unveil surprising connections with classic mathematical results, concerning the limiting shape of random Young diagrams, as well as with similar problems in classical statistical physics.

Based on the mapping, we can trace back the observed ergodicity breaking in the dynamics of the interface in $2d$ to the Wannier-Stark localization of the single-particle eigenstates of the dual fermionic theory. Surprisingly, we find that, even moving away from the limit $J\to\infty$ in a perturbation theory in $g/J\ll 1$, the emerging many-particle eigenstates of the resulting perturbative Hamiltonian are \emph{Stark many-body localized} (MBL)~\cite{Schulz2019Stark,Rafael2019Bloch}: thus, they display the typical MBL phenomenology~\cite{Basko2006Metal,de2013ergodicity,huse2014phenomenology,ros2015integrals,luitz2015many}, which carries over to the $2d$ quantum Ising model. As several works have questioned the existence of MBL in more than one spatial dimension~\cite{de2017many,de2017stability} (even in the disordered version of the model studied here~\cite{Balducci2022Slow}), the present case provides a valuable example of a mechanism by which the non-ergodic dynamics of a one-dimensional model renders the dynamics of the dual two-dimensional model non-ergodic. Moreover, the phenomenology observed here reminds of the confinement that takes place in $1d$~\cite{Mazza2019Suppression,Lerose2020Quasilocalized}.

It is interesting to remark also that the $2d$ quantum Ising model displays strong stability of magnetic domains even when, in the absence of external fields, a Floquet dynamics is considered, characterized by imperfect stroboscopic single-spin kicks~\cite{Santini2022Clean}. Therefore, the interest in this model is renewed also by the possibility of probing different mechanisms for the breakdown of ergodicity, even if disorder-induced MBL is not present.

The discussion of the dynamics of an infinitely extended smooth interface (separating two semi-infinite domains) can be used as a starting point to investigate the case of finite but large ``bubbles'' of one phase, surrounded by the other. In truth, this was the original motivation of the work, being it related to the problem of the decay of a false vacuum. In particular, we discuss and estimate the relevant timescales which are involved in the possible ``melting'' of the bubble, the complete description of which is however beyond the scope of the present work.

The rest of the presentation is organized as follows. In Sec.~\ref{sec:model} we briefly introduce the Ising model, discussing how it reduces to a so-called ``PXP'' model in the limit of strong coupling (Sec.~\ref{sec:constraints}), for which Hilbert space fragmentation is expected to occur (Sec.~\ref{sec:infinite_J_fragmentation}). 
In Sec.~\ref{sec:infinite_coupling} we focus on the dynamics of the model in the infinite-coupling limit. In particular, in Sec.~\ref{sec:strip} we study strip-like initial configurations, while in Sec.~\ref{sec:Lipschitz} we consider more generic initial states, characterized by the presence of a smooth and infinite interface separating spins with opposite orientation. In Sec.~\ref{sec:Lipschitz-cont} we describe the continuum limit of the latter, and the connections with a semiclassical limit for the single-particle dynamics. 
In Sec.~\ref{sec:corner} we focus on a subset of initial configurations belonging to the general class discussed in Sec.~\ref{sec:Lipschitz}, i.e.\  a corner-shaped interface: in Sec.~\ref{sec:av_interface} we determine the average shape of such interface during the dynamics, while in Sec.~\ref{sec:fluct_interface} we study its fluctuations. 
In Sec.~\ref{sec:entanglement} we focus on the time evolution of the entanglement, discussing the computation of the entanglement entropy. 
In Sec.~\ref{sec:Young_diagrams} we show the connection between the unitary dynamics of the interface of a corner and some known results concerning the phenomenon of Plancherel measure concentration in random Young diagrams.
Moving to Sec.~\ref{sec:integrability_breaking}, we discuss how the emergent integrability can be broken, either in a domain of finite size (Sec.~\ref{sec:bubble}) or when the ferromagnetic coupling is no longer assumed to be infinitely large (Sec.~\ref{sec:1st_order_corrections} and~\ref{sec:Stark_MBL}), giving also a comparison between the lattice and the field theoretic dynamics of false vacuum bubbles (Sec.~\ref{sec:false_vacuum}). Finally, in Sec.~\ref{sec:conclusions} we present our conclusions and outlook.

Part of the work presented here was briefly reported in  Ref.~\cite{Balducci2022Localization}.

\section{The Model}
\label{sec:model}

\begin{figure}
    \centering
    \includegraphics[width=0.9\columnwidth]{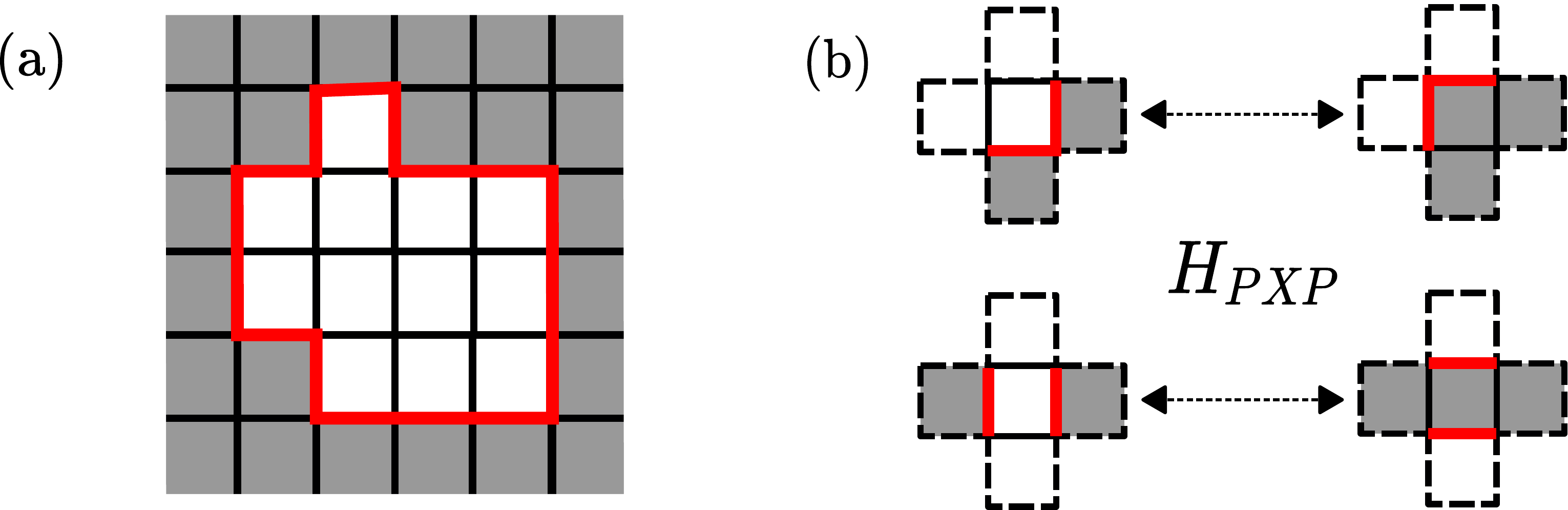}
    \caption{(a) Example of a ``convex'' (in the sense defined in Sec.~\ref{sec:bubble}) bubble of ``up'' spins ($\uparrow = \gsquare$) in a sea of ``down'' spins ($\downarrow = \Box$). Here each spin is represented by the surrounding square plaquette in the dual lattice. The side of a plaquette separating neighbouring spins with the same or opposite orientation is marked in black or red, respectively, the latter corresponding to a portion of a domain wall.
    (b) Example of transitions allowed  at the leading order in the coupling $J$, i.e.~due to the term $\propto g$ in $H_{\mathrm{PXP}}$, see Eq.~\eqref{eq:H_PXP_graphical}. Flipping the central spin makes the part highlighted in red of the domain wall move, in the corresponding plaquette, as represented in the figure. The remaining possible moves (not displayed) are obtained by considering all the configurations of the central spin and its neighbours, with the constraint that two neighbours are up and two down.}
    \label{fig:PXP_Ham}
\end{figure}

As anticipated in the Introduction, we are interested in the dynamics of the quantum Ising model on a two-dimensional square lattice. The Hamiltonian reads
\begin{equation}
    \label{eq:Ising2d_ham}
    H_{\mathrm{Is}} = -J\sum_{\langle ij \rangle} \sigma_i^z \sigma_j^z - g \sum_i \sigma_i^x - h \sum_i \sigma_i^z,
\end{equation}
where $\sigma^{x,y,z}_i$ are Pauli matrices acting on a lattice site $i\in\mathbb{Z}^2$, $\langle i j \rangle$ indicates the restriction of the sum to nearest neighbors, $g$ and $h$ are the strength of the transverse and longitudinal magnetic fields, respectively, and $J>0$ is the ferromagnetic coupling. We set $g>0$, while we let $h$ take both positive and negative values: the sign of $h$, indeed, will be relevant in Sec.~\ref{sec:1st_order_corrections}.

In thermal equilibrium at temperature $T$, this model displays a quantum phase transition at $T=0$ and $h=0$, belonging to the universality class of the classical 3$d$ Ising model: upon decreasing $g$ below a critical value $g_c$, it passes from a quantum paramagnet to a quantum ferromagnet, characterized by two degenerate, magnetized ground states spontaneously breaking the $\mathbb{Z}_2$ symmetry. Upon increasing $T$, the ferromagnetic phase survives up to a finite critical temperature $T_c$ (depending on $g$ and $J$), since the energetic cost of creating domains with reversed magnetization increases upon increasing their perimeter (as entailed by Peierls' argument). At $g=0$, the model becomes the $2d$ classical Ising model, therefore displaying the corresponding critical properties. These critical properties also characterize the transition occurring on the line of thermal critical points, which joins the classical model at $g=0$ to the quantum critical point at $T = 0$. The longitudinal field $h\neq 0$ breaks explicitly the $\mathbb{Z}_2$ symmetry of the two possible ground states, lifting their degeneracy. Accordingly, the model at $T=0$ and $g<g_c$ undergoes a first-order quantum phase transition as $h$ crosses $0$. As discussed in the Introduction, one expects that highly non-equilibrium false vacuum states exhibit a slow decay, through the nucleation of bubbles of characteristic size related to the inverse decay rate. With this background motivation in mind, below we will be interested in the fate of such bubbles, and more generally of interfaces, under the subsequent, coherent unitary evolution.

Studying the dynamics of $2d$ interacting models constitutes \emph{a priori} a formidable task: numerical methods are limited to very small system sizes or very short times. In addition, analytical tools are restricted to near-equilibrium conditions, or generally involve uncontrolled approximations, as dynamical mean-field theory~\cite{Aoki2014Nonequilibrium} or kinetic equations~\cite{Sphon2007Kinetic}. Despite these shortcomings, insight can be obtained from suitable limits. While the extreme paramagnetic regime  $J \ll |h|,g$ reduces to a set of weakly interacting ``magnonic'' excitations, the strongly-coupled ferromagnetic regime  $J \gg |h|,g$ retains great part of the interacting nature of the problem. It is the purpose of this work to show that, in such strong-coupling limit, there exists a relevant class of highly excited, non-thermal initial states, the dynamics of which is amenable of analytical treatment. In particular, in the next Sections we show that the formal limit $J\to\infty$ of infinitely strong ferromagnetic coupling actually renders a highly non-trivial constrained dynamical problem, characterized by a fragmented Hilbert space.

\subsection{Constrained dynamics in the strong-coupling limit}
\label{sec:constraints}

Starting from this Section, and throughout this work, we will consider the strong-coupling limit $J \gg |h|,g$. In practice, we start by formally taking $J = +\infty$, while later on in Sec.~\ref{sec:1st_order_corrections} we will relax this assumption. In this limit it is particularly convenient to study the problem in the basis of the eigenstates $\bigotimes_i |\!\uparrow\!\!/\!\!\downarrow\rangle_i$ of $\sigma^z_i$ at each lattice site $i$, with $\sigma^z_i \sket{\!\uparrow}_i =  \sket{\!\uparrow}_i$ and $\sigma^z_i \sket{\!\downarrow}_i = - \sket{\!\downarrow}_i$. At the leading order in $J$, the model is actually diagonal (i.e., classical) in this basis and, up to a constant, the energy of each of these eigenstates is given by $2Jl$, where $l$ is the number of distinct pairs of neighbouring spins with opposite orientation. Accordingly, the Hilbert space $\mathcal{H}$ at infinite coupling is fragmented into dynamically independent sectors with $\mathcal{H} = \bigoplus_l \mathcal{H}_l$, each sector $\mathcal{H}_l$ being identified by $l$~\cite{Yoshinaga2021Emergence,Balducci2022Localization}. Being $J=+ \infty$, in fact, no transitions are actually allowed from a state in $\mathcal{H}_l$ to one in $\mathcal{H}_{l'}$, unless $l = l'$, since the energy difference between them would be infinite. Note that, equivalently, $l$ measures the total length of the domain walls which are present on the lattice, separating the regions with spins $\sigma_i^z = +1$ from those with $\sigma_i^z = -1$. Accordingly, in the  limit $J \to \infty$, \emph{dynamical constraints} emerge, in the form of a \emph{perimeter constraint} on the bubbles of spins aligned along the same direction. Stated more formally, the domain-wall length operator
\begin{equation}
    D=\frac{1}{2} \sum_{\langle i,j \rangle} (1 - \sigma^z_i \sigma^z_j ),
\label{eq:D-dl}
\end{equation} 
is exactly conserved by $H_{\mathrm{Is}}$ in the $J\to \infty$ limit.

As a consequence of the perimeter constraint, the dynamics of the model can be effectively studied by focusing on each sector $\mathcal{H}_l$ separately, thereby reducing significantly the complexity of the problem. Let us start by determining the reduced Hamiltonian in $\mathcal{H}_l$ by elementary reasoning. Since the total domain-wall length must be conserved, the only spins that can be flipped by the term $\propto g$ in Eq.~\eqref{eq:Ising2d_ham} are those that just displace an existing domain wall. In practice, these spins are characterized by having two neighbours up ($\uparrow$) and two neighbours down ($\downarrow$), such that their flipping does not change the number of distinct pairs of neighbouring spins with opposite orientation, i.e., the length of the domain wall in the associated plaquettes. Considering the $(4\times 3)/2$ possible configurations of the four spins $\mathrm{L}i/ \mathrm{R}i/ \mathrm{U}i/ \mathrm{D}i$ which satisfy this constraint and are, respectively, left/right/above/below a site $i \in \mathbb{Z}^2$ with a certain spin orientation, one easily gets convinced that the only allowed transitions are those generated by the following reduced Hamiltonian:
\begin{equation}
\label{eq:H_PXP}
\begin{split}
    H_{\mathrm{PXP}} = - h \sum_i\sigma_i^z &\\
    - g\sum_{i} \big( &P_{\mathrm{L}i}^{\uparrow} P_{\mathrm{D}i}^{\uparrow} \sigma_i^x P_{\mathrm{R}i}^{\downarrow} P_{\mathrm{U}i}^{\downarrow} + P_{\mathrm{L}i}^{\uparrow} P_{\mathrm{D}i}^{\downarrow} \sigma_{i}^x P_{\mathrm{R}i}^{\downarrow} P_{\mathrm{U}i}^{\uparrow} \\[-2.2mm]
    +& P_{\mathrm{L}i}^{\downarrow} P_{\mathrm{D}i}^{\downarrow} \sigma_{i}^x P_{\mathrm{R}i}^{\uparrow} P_{\mathrm{U}i}^{\uparrow} + P_{\mathrm{L}i}^{\downarrow} P_{\mathrm{D}i}^{\uparrow} \sigma_{i}^x P_{\mathrm{R}i}^{\uparrow} P_{\mathrm{U}i}^{\downarrow}\\
    +&P_{\mathrm{L}i}^{\uparrow} P_{\mathrm{D}i}^{\downarrow} \sigma_{i}^x P_{\mathrm{R}i}^{\uparrow} P_{\mathrm{U}i}^{\downarrow}+ P_{\mathrm{L}i}^{\downarrow} P_{\mathrm{D}i}^{\uparrow} \sigma_{i}^x P_{\mathrm{R}i}^{\downarrow} P_{\mathrm{U}i}^{\uparrow}\big),
\end{split}
\end{equation}
where we introduced the projectors
\begin{equation}
\label{eq:projectors}
    P_i^{\uparrow} := \frac{1 + \sigma_i^z}{2} = \sket{\! \uparrow}_i {}_i \sbra{\uparrow \!}, \quad
    P_i^{\downarrow} := \frac{1 - \sigma_i^z}{2} = \sket{ \! \downarrow}_i {}_i \sbra{\downarrow \! }.
\end{equation}
The term $\propto h$ in Eq.~\eqref{eq:Ising2d_ham}, being diagonal in $\sigma^z_i$, is instead unaffected. One can recognize that Eq.~\eqref{eq:H_PXP} has the structure of a so-called PXP Hamiltonian~\cite{Turner2018Quantum}.

The elementary procedure outlined above can be viewed as the first step of a systematic elimination, from a Hamiltonian with large energy gaps, of highly non-resonant transitions. This is formally implemented by an order-by-order unitary transformation known as Schrieffer-Wolff transformation~\cite{Schrieffer1966Relation}. In Sec.~\ref{sec:1st_order_corrections} we will be concerned with the possible additional contributions to Eq.~\eqref{eq:H_PXP} due to higher-order corrections $O(J^{-1})$.

We stress here that the constrained Hamiltonian in Eq.~\eqref{eq:H_PXP}
is actually similar to the one describing strongly interacting Rydberg atom arrays~\cite{schauss2012observation,Ebadi2021Quantum}. 
In this case, each spin-$1/2$ describes a trapped neutral atom, which can be in either its ground state ($\downarrow$) or in a highly excited Rydberg state ($\uparrow$).
The basic model Hamiltonian that describes a lattice of such strongly interacting atoms reads~\cite{schauss2012observation}
\begin{equation}
\label{eq_basicRyd}
    H_{\mathrm{Ryd}} = \Delta \sum_i  n_i
    + \Omega \sum_{i}   \sigma_i^x +
    \sum_{i,j} V_{ij} n_i n_j
\end{equation}
where $n_i=(1+\sigma^z_i)/2$ counts the local number of atoms excited to the Rydberg state while the interaction $V_{ij}$ is very strong for neighboring sites and it decays rapidly as the distance $|i-j|$ increases. Upon rearranging the various terms, Eq.~\eqref{eq_basicRyd} may be viewed as a $2d$ quantum Ising model; the strong coupling $V_{ij}$, however, couples here to the operator $n_i n_j$ rather than to $\sigma^z_i \sigma^z_j$. When this nearest-neighbor interaction becomes larger than all the other energy scales--- as it happens in the so-called regime of Rydberg blockade---its dynamics is described by an effective constrained Hamiltonian,
\begin{equation}
    \label{eq:blockade}
    H^0_{\mathrm{Ryd}} = \frac \Delta 2 \sum_i  \sigma_i^z
    + \Omega \sum_{i}  P_{\mathrm{L}i}^{\downarrow} P_{\mathrm{D}i}^{\downarrow} \sigma_i^x P_{\mathrm{R}i}^{\downarrow} P_{\mathrm{U}i}^{\downarrow}.
\end{equation}
which is obtained from Eq.~\eqref{eq_basicRyd} by setting $V_{ij}\to\infty$ for neighboring atoms $\langle ij \rangle$ and $V_{ij}=0$ otherwise. In this case, pairs of neighboring excited atoms are completely frozen, and an atom can flip only if all its four neighbors are in the ground state, which is expressed by the last term in Eq.~\eqref{eq:blockade}. The Hamiltonian in Eq.~\eqref{eq:H_PXP}, instead, imposes a different form of the constraint, which implements the local perimeter-conserving motion of domain walls. It is interesting to note, however, that the two constraints differ only by a strong longitudinal field term, which can be adjusted to transform one into the other. Specifically, by identifying $V\equiv-4J$, it is sufficient to take a single-atom energy level detuning  $\Delta\equiv 2J+h$ to obtain the Ising model~\eqref{eq:Ising2d_ham} and hence, in the regime of Rydberg-blockade, the effective  Hamiltonian in Eq.~\eqref{eq:H_PXP}~\footnote{We note, however, that this might be problematic at experimental level, as the Rydberg interactions are very sensitive to the precise position of the trapped atoms, resulting in unwanted noisy fluctuations of the longitudinal field. We thank Hannes Pichler for this clarification (private communication).}. 

The Hamiltonian in Eq.~\eqref{eq:H_PXP} can be alternatively written via a shorthand notation, which describes graphically the transitions induced on the part of domain wall (in red) existing in the square plaquette surrounding a spin (i.e. the dual lattice), due to its allowed flipping (see also Fig.~\ref{fig:PXP_Ham}):
\begin{multline}
    \label{eq:H_PXP_graphical}
    H_{\mathrm{PXP}} = - h \sum_i \sigma_i^z  \\
    - g \sum_i \Big( \sket{\ulmove}_i{}_i\sbra{\drmove} + \sket{\dlmove}_i{}_i\sbra{\urmove} + \sket{\hhmove}_i{}_i\sbra{\vvmove} + \mathrm{H.c.} \Big).
\end{multline}
Here, the transitions due to the coupling $g$ are apparent: either a domain wall corner is moved across the diagonal of a plaquette ($\ulmove \leftrightarrows \drmove$ or $\dlmove \leftrightarrows \urmove$), or two parallel segments of the domain wall are recombined across opposite sides of the plaquette ($\hhmove \leftrightarrows \vvmove$). These moves guarantee the conservation of the domain wall length.

\subsection{Hilbert space fragmentation}
\label{sec:infinite_J_fragmentation}

The convenient notation of Eq.~\eqref{eq:H_PXP_graphical} makes it possible to analyze the fate of the dynamics of large portions of the $2d$ lattice in various cases. For instance, consider multiple, distant spins oriented up, i.e.\ with $\sigma^z = +1$, embedded in a sea of oppositely aligned spins, with $\sigma^z = -1$. This configuration is fully frozen, as no allowed transition can shift any of the domain walls. Thus, all of these states are eigenstates of the constrained Hamiltonian~\eqref{eq:H_PXP_graphical}. This simple example---easily generalizable to many others~\cite{Yoshinaga2021Emergence}---shows that individual sectors $\mathcal{H}_l$ are, in general, further heavily fragmented. More formally, one can introduce the notion of \emph{Krylov subspace} of a state $\ket{\psi_0}$: by definition, it is the subspace of $\mathcal{H}$ spanned by the set of vectors $\{\ket{\psi_0}, H\ket{\psi_0}, H^2 \ket{\psi_0}, \dots\}$, where $H$ is the Hamiltonian of the system. With this definition, one recognizes that the Krylov sector of a state $\ket{\psi} \in \mathcal{H}_l$ may not coincide with the full $\mathcal{H}_l$, but instead represent a finer shattering. A detailed study of the Krylov sectors of the model under consideration was presented in Ref.~\cite{Hart2022Hilbert}; in this work, instead, we will be concerned mainly with the \emph{dynamical effects} of the fragmentation on some physically relevant states. This is what we set out to study in the next Section.

\section{Infinite-coupling dynamics for strips and smooth domain walls}
\label{sec:infinite_coupling}

In the previous Section we have argued that, in the limit of large $J$, the dynamics of the $2d$ quantum Ising model simplifies significantly, because of the presence of emergent constraints. Here, we show that this simplification is really substantial in some particular cases, as it leads to simple \emph{one-dimensional effective models}. 

From Eq.~\eqref{eq:H_PXP_graphical}, one can see that the first two terms ($\sket{\ulmove}_i{}_i\sbra{\drmove} + \mathrm{H.c.}$ and $\sket{\dlmove}_i{}_i\sbra{\urmove} + \mathrm{H.c.}$) correspond to the translation of a domain wall, while the last one ($\sket{\hhmove}_i{}_i\sbra{\vvmove} + \mathrm{H.c.}$) cuts two nearby portions of domain wall into two halves and recombines those belonging to different portions. If the initial condition has a geometry that allows only one of the two types of transitions, then it is possible to gain further analytical control on the dynamics. In particular, we show in Sec.~\ref{sec:strip} that initial conditions consisting of a thin, pseudo-1$d$ domain are only affected by interface-recombining moves. This allows us to make a connection with 1$d$ PXP and confining Ising models. In Sec.~\ref{sec:Lipschitz}, instead, we show that if the $2d$ lattice is cut by a single, Lipschitz-continuous interface (this notion will be clarified further below), then its dynamics can be studied via an effective 1$d$ model of non-interacting fermions in a linear potential. Its emergent integrability allows us to predict the $2d$ evolution exactly, and to describe precisely how ergodicity is broken.

\subsection{Strip-like configurations}
\label{sec:strip}

In this Section, we consider a class of initial configurations that are essentially one-dimensional. As it was also pointed out in Ref.~\cite{Yoshinaga2021Emergence}, for this type of states it is possible to establish an explicit connection with $1d$ PXP models. We show here that, when the initial configuration $\ket{\Psi_0}$ has no overlap with scarred states \footnote{Quantum many body scars denote special eigenstates of the spectrum that does not satisfy the eigenstate thermalization hypothesis. This means that the expectation values of observables evaluated on such states does not attain the thermal value, even if their energy density corresponds to infinite temperature states.}, it is possible to calculate the asymptotic magnetization of the bubble.

We focus on an initial condition consisting of a linear strip of $L$ consecutive down spins ($\downarrow$), along one of the principal lattice axes, surrounded by up spins ($\uparrow$). In the Krylov sector of this configuration (see Sec.~\ref{sec:infinite_J_fragmentation} above), and in the absence of longitudinal magnetic field (i.e.~for $h=0$), the PXP Hamiltonian~\eqref{eq:H_PXP} reduces to the one-dimensional PXP Hamiltonian familiar from tilted bosonic traps~\cite{Fendley2004Competing}, one-dimensional Rydberg-blockaded arrays~\cite{Bernien2017Probing}, or dimer models~\cite{Chepiga2021Kibble}. Indeed, due to the perimeter constraint, neither the spins outside the initial strip nor those at its two ends can be flipped by $H_{\mathrm{PXP}}$; accordingly, the only dynamical degrees of freedom are the internal spins initially set to be down. This reduces the full, $2d$ dynamics to an effectively 1$d$ dynamics.

For convenience, we label the accessible basis states by the corresponding 1$d$ configuration of the spins in the strip; the initial state $\ket{\Psi_0}$ is therefore denoted by $\ket{\Psi_0} = \ket{\downarrow \downarrow \dots \downarrow}$. Assuming for the moment $h=0$, the Hamiltonian~\eqref{eq:H_PXP} reduces to
\begin{equation}
\label{eq:H_PXP_1d}
    H_{\mathrm{PXP,1d}} = - g \sum_{j=2}^{L-1} P_{j-1}^\downarrow \sigma_j^x P_{j+1}^\downarrow,
\end{equation}
as the spins above and below the strip are fixed to be up. Above, we are also taking into account that the first and last spin of the strip cannot be flipped. 

\begin{figure}[t]
    \centering
    \includegraphics[width=0.9\columnwidth]{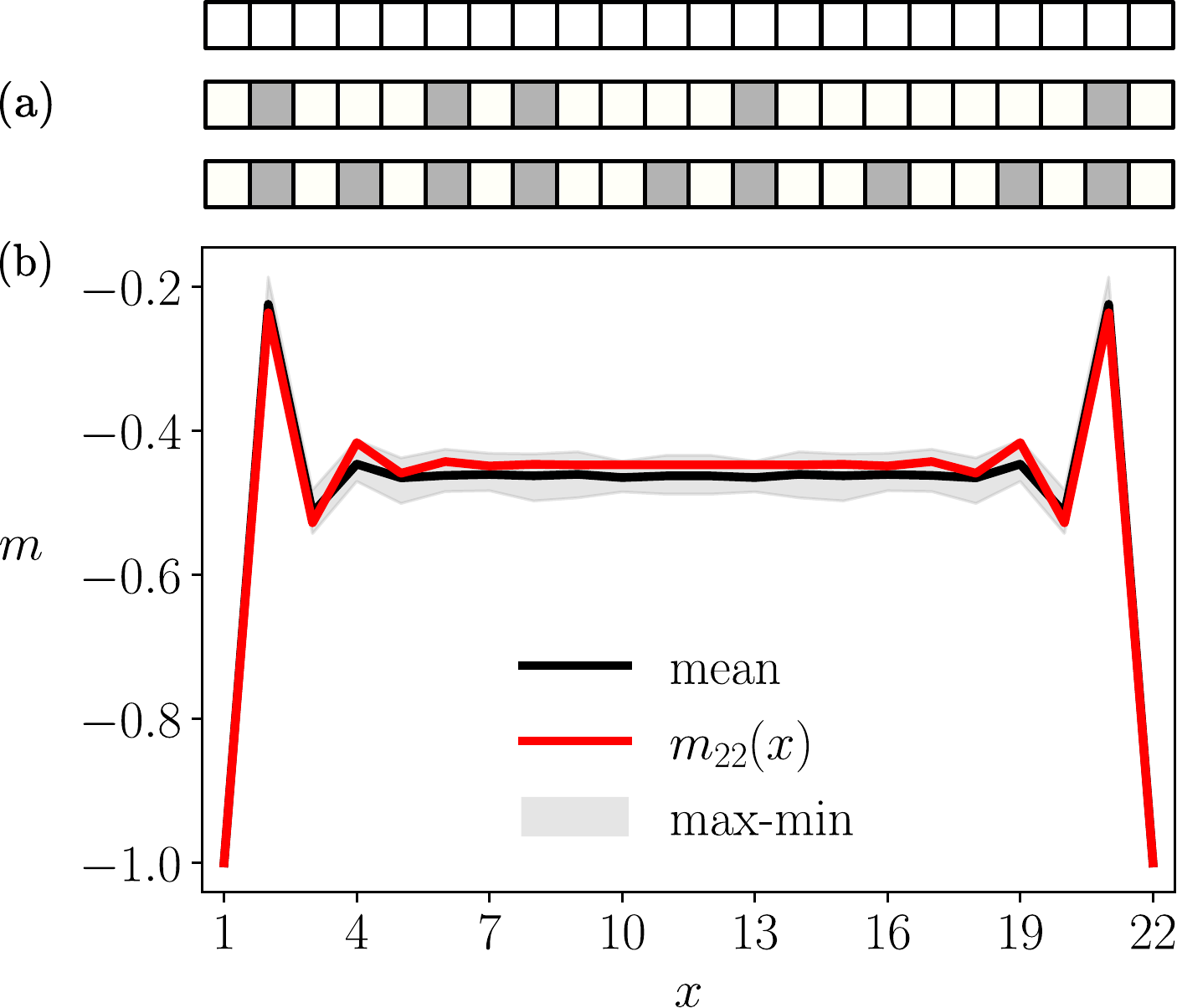}
    \caption{(a) The top row represents the initial state of the strip $\sket{\Psi_0}$; the middle row shows an example of configuration which can be dynamically reached from $\sket{\Psi_0}$; the bottom row displays a completely fragmented configuration.
    (b) Magnetization along the strip of panel (a) at long times. The comparison between the analytical prediction ($m_{22}(x) = \av{m_L(x)}$, in red, corresponding to Eq.~\eqref{eq:magn1}) and the numerical results for the magnetization is reported. The numerical analysis is performed by unitarily evolving the initial state. The plot shows the minimum and maximum magnetization for $5000<t<10000$ (shaded gray area) and the magnetization for $t=10000$ (black). One can see a good agreement between the numerical simulations and the analytical prediction, showing that the classical sampling introduced in the text is effective in describing the infinite-temperature magnetization. }
    \label{fig:StripData}
\end{figure}

Because of the constraints, not all 1$d$ configurations are dynamically accessible: for example, those containing two or more consecutive spins up are not. This implies that the spins adjacent to a spin up are down, and that completely fragmented configurations consist of singlets or pairs of spins down, separated by single spins up, see Fig.~\hyperref[fig:StripData]{\ref{fig:StripData}a}. Denoting by $l$ the number of spins which are reversed compared to the initial configuration, the number $C(L,l)$ of accessible basis states in the strip of length $L$ satisfies the recursion relation (see also App.~\ref{app:sec:strip})
\begin{equation}\label{eq:recursion_strip}
    C(L,l)=C(L-1,l)+C(L-2,l-1),
\end{equation}
which has solution
\begin{equation}\label{eq:numConfig}
    C(L,l)=\binom{L-l-1}{l},
\end{equation}
once the initial condition $C(L,0)=1$ for all $L$ is enforced.
The maximum number $l_{\rm max}$ of spins that can be flipped satisfying the perimeter constraint is
\begin{equation}\label{eq:n_max}
    l_{\rm max}=\bigg\lceil\frac{L-2}{2}\bigg\rceil,
\end{equation}
and the total number of accessible configurations is therefore given by 
\begin{equation}
    F_L=\sum_{l=0}^{l_{\max}}C(L,l),
\end{equation}
i.e.~by the $L$-th Fibonacci number~\cite{WolframFibonacci}. 

It is worth recalling that PXP Hamiltonians exhibit quantum many-body scars~\cite{Turner2018Quantum}, i.e.\ particular eigenstates that violate the eigenstate thermalization hypothesis~\cite{Deutsch1991Quantum,Srednicki1994Chaos}. The number of such eigenstates increases only algebraically upon increasing the system size, making them very rare in the many-body spectrum. However, they profoundly affect the dynamical properties of particular initial configurations: for instance, the N\'eel state $\ket{\mathbb{Z}_2} = \ket{\downarrow \uparrow \downarrow \uparrow \dots}$ exhibits remarkable long-lived revivals, as discovered in early experimental explorations~\cite{Turner2018Weak}. While it has become clear that these non-thermal eigenstates slowly disappear in the large-size limit of the PXP model, their ultimate origin is presently unclear, despite significant research efforts, and is the subject of an active ongoing debate~\cite{Serbyn2021Quantum}. On the other hand, the initial state $\ket{\Psi_0}$ we consider here is not significantly affected by quantum many-body scars~\cite{Turner2018Quantum,Serbyn2021Quantum}. Accordingly, it is expected that the magnetization profile along the chain at long times is compatible with an assumption of ergodicity, i.e.~that all allowed configurations (having the same expectation value of the energy) will be occupied with uniform probability. Under this assumption, the long-time average magnetization $\av{m_L(j)}$ at position $j=1, \ldots, L$ along the strip of length $L$ is expected to be given by
\begin{equation}
\label{eq:magn1}
    \av{m_L(j)} = 2\frac{F_{L-j}F_{j-1}}{F_L}-1,
\end{equation}
as detailed in App.~\ref{app:sec:strip}. The explicit expression of the Fibonacci numbers~\cite{WolframFibonacci},
\begin{equation}
\label{eq:FN}
F_n = \frac{\phi^n - (-\phi)^{-n}}{2\phi-1}
\end{equation}
in terms of the golden ratio $\phi = (1+\sqrt{5})/2$, allows us to determine the resulting magnetization profile $\av{m_L(j)}$. We compare it with numerical simulations for short strips in Fig.~\ref{fig:StripData}, showing fairly good agreement with the assumption of ergodicity. The magnetization, as expected, is fixed at the boundaries of the strip, due to the fact that fluctuations cannot occur there, while its absolute values decreases upon moving away from the boundaries. In particular, the value $\av{m_{\infty, \text{bulk}}}$ of the magnetization  $\av{m_L(j)}$ in the middle of an infinitely long strip can be easily obtained by taking first the limit $L \to \infty$ and then $j \to \infty$ in Eq.~\eqref{eq:magn1}, finding (see App.~\ref{app:sec:strip})
\begin{equation}
    \av{m_{\infty, \text{bulk}}} = \frac{2}{(2\phi-1)\phi} - 1 = -\frac{1}{\sqrt{5}},
\label{eq:m-bulk}
\end{equation}
where we used Eq.~\eqref{eq:FN}.
The (alternating-sign) approach of $\av{m_\infty(j)}$ to $\av{m_{\infty, \text{bulk}}}$ upon increasing $j$ turns out to be exponential, with a rather short characteristic length $\xi_b = - 1/\ln|1-\phi^{-1}| \simeq 1.04$. The derivation of this fact is provided again in App.~\ref{app:sec:strip}, see Eq.~\eqref{eq:Msemi-inf}.

The solution presented above applies to the case of a single strip of reversed spins running along one of the principle lattice axes. In the presence of more than one strip (possibly having different orientations), the same results apply to each strip separately as long as the spins belonging to two different strips do not have a common nearest-neighbour. In fact, in case they have one, a change of its orientation might cause the interfaces of the two strips to merge and, due to the resulting shape, the term $\sket{\ulmove}_i{}_i\sbra{\drmove} + \sket{\dlmove}_i{}_i\sbra{\urmove} + \mathrm{H.c.}$ in the effective Hamiltonian \eqref{eq:H_PXP_graphical} would contribute to the dynamics as well. In particular, it is easy to realize that the initial condition $\sket{\Psi_0}$ discussed above is dynamically connected with the configuration consisting of the largest rectangular ``envelope'', which contains all initial strips with at least one common nearest-neighbour. While the dynamics in this case turns out to be highly non-trivial, in Sec.~\ref{sec:corner} we will focus on what happens to one of the corners of this rectangular envelope when it is sufficiently extended.

We conclude this Section by noting that what we have done, essentially, was to compute local observables in the infinite-temperature ensemble within the Krylov sector of the initial configuration $\sket{\Psi_0}$, instead of computing the expectation values on $\sket{\Psi_0(t)}$. The two procedures are equivalent, since the initial state $\sket{\Psi_0}$ lies in the middle of the spectrum (and thus is an infinite-temperature state)~\footnote{That $\sket{\Psi_0}$ lies in the middle of the spectrum follows from the fact that, first, it holds $\bra{\Psi_0} H \sket{\Psi_0}=0$; and second, that the spectrum is symmetric around zero ($H_{\mathrm{PXP,1d}}$ commutes with the space reflection operator $I$, and anti-commutes with the spectral reflection operator $C=\prod_j \sigma^z_j$~\cite{Schecter2018Manybody}).}, and the 1$d$ PXP model is ergodic~\cite{Turner2018Quantum,Serbyn2021Quantum}.

\subsection{Smooth domain walls on the lattice}
\label{sec:Lipschitz}

In the previous Section we considered strip-like initial configurations, the dynamics of which involved only the operators $\sket{\hhmove}\sbra{\vvmove} + \text{H.c.}$ of Eq.~\eqref{eq:H_PXP_graphical}, i.e.\ only domain-wall-breaking transitions. We now turn to a different family of initial states for which, instead, the only involved operators are $\sket{\dlmove}\sbra{\urmove} + \mathrm{H.c.}$ or $\sket{\ulmove}\sbra{\drmove} + \mathrm{H.c.}$: thus, solely domain-wall-moving transitions are generated.

\begin{figure}[t]
    \centering
    \includegraphics[width=0.9\columnwidth]{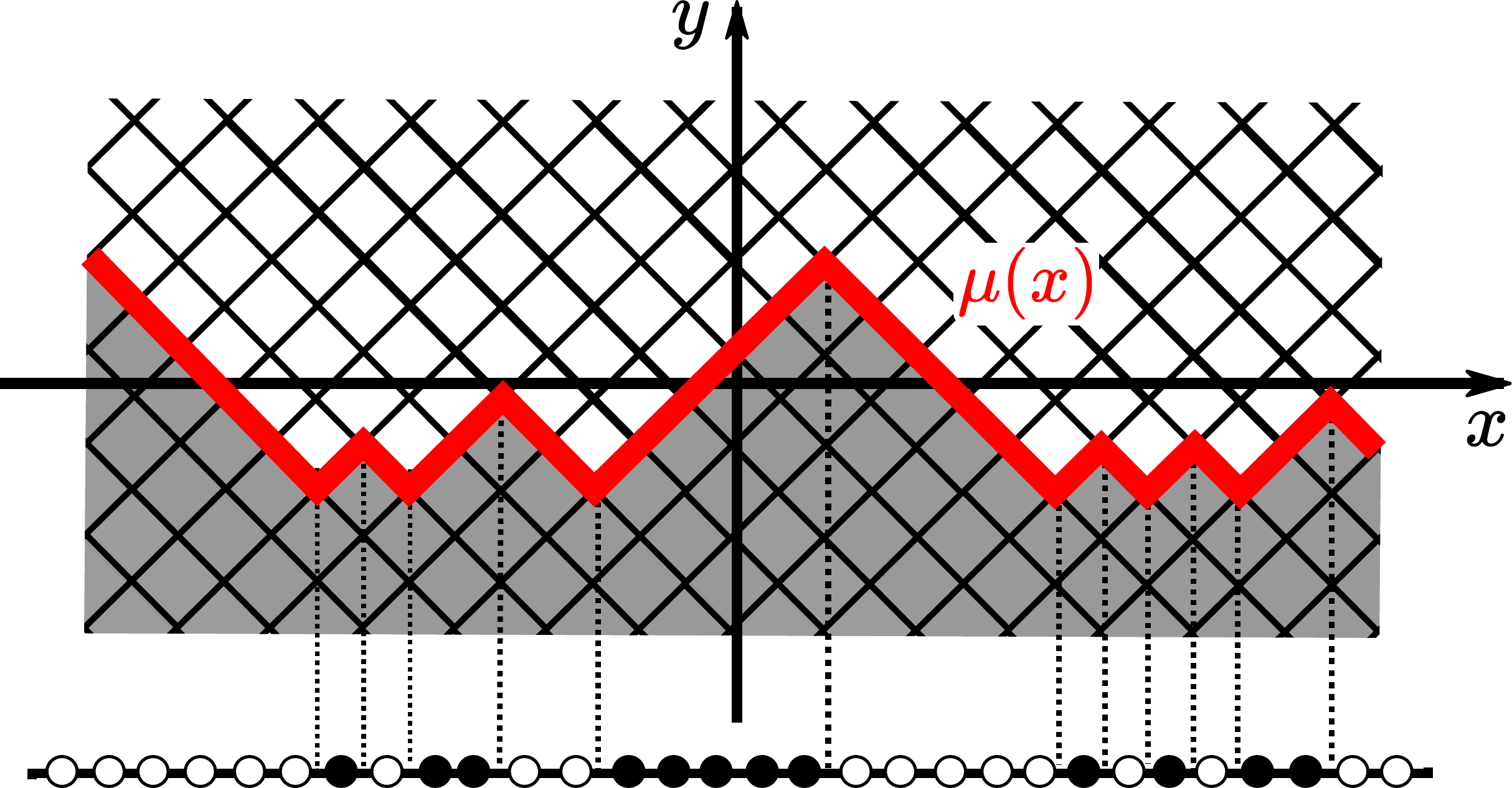}
    \caption{Graphical representation of the mapping from the $2d$ Lipschitz interface to the $1d$ fermionic chain. Moving from left to right, each downward segment corresponds to an empty site on the chain, while an occupied site is associated to each upward line. For completeness notice that, in doing the projection, the lattice spacing on the chain is reduced by a factor $\sqrt{2}$ compared to the original one on the $2d$ lattice. }
    \label{fig:interface_mapping}
\end{figure}

For later convenience, let us rotate by a $\pi/4$ angle with respect to the vertical and horizontal directions the square lattice on which the model is defined, such that the lattice axes are oriented along the diagonals of the quadrants of the standard coordinate system, as shown in Fig.~\ref{fig:interface_mapping}. Then, let us consider an interface separating a domain of spins up ($\uparrow  = \gsquare$) from one of spins down ($\downarrow = \Box$), highlighted in red in Fig.~\ref{fig:interface_mapping}. We require that such an interface varies only slowly, so that it can be thought of as the graph of a function in the rotated frame, see Fig.~\ref{fig:interface_mapping}. More precisely, the interface profile $\mu(x)$ should be described by a quantum superposition of functions $\mu_0: \mathbb{Z} \to \mathbb{Z}$ which are \emph{Lipschitz-continuous} on the lattice, i.e.\
\begin{equation}
    \label{eq:Lipschitz}
    |\mu_0(x)-\mu_0(y)| \leq |x-y| , \qquad \forall x,y \in \mathbb{Z}.
\end{equation}
Let us remark that, since only the operators $\sket{\dlmove}\sbra{\urmove} + \mathrm{H.c.}$ or $\sket{\ulmove}\sbra{\drmove} + \mathrm{H.c.}$ of the Hamiltonian in Eq.~\eqref{eq:H_PXP_graphical} act on these configurations, the Krylov sector of a Lipschitz state contains only Lipschitz states. Accordingly, the unitary dynamics starting from such configurations involves only Lipschitz states and their superpositions, and cannot generate kinks or overhangs of the interface. Two-dimensional initial states of this type, other than being rather generic in the context of interface dynamics, are interesting because they can be alternatively described as states of a corresponding \emph{one-dimensional} system. The mapping simply consists in associating to each downward segment of the interface an empty site on the $1d$ chain, and to each upward segment a site occupied by a particle (see Fig.~\ref{fig:interface_mapping}), following the interface line from left to right. In practice, this mapping amounts at a differentiation: in fact, one associates an empty (resp.\ occupied) site if the domain-wall derivative is negative (resp.\ positive). As a consequence, the interface profile $\mu(x)$ can be reconstructed by ``integrating'' the density profile $n(x)$ on the chain~\cite{Fujimoto2020Family-Vicsek,Fujimoto2021Dynamical,Fujimoto2022Impact,Jin2020Stochastic}:
\begin{equation}
\label{eq:mu_from_n}
    \mu(x) = \sum_{y \leq x} \left[2n(y) - 1\right] + \mathrm{const.}
\end{equation}

The mapping described above works also in a classical setting, where a fluctuating interface induces on the 1$d$ particles an effective dynamics, as the simple exclusion processes~\cite{Rost1981Non,Liggett1999Stochastic,Schutz2001Exactly} (more on this at the beginning of Sec.~\ref{sec:corner}). In the quantum setting, the statistics of the particles plays a fundamental role. For the case under consideration---i.e.\ the $2d$ quantum Ising model---these particles have to be hard-core bosons, because at most one particle can be present at a lattice site, and those at different sites commute. Applying a Jordan-Wigner transformation, these hard-core bosons can be equivalently represented as fermions. From now on we will adopt this more convenient representation. 

Having set up the mapping between the accessible basis states of the two systems, we can proceed to determine the 1$d$ Hamiltonian on the chain, corresponding to the $2d$ PXP Hamiltonian \eqref{eq:H_PXP_graphical}. With a bit of reasoning, one notices that each allowed spin flip in $2d$ (which induces one of the  transitions  $\downflip \rightleftarrows \upflip$ in the interface of Fig.~\ref{fig:interface_mapping}) corresponds to a fermion hop along the chain. At the same time, in the presence of the longitudinal magnetic field $h\neq 0$, each spin flip in $2d$ contributes with a $\mp 2h$ energy difference depending on the corresponding upward/downward direction of the domain-wall transition $\downflip \rightleftarrows \upflip$, and therefore every fermion hop must be accompanied by the same energy change. This is achieved by introducing a linear potential in the $1d$ Hamiltonian such that a particle jumping to the right (resp.\ left) gains (resp.\ loses) an energy $2h$. The same procedure applied to off-diagonal elements fixes the hopping term of the chain, leading to the fermionic Hamiltonian
\begin{equation}
\label{eq:H_fermions}
    H_{\mathrm{F}} = - g \sum_x \left(\psi_{x}^{\dagger}\psi_{x+1} + \mathrm{H.c.} \right) + 2 h\sum_x x \,\psi_{x}^{\dagger}\psi_{x} ,
\end{equation}
defined up to a constant related to the choice of the origin of $x$.

Equation~\eqref{eq:H_fermions} is the well-known Wannier-Stark Hamiltonian~\cite{grosso2013solid}. It is diagonalized by the unitary transformation 
\begin{equation}
\label{eq:b_m}
    b_m = \sum_{x\in\mathbb{Z}} J_{x-m}\left(\gamma\right)\psi_x,
\end{equation}
where $\gamma:=g/h$ and $J_\nu(z)$ is the Bessel function of the first kind, yielding
\begin{equation}
    H_{\mathrm{F,diag}} = 2 h\sum_{m\in\mathbb{Z}} \, m \, b^{\dagger}_m b^{\phantom{\dagger}}_m.
\end{equation}
The energy spectrum is thus given by a set of equally spaced levels $E_m = 2 hm$, insensitive to $g$. We anticipate that this feature will be important in the discussion about non-ergodicity, further below in Secs.~\ref{sec:1st_order_corrections}--\ref{sec:Stark_MBL}.

In terms of the functions introduced above we are now able to predict the dynamics of any Lipschitz initial state $\ket{\Psi_0}$. In fact, such a state can be expressed as
\begin{equation}
\label{eq:initial_state}
    \ket{\Psi_0}=\prod_k \psi^{\dagger}_{x_k}\ket{0}
\end{equation}
on the chain, where the sequence $\{x_k\}_{k \in \mathbb{Z}}$ contains the sites occupied at $t=0$ and $\ket{0}$ is the vacuum of the $1d$ chain. The time evolution of the operators $b_m$ is simply given by $b_m(t)=b_m(0)e^{-2ihm t}$, and thus
\begin{equation}
\begin{aligned}
\label{eq:psi_t}
    \psi_x(t) &= \sum_{m,y} J_{x-m}(\gamma) J_{y-m}(\gamma) e^{-2ihmt} \psi_y(0)\\
    &= \sum_{y} e^{-i(x+y)ht} i^{x-y} J_{x-y}(\omega_t) \psi_y(0),
\end{aligned}
\end{equation}
where we introduced
\begin{equation}
    \label{eq:omega_t}
    \omega_t := 2|\gamma \sin(ht)| = 2 \left| \frac{g}{h} \sin(ht) \right|,
\end{equation}
and used the completeness relation of the Bessel functions, Eq.~\eqref{eq:completeness_Bessel}. Similarly, by using the previous expression and by calculating some Wick contractions, one can determine the evolution of the average of the density $n(x,t) = \psi^\dagger_x(t)\psi_x(t)$, i.e.
\begin{alignat}{3}
    \av{n(x,t)}
    &= \sum_{y,z} && \sum_k J_{x_k-y}(\gamma) J_{x_k-z}(\gamma) \nonumber \\
    & &&\times J_{x-y}(\gamma)J_{x-z}(\gamma)e^{-2iht(z-y)}\nonumber\\
    \label{eq:av_density}
    &= \sum_k && J^2_{x_k-x} (\omega_t),
\end{alignat}
where we used the property in Eq.~\eqref{eq:sumBessel}, and defined the averages over the initial state
\begin{equation}
\label{eq:def_average}
    \av{\cdots}
    := \bra{\Psi_0} \cdots \ket{\Psi_0}.
\end{equation}
As discussed above, $\av{n(x,t)}$ corresponds to the average slope of the quantum-fluctuating interface in the original $2d$ system, and therefore describes its evolution \footnote{It is interesting to note that, by time-reversal symmetry, Eq.~\eqref{eq:av_density} can also be interpreted as the total probability of finding a single particle starting at $x$ at time $0$, in the subset of positions $\{x_k\}$ at time $t$.}. Moreover, the expression for $\av{n(x,t)}$ in Eq.~\eqref{eq:av_density} clearly shows that the dynamics in the cases $h = 0$ and $h \neq 0$ are simply connected by the minimal substitution $2 g |t| \to \omega_t$ (and that the latter, in particular, is independent of the sign of $h$).

Equation~\eqref{eq:av_density}, and its dependence on time via $\omega_t$ (Eq.~\eqref{eq:omega_t}), imply that the dynamics on the chain---and therefore the full $2d$ dynamics---is periodic with period $\pi/|h|$: this is due to the \emph{Bloch oscillations}~\cite{grosso2013solid}, which localize each fermion near its initial position $x_k$. In fact, $J^2_{x_k-x}(\omega_t)$ in Eq.~\eqref{eq:av_density} decays exponentially fast upon increasing $|x-x_k|$ beyond $\omega_t$, and therefore each fermion, during the evolution, explores a region of space of amplitude (in units of the lattice spacing)
\begin{equation}
    \label{eq:lloc}
    \ell \simeq |g/h|.
\end{equation}
This perfect localization is a feature of the $J=+\infty$ limit, and of the presence of a nonzero longitudinal field $h$ which makes $\omega_t$ a periodic function of time. If, instead, $h \to 0$, one finds that $\omega_t=2g|t|$. Accordingly, the dynamics of the $1d$ system becomes ballistic, as the underlying fermionic excitations are free to move. Note that the presence of $h$ induces periodicity in the time evolution already at the level of Eq.~\eqref{eq:psi_t}, which is the solution of the Heisenberg equation for $\psi_x$. We emphasize again that such periodicity is both due to the external field $h$ \emph{and} the presence of the lattice. In Secs.~\ref{sec:1st_order_corrections} and \ref{sec:Stark_MBL} we will investigate the extent to which the localization is preserved at finite but large $J$, and nonzero $h$.

In general, $\av{n(x,t)}$ cannot be calculated in closed form from Eq.~\eqref{eq:av_density} for an arbitrary initial condition $\av{n(x,t=0)} = \sum_k \delta_{x_k,x}$. However, in some special cases this can be done. As an example, consider an initial state consisting of a sequence of fermions alternated by $s-1$ empty lattice sites, with $x_k = sk$ for some $s\in \mathbb{N}$ with $s \geq 1$. This corresponds to a domain wall in the $2d$ lattice which is almost flat, with an approximate slope $-(s-2)/s$. With this initial state, the average profile $\mu(x,t)$ can be determined from the average number density $\av{n(x,t)}$ on the chain (see Eq.~\eqref{eq:av_density}):
\begin{equation}
\begin{aligned}
    \label{eq:average_number_zigzag}
    \av{n(x,t)} &= \sum_k J_{sk-x}^2(\omega_t)\\
    &= \frac{1}{s} \sum_{0\leq n < s} e^{2 i x n \pi/s} J_0\left( 2\omega_t \sin \frac{n\pi}{s} \right),
\end{aligned}
\end{equation}
where the last equality follows from the integral representation of the Bessel functions, Eq.~\eqref{app:eq:bessel_sum_multiples}. At spatial scales much larger than the (unit) lattice spacing, i.e.\ for $|x|\gg 1$, the expression above implies  
\begin{equation}
    \label{eq:av-1sus}
    \av{n(x,t)} \simeq 1/s,
\end{equation}
because only the term with $n=0$ contributes to the sum for large $|x|$, due to the oscillating exponentials of the remaining terms. This result is expected, as the value $1/s$ actually corresponds to the average occupation along the chain in the initial condition which, up to lattice effects, does not evolve in time. After summing Eq.~\eqref{eq:average_number_zigzag} over space, as prescribed by Eq.~\eqref{eq:mu_from_n}, one obtains the average shape of the interface. In the limit $|x|\gg 1$ this corresponds to $\av{\mu(x,t)} \simeq x(2/s  - 1) + \mbox{const.}$, i.e.\ to a time-independent flat interface, with the slope fixed by the initial condition. Accordingly, up to lattice effects, flat interfaces in the $2d$ system do not evolve, independently of the underlying lattice: this actually suggests that a proper continuum limit of this lattice dynamics might emerge, as we discuss in more detail in the next Section.

\subsection{Smooth domain walls on the continuum and the semi-classical limit}
\label{sec:Lipschitz-cont}

We now explore how to modify the parameters of the fermionic model discussed above in a way such that, after reinstating the lattice spacing $a$, a non-trivial continuum limit of the dynamics of the particle density, or of the corresponding (Lipschitz) interface, is obtained as $a \to 0$. In particular, in Sec.~\ref{suc:continuum} we derive the dynamics of the fermion density and of the Lipschitz interfaces on the continuum, while in Sec.~\ref{sec:semiclass} we provide a physical interpretation of this dynamics in terms of a semiclassical picture.

\subsubsection{Dynamics on the continuum}
\label{suc:continuum}

We begin by noting that Eq.~\eqref{eq:av_density} can also be rewritten in an integral form as
\begin{equation}
\label{eq:av_number}
    \av{n(x,t)} = \int dy \, \rho(y) J_{y - x}^2(\omega_t),
\end{equation}
where we introduced the initial density
\begin{equation}
\label{eq:rho_comb}
    \rho(y) = \sum_k \delta(y-x_k).
\end{equation}
To discuss the continuum limit of the these expressions, it is convenient to introduce an absolute value in the index of the Bessel function in Eq.~\eqref{eq:av_number}, owing the symmetry in Eq.~\eqref{eq:sym-Jn-1}: $J^2_{y-x} \rightarrow J^2_{|y-x|}$. The above expressions are valid in full generality, for any Lipschitz initial state on the lattice, completely specified by $\rho(y)$. In taking the continuum limit as we will describe below, this comb-like function eventually turns into a smooth function, which is obtained by properly rescaling the coordinates with the lattice spacing.

The continuum limit is expected to provide accurate predictions at large distances and long times if, correspondingly, the typical amplitude $\ell$ of the Bloch oscillations given by Eq.~\eqref{eq:lloc} (in units of the lattice spacing $a$) becomes large on the lattice scale, but attains a finite value when measured in actual units, i.e.\ if $\ell a$ is finite as the formal continuum limit $a\to0$ is taken. According to Eq.~\eqref{eq:lloc}, this is obtained by assuming $h\sim a$ and therefore $\gamma \sim a^{-1}$, see the definition of $\gamma$ after Eq.~\eqref{eq:b_m}. Equivalently, the same goal can be achieved by requiring that $g \sim a^{-1}$. Moreover, as the dependence of the relevant quantities such as $\av{n(x,t)}$ and $\av{\mu(x,t)}$ on time $t$ is only via $\omega_t$ (Eq.~\eqref{eq:omega_t}), which involves the product $ht$, a non-trivial limit is obtained by considering long times, with $t \sim a^{-1}$ as $a\to 0$, but such that $h t$ remains constant. In turn, this implies that $\omega_t \to \omega_t/a$ in Eq.~\eqref{eq:av_number}, see also Eq.~\eqref{eq:omega_t}. The scaling $h \sim a$ actually corresponds to effectively diminishing the strength of $h$ with respect to $g$, making it easier for fermions to move. In practice, it can be obtained by introducing a factor $a$ in front of the linear potential in the Hamiltonian~\eqref{eq:H_fermions}. This can be understood, in an equivalent manner, as the requirement that the external potential generated by a (finite) constant field $E$ must be proportional to the physical position $X:= x a$ in the continuum: if $V(X)=-EX=-Eax$, where $x\in{\mathbb Z}$ labels the lattice site, then one readily recognizes that $h = E a\propto a$.

Quite generically, it is possible to infer the continuum limit of the density of fermions starting from Eqs.~\eqref{eq:av_number} and \eqref{eq:rho_comb}. In fact, after reinstating the lattice spacing $a$ and introducing the actual coordinate $X$ as above (and analogously $Y:= a y$), one can write
\begin{equation}
\label{eq:n-cont-gen-1}
    \av{n(X,t)} = \int \frac{dY}{a}\, \hat\rho(Y) J^2_{|X-Y|/a}(\omega_t/a),
\end{equation}
Above, with a slight abuse of notation we use the same notation for the density on the continuum $\av{n(X=a x,t)}$ and on the lattice $\av{n(x,t)}$. Moreover, we introduced
\begin{equation}
    \hat\rho(Y) := a \sum_k \delta(Y- a x_k) = \rho(Y/a)
    \label{eq:def-hrho}
\end{equation}
it is the initial density of fermions in the actual coordinates. 
As $a \to 0$, the comb-like function $\hat\rho(Y)$ attains a more regular dependence on $Y$---with $0\le \hat\rho(Y)\le 1$ due to the fermionic nature of the particles on the chain---, and we can use Eq.~\eqref{eq:J2-continuum} to determine the continuum limit of the kernel $J^2_{|y-x|}$. Then, Eq.~\eqref{eq:n-cont-gen-1} in the limit $a\to 0$ can be written as
\begin{equation}
\begin{aligned}
\label{eq:n-cont-gen-2}
    \av{n(X,t)} \stackrel{a\to 0}{\longrightarrow} & \int_{-\infty}^{+\infty} dY \, \hat\rho(Y)\, \frac{\theta(\omega_t-|X-Y|)}{\pi\omega_t\sqrt{1-|X-Y|^2/\omega_t^2}}\\ = & \int_{-1}^{1} dz\,\frac{\hat\rho(z\omega_t + X)}{\pi\sqrt{1-z^2}},
\end{aligned}
\end{equation}
where $\theta(x)$ is the Heaviside step function: $\theta(x\ge0)=1$ and  $\theta(x<0)=0$ . In Sec.~\ref{sec:semiclass} below we provide an interpretation of this expression in terms of the semiclassical limit of the fermion dynamics. 

It is worth noticing that the kernel $1/(\pi\sqrt{1-z^2})$, which appears in the previous equation, is normalized to $1$ in the interval $|z|\le 1$ in such a way that, for $t=0$, one recovers $\av{n(X,t=0)} \to \hat\rho(X)$. From this expression it is also apparent that any initial condition of the fermions on the lattice, which translates into a space-independent $\hat\rho$ on the continuum, does not actually evolve in the continuum limit. This is the case, for example, of the initial condition considered at the end of Sec.~\ref{sec:Lipschitz}, with $x_k = s k$ and $s=1, 2, \ldots$, for which (see Eq.~\eqref{eq:def-hrho})
\begin{equation}
    \hat\rho(Y) = a \sum_{k\in \mathbb{Z}} \delta(Y - a s k). 
\end{equation}
In the continuum limit, the sum above turns into an integral, i.e.\ $\sum_{k\in \mathbb{Z}} f(ak) \stackrel{a\to 0}{\longrightarrow} a^{-1} \int d\xi\, f(\xi)$ and therefore
\begin{equation}
    \hat\rho(Y) \stackrel{a\to 0}{\longrightarrow} \int_{-\infty}^{+\infty}\!\!d\xi\, \delta(Y-s \xi) = \frac{1}{s}.
\end{equation}
By inserting this density on the continuum in Eq.~\eqref{eq:n-cont-gen-2}, one readily finds Eq.~\eqref{eq:av-1sus}. Note that, while on the lattice we considered integer values of $s$, in the continuum limit $s$ can take any value $s\ge 1$, which corresponds to having an initial average density $1/s$ of fermions on the lattice. 

The linear and translationally-invariant structure of the relationship between the initial fermion density $\av{n(X,t=0)}= \hat\rho(X)$ and its value $\av{n(X,t)}$ at a later time carries over to the corresponding average positions $\av{\mu(X,t)}$ of the interface, given that $d\av{\mu(X,t)}/dX = 2 \av{n(X,t)}-1$. This can be seen with an integration by parts after having expressed $\av{n(X,t)}$ as the derivative of $\av{\mu(X,t)}$ on both sides of Eq.~\eqref{eq:n-cont-gen-2}. Accordingly, in the continuum
\begin{equation}
    \av{\mu(X,t)} = \int_{-1}^{1} dz\, \frac{\mu_0(z\omega_t + X)}{\pi\sqrt{1-z^2}},
    \label{eq:mu-cont-gen-1}
\end{equation}
where $\mu_0(X)$ stands for the initial condition. Note that the fermionic constraint on the possible values of $\hat\rho$ translates into the request that $|d\av{\mu_0(X)}/dX|\le 1$, as it is for a Lipschitz function on the continuum. 
On the lattice, on the other hand, even if there is a clear relation between the initial configuration of the chain and of the interface (given by the mapping), the linearity is not present because of the fermionic nature of the particles. Indeed, while in the continuum one can multiply the particle density by a constant $\kappa$ as long as $0 \le \rho_0, \, \kappa \rho_0 \le 1$, the same cannot be done \emph{locally} on the lattice.

Due to the positivity of the kernel it is also rather straightforward to show that if the initial condition $\mu_0(X)$ is a Lipschitz function with a certain constant, then the same applies to the evolved function $\av{\mu(X,t)}$. As anticipated above, Eq.~\eqref{eq:mu-cont-gen-1} clearly shows that any flat initial profile $\mu_0(X) = \alpha X + X_0$ does not evolve in time (with a possible dynamics occurring solely at the lattice scale). 

More generally, if the variation of the intial interface $\mu_0$ occurs on a length scale $\ell_0$ much larger than $\omega_t$, the function $\mu_0(Y=z\omega_t+X)$ on the r.~h.~s.~of Eq.~\eqref{eq:mu-cont-gen-1}, can be expanded around $Y=X$ (recall that $|z|\le 1$) and one finds that
\begin{multline}
\av{\mu(X,t)} = \mu_0(X) + \frac{\omega_t^2}{4}\mu''_0(X) \\ + \frac{3}{8} \frac{\omega_t^4}{4!} \mu_0^{(4)}(X) +  O\left((\omega_t/\ell_0)^6\right).
\end{multline}
This implies, inter alia, that a locally quadratic portion of the profile is simply shifted upward or downward depending on the sign of its curvature. 

As an explicit application of Eq.~\eqref{eq:mu-cont-gen-1}, consider the case in which the initial interface is described on the continuum by $\mu_0(X)= A \sin (\kappa X)$, with $|\kappa A| \le 1$ for the Lipschitz condition to hold. From Eq.~\eqref{eq:mu-cont-gen-1}, one readily infers that  $\av{\mu(X,t)} = A J_0(\kappa\omega_t) \sin(\kappa X)$, i.e.\ the shape of the boundary is not affected by the dynamics but its amplitude is periodically modulated. Generalizing this result, the linearity of the relationship between $\mu_0$ and $\mu$ allows us to conclude that if the initial profile has a spatial Fourier transform $\tilde \mu_0(k)$ on the continuum, then $\mu(X,t)$ has $\tilde\mu(k,t) = J_0(k\omega_t)\tilde \mu_0(k)$ as its Fourier transform in $X$. This means that if the spatial average of the interface height is initially finite, i.e. $\tilde \mu_0(k=0)$ is finite, then this average is not affected by the dynamics because $\tilde \mu(k=0,t) = \tilde \mu_0(k=0)$.

\subsubsection{Semiclassical limit}
\label{sec:semiclass}

Equation \eqref{eq:n-cont-gen-2} allows one to predict on the continuum the average fermion density $\av{n(x,t)}$ in terms of its initial value $\av{n(x,t=0)} = \hat\rho(X)$ for $a\to 0$. Interestingly enough, the same expression can be derived starting directly from a semiclassical model for the evolution of the effective excitations at the interface.

To see this more explicitly, consider the case of a single fermion evolving with Eq.~\eqref{eq:H_fermions} and take the classical limit of its Hamiltonian, which is given by (see e.g.\ Ref.~\cite{Lerose2019Quasilocalized}) 
\begin{equation}
    \label{app:eq:ham_classical}
    {\mathscr H}(p,q) = -2g \cos{p} + 2hq,
\end{equation}
in the phase space $(q,p)\in {\mathbb R} \times [0,2\pi)$, with $q$ the coordinate of the particle and $p$ the conjugated momentum. Consequently, the equations of motion are
\begin{align}
    \dot{q}(t) &= \partial_p {\mathscr H}= 2g \sin{p(t)}, \\
    \dot{p}(t) &= -\partial_q{\mathscr H} = -2h,
\end{align}
that lead to
\begin{align}
    \label{app:eq:p}
    p(t) &= (- 2ht + p_0) \ \mbox{mod}\ 2\pi, \\
 \label{app:eq:q}
    q(t) &= q_0 - \frac{2g}{h}\sin{(ht)}\sin{(ht-p_0)},
\end{align}
where $q_0$ and $p_0$ indicate the initial values of $q(t)$ and $p(t)$, respectively.

Since we are dealing with non-interacting fermions, in the classical analog we can consider a single particle located at a certain position $q_0$ at time $t=0$. As a consequence of the uncertainty principle, the momentum $p_0$ of the particle will be distributed uniformly over the interval $[0,2\pi)$, with a uniform probability density: $P(p_0)=1/(2\pi)$. Accordingly, at a certain time $t$ the position $q(t)$ of the particle will have a distribution centered around $q_0$, with an amplitude $|2 g\sin(ht)/h| = \omega_t$. The resulting distribution $P(q(t))$ is obtained by inverting Eq.~\eqref{app:eq:q}, yielding
\begin{equation}
    \label{app:eq:distr_q}
    P(q(t)) = \frac{\theta(\omega_t-|q(t)-q_0|)}{\pi \omega_t \sqrt{1-[q(t)-q_0]^2/\omega_t^2}}.
\end{equation}
This is exactly the same kernel of Eq.~\eqref{eq:n-cont-gen-2}, with the identification $q_0 \to Y$ and $q(t) \to X$. The procedure just outlined is very reminiscent of the hydrodynamics approach to free fermions~\cite{Scopa2022Exact} which, however, uses the Wigner function to extract the quantities of interest.

\section{Infinite-coupling dynamics for an infinite corner}
\label{sec:corner}

\begin{figure}[t]
    \centering
    \includegraphics[width=0.9\columnwidth]{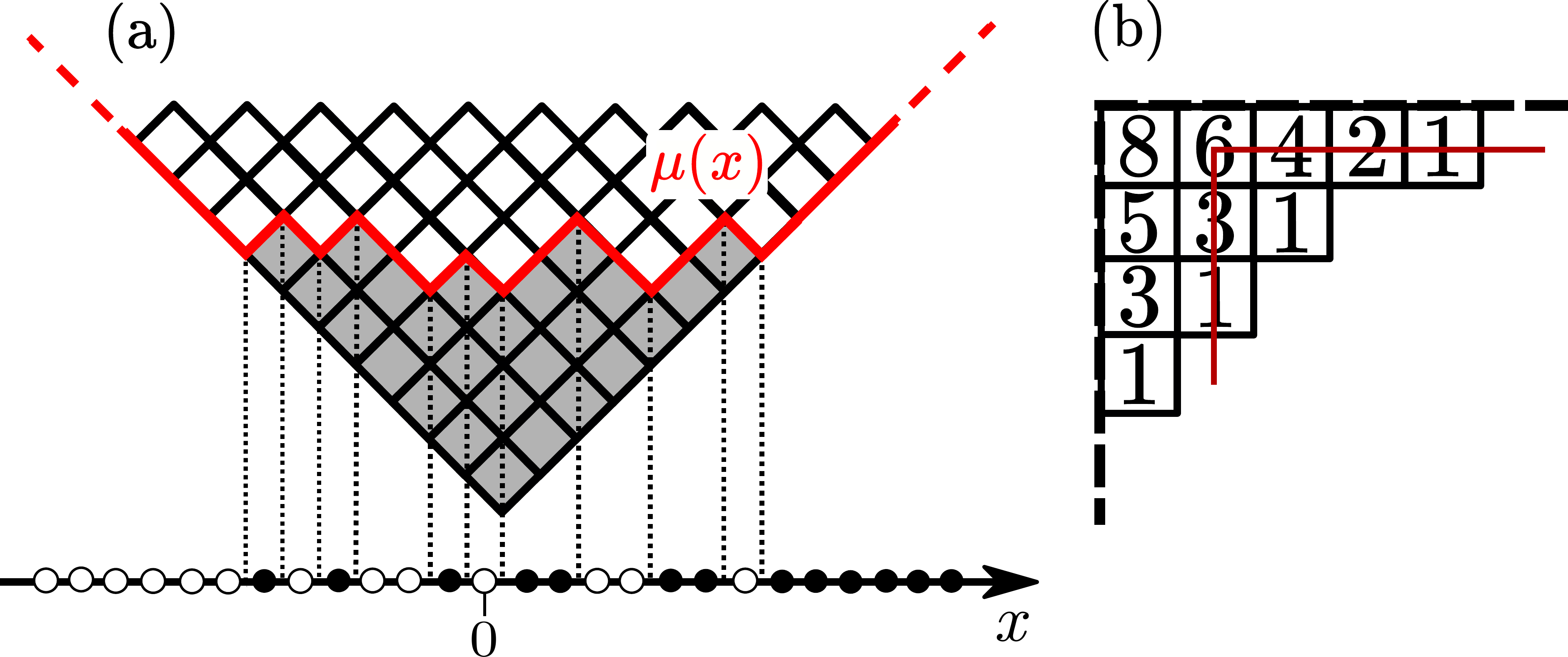}
    \caption{(a) Visual representation of the mapping introduced Sec.~\ref{sec:Lipschitz}, applied to the corner configuration. The squares highlighted in gray correspond to the ``eroded area'', i.e.\ to spins belonging to the corner which have been flipped from $\downarrow = \Box$ to $\uparrow = \gsquare$ due to the dynamics, forming a Young diagram. The interface $\mu(x)$ is highlighted in red.
    (b) An example of Young diagram with the hook $\mathfrak{h}(\square)$ indicated for each of its elements $\square$. The hook of a box $\square$ is obtained by summing the number of boxes below it and to its right (highlighted in red for a specific box in the figure), plus one (corresponding to the box itself).}
    \label{fig:mapping_lattice}
\end{figure}

In the previous Section we have studied two particular cases---the strip-like configuration and a Lipschitz interface---for which the dynamical constraints emerging at infinite $J$ significantly simplify the evolution of the interface, which then can be described in terms of an equivalent 1$d$ model. In this Section, we specialize the generic case discussed in Sec.~\ref{sec:Lipschitz}, by considering an interface shaped as in Fig.~\hyperref[fig:mapping_lattice]{\ref{fig:mapping_lattice}a}, which is composed of two straight lines (parallel to the lattice directions) and a single, right-angled corner. This interface is Lipschitz-continuous in the sense of Eq.~\eqref{eq:Lipschitz}, and therefore the approach outlined in Sec.~\ref{sec:Lipschitz} can be applied.

The case of a corner-shaped interface is particularly instructive, because of several connections to other fields of physics and mathematics:
\begin{enumerate}[wide=0pt]
    \item  Its evolution can be thought of as the quantum counterpart of \emph{corner growth models} studied in classical, non-equilibrium statistical mechanics~\cite{richardson1973random,Rost1981Non,Krug1991Kinetic,Spohn2006Exact,Krapivsky2004Dynamics}. These models describe the process of erosion of crystals; the case considered here extends the investigation of the melting phenomenon to \emph{quantum} crystals~\cite{Dijkgraaf2009Quantum,araujo2021quantum}. In fact, while a flat interface (of the type considered in Sec.~\ref{sec:Lipschitz}) can only fluctuate around its initial position, the corner configuration can be eroded indefinitely---if no other localization mechanism is present, as we will discuss below (see also Ref.~\cite{Balducci2022Slow}). However, in comparing the quantum to the classical case one should bear in mind that, for the quantum model under consideration, the addition/removal of a block from the corner (i.e.\ a spin flip) is always a \emph{coherent} process, while in the classical problems the removed blocks ``dephase'' in the liquid state before being possibly reattached to the solid. 
    
    It is also interesting to notice the following feature. According to the stochastic dynamics, which is usually implemented for the \emph{classical} Ising model (corresponding to Eq.~\eqref{eq:Ising2d_ham} with $g=0$), the possible transitions between different spin configurations occur with a rate which is biased by $\exp(-\Delta E/T)$ (in a specific way that depends on the algorithm), where $\Delta E$ is the energy difference between the final and the initial configuration and $T$ the temperature of the bath. This implies, as expected on physical ground, that at zero temperature $T=0$ the possible transitions are those with $\Delta E \le 0$. Assuming that the stochastic dynamics proceeds via randomly flipping single spins (as the coupling $\propto g$ does in the quantum case), this implies that the allowed classical spin moves can be represented analogously to Eq.~\eqref{eq:H_PXP_graphical} as 
    \begin{enumerate}
        \item $\ulmove \leftrightarrows \drmove$, $\dlmove \leftrightarrows \urmove$ and $\hhmove \leftrightarrows \vvmove$ for the fully reversible transitions with $\Delta E =0$ (or, more generally, $\Delta E = o(J)$ for $h=0$). These are the moves contained in Eq.~\eqref{eq:H_PXP_graphical}.
        
        \item $\treb \rightarrow \unb$, its spatial $\pi/2$ rotations, and 
        $\quatb \rightarrow \cdot\ $. These moves, occurring as indicated by the arrows, are not reversible and correspond to $\Delta E < 0$, with $\Delta E = O(J)$.
    \end{enumerate}
    Moves of type (b) are not present in Eq.~\eqref{eq:H_PXP_graphical}. However, it is easy to realize that, when considering an initial state with an interface in the form of a corner or, more generally of a Lipschitz function, these moves as well as the third type of moves in (a) are inconsequential, making the classical and the quantum dynamics actually explore the same set of configurations. In a heuristics sense, they share the same Krylov space of configurations in the $\sigma^z$-basis. As a consequence, the mapping discussed in Sec.~\ref{sec:Lipschitz} and in Fig.~\ref{fig:interface_mapping} for the quantum interface can be applied also to the classical interface. This was done, e.g., in Refs.~\cite{Krapivsky2004Dynamics,Krapivsky2013Limiting,Krapivsky2012Limiting,Krapivsky2021Stochastic}. The corresponding \emph{classical} model is characterized by the classical equivalent of the fermionic statistics, i.e.\ by the constraint of exclusion in the occupation number of each lattice site which can be at most one, making it belonging to the general class of \emph{simple exclusion processes} (SEPs) \cite{Rost1981Non,Liggett1999Stochastic,Spohn2006Exact}. In the absence of the external field $h=0$, the only allowed transitions starting from a corner (see Fig.~\hyperref[fig:mapping_lattice]{\ref{fig:mapping_lattice}a} for the conventions) are $\downflip \rightarrow \upflip$ and its reversed $\upflip \rightarrow \downflip$, corresponding to flipping a spin inside a corner from its two possible initial states. Such moves have the same rate, and therefore each classical particle in $1d$ attempts jumps to the left or to the right empty neighbouring sites with the same rate, resulting in the so-called \emph{symmetric simple exclusion process} (SSEP). Due to the intrinsic (unbiased) diffusive nature of ther dynamics, the growth of the interface turns out to be diffusive, while it is ballistic in the quantum case, as discussed further below. For $h\neq 0$, on the other hand, the transition $\downflip \rightarrow \upflip$ and its reversed $\upflip \rightarrow \downflip$ occur with different rates, depending on the sign and magnitude of $h/T$. In particular, for $T=0$ it turns out that the only allowed moves are $\upflip \rightarrow \downflip$ for $h<0$ and  $\downflip \rightarrow \upflip$ for $h>0$.  This corresponds to the classical particle jumping only towards the empty neighbouring site to the left or to the right depending on having $h>0$ or $h<0$, i.e.\ to the so-called \emph{totally asymmetric simple exclusion process} (TASEP). This model turns out to displays generically a ballistic growth (see, e.g.\ Ref.~\cite{Krapivsky2004Dynamics}), while the quantum dynamics is actually localized for $h \neq 0$. In addition, also the resulting limit shape is different: we discuss this aspect in more detail further below in this Section.

    \item Each configuration which is dynamically connected to the corner corresponds to a \emph{Young diagram}, as detailed in Sec.~\ref{sec:Young_diagrams}. In particular, we will show an interesting connection between two seemingly unrelated measures on Young diagrams: the probability density of the quantum-fluctuating interface, which naturally emerges in the context of the $2d$ Ising model, and the Plancherel measure, commonly studied in representation theory~\cite{Logan1977Variational,Vershik1977AsymPlancherel,Vershik1985AsymMaximal,Okounkov2000Random,Borodin2000Asymptotics,Okounkov2001Infinite,Okounkov2006Quantum}. 

    \item Lastly, it is worth mentioning that the case of a Lipschitz interface, and in particular of a corner, the mapping to free fermions points to an explicit form of \emph{holography}: a two-dimensional quantum problem in strong-coupling limit is mapped to a free, simpler problem in one less spatial dimension. This is reminiscent of the AdS/CFT duality~\cite{maldacena1999large,witten1998anti,gubser1998gauge}: the interface in the Ising model is the string in two spatial dimensions (plus time as an additional coordinate), while the non-interacting fermions on the chain are the dual field theory. When the string tension $J$ is large, the corresponding field theory is free. When the string tension decreases, the field theory becomes interacting and, in our case, no longer integrable.
    
    However, in order to discuss the melting of a bubble and not of a simple corner, one has necessarily to introduce a more complicated theory of fermions, possibly with many species. It worth noticing that going back further in time, one finds other connections between the Ising model and string theory, for instance the conjecture that the $3d$ Ising model should be dual to a weakly-coupled string theory~\cite{Fradkin1980,POLYAKOV1981211} (for a recent discussion see Ref.~\cite{Iqbal2020}), although that is supposed to hold only at the critical point. 
   
\end{enumerate}

Before passing on, we finally notice that the initial condition for the $2d$ Ising model discussed in this Section actually corresponds to a single domain wall on the fermionic chain, which separates the filled part of the chain from the empty one. Let us emphasize that, for $h \neq 0$, this initial configuration is close to the boundary of the spectrum of the Hamiltonian within the Krylov sector it belongs to. Indeed, such configuration maximizes (or minimizes, depending on the sign of $h$) the expectation value of the Hamiltonian, being the state of maximal area of its Krylov sector. While in the limit $J=\infty$ this observation in marginal, as the system is integrable (thus any initial configuration leads to a non-ergodic behavior), it becomes relevant at finite $J$, where the behavior of states at the middle of the spectrum can be also qualitatively different from the ones at the edges. This observation will be relevant in Secs.~\ref{sec:1st_order_corrections}--\ref{sec:Stark_MBL} when discussing the finite-$J$ corrections.

\subsection{Average of the interface and its continuum limit}
\label{sec:av_interface}

In the language of Sec.~\ref{sec:Lipschitz}, the corner-shaped initial state corresponds on the fermionic chain to
\begin{equation}
\label{eq:domain_wall}
    \ket{\Psi_0}=\prod_{x>0} \psi^{\dagger}_x \ket{0},
\end{equation}
with a domain wall separating the empty half-chain for $x\le 0$ from the completely filled one at $x\ge 1$. In the language of electronics, this would be called a ``maximum voltage bias'' Fermi sea. By applying the approach previously illustrated (in particular Eq.~\eqref{eq:av_density}), one easily finds that the average density profile on the chain is given by
\begin{equation}
\label{eq:corner_av_n}
    \av{n(x,t)} = \sum_{y<x} J^2_y \left(\omega_t \right).
\end{equation}
Summing over space (see Eq.~\eqref{eq:mu_from_n}), one obtains the average interface profile
\begin{equation}
\label{eq:corner_av_shape}
    \av{\mu(x,t)} = 2\sum_{y\leq x} (x-y) J^2_y \left(\omega_t \right) - x,
\end{equation}
which, as anticipated, displays periodic oscillations with period $\pi/|h|$ at each position $x$.

As discussed in Sec.~\ref{sec:Lipschitz-cont}, in order to determine the continuum limit of Eqs.~\eqref{eq:corner_av_n} and \eqref{eq:corner_av_shape} it is then sufficient to replace $\gamma$ by $\gamma/a$ and therefore $\omega_t$ by $\omega_t/a$ in Eqs.~\eqref{eq:corner_av_n}--\eqref{eq:corner_av_shape}, after reinstating the lattice spacing $a$. Then the limit $a\to 0$ can be determined as explained in Sec.~\ref{sec:Lipschitz-cont}. Alternatively, one can specialize the general prediction in Eq.~\eqref{eq:n-cont-gen-2} to
the corner considered above, which corresponds to having, in the continuum,
\begin{equation}
\begin{split}
    \hat\rho(Y) &= a \sum_{k=0}^{+\infty} \delta(Y - a k) \\
    & \stackrel{a\to 0}{\longrightarrow} \int_0^{+\infty}\!\!d\xi\, \delta(Y-\xi) = \theta(Y),
\end{split}
\end{equation}
i.e.\ a homogeneous spatial density of fermions equal to $1$ for $Y\ge 0$ and an empty lattice for $Y<0$. A straightforward integration leads to
\begin{equation}
\label{eq:num_density}
    \av{n(X,t)} =
    \begin{cases}
        0      & \mbox{for} \quad X \leq - \omega_t, \\[1mm]
    \displaystyle \frac{1}{2} + \frac{1}{\pi}\arcsin (X/\omega_t)
        & \mbox{for} \quad |X| < \omega_t, \\[1mm]
        1       & \mbox{for} \quad  X \ge \omega_t,
    \end{cases}
\end{equation}
which, for $h=0$, agrees with the prediction of Ref.~\cite{Antal1999Transport} for free fermions. Integrating over $X$, one finds
\begin{equation}
\label{eq:mu_Omega}
    \av{\mu(X,t)} = \omega_t \, \Omega(X/\omega_t),
\end{equation}
with
\begin{equation}
\label{eq:Omega}
    \Omega(v)=
    \begin{cases}
        |v| & \mbox{for} \quad |v| \geq 1, \\
        \frac{2}{\pi} \left(\sqrt{1-v^2} + v \arcsin v \right) & \mbox{for} \quad |v| < 1.
\end{cases}
\end{equation}
Alternatively, this expression can be derived directly from Eq.~\eqref{eq:mu-cont-gen-1}, by using $\mu_0(X) = [2 \theta(X)-1] X$ as the initial condition. 

Equations~\eqref{eq:mu_Omega} and \eqref{eq:Omega} can be easily generalized to the case of a Lipschitz corner in which, however, the slopes of the interface in its two sides are not the same. On the lattice, this corresponds to having a certain average density of fermions on the left of the origin and a different one on its right. In fact, consider an initial profile which is linear for both $X<0$ and $X>0$, but with two different slopes $\alpha_-$ and $\alpha_+$, respectively, and that fulfils $\mu_0(X=0)=0$. Such a profile must take the form
\begin{equation}
    \mu_0(X) = [\alpha_- \theta(-X) + \alpha_+  \theta(X)]X.
\end{equation}
The right-angled corner considered above corresponds to $\alpha_\pm = \pm 1$. With this $\mu_0(X)$,  Eq.~\eqref{eq:mu-cont-gen-1} implies that
\begin{equation}
    \av{\mu(X,t)} = \frac{\alpha_++\alpha_-}{2} X + \frac{\alpha_+ - \alpha_-}{2} \omega_t \Omega(X/\omega_t),
\end{equation}
where $\Omega(v)$ is given by Eq.~\eqref{eq:Omega}. In fact, this expression simply follows from the linearity of the equation and from the result reported above for the right-angled corner.

Remarkably, the function $\Omega(v)$ in Eq.~\eqref{eq:Omega} first appeared in the context of random Young diagrams~\cite{Logan1977Variational,Vershik1977AsymPlancherel,Vershik1985AsymMaximal}; we will elaborate more on this point in Sec.~\ref{sec:Young_diagrams}. Here, instead, we comment on the connection between the dynamics studied above and the \emph{classical} melting processes which were mentioned at point 1. of the introduction to this Section. In fact, in the cases of the SSEP or the TASEP, the stochastic dynamics starting from a completely filled half-line---corresponding to the dynamics at zero temperature of a corner in the Ising model---can be solved, obtaining the large-time behaviour of the density of particles (briefly reported in App.~\ref{app:sec:classical}). As anticipated at the beginning of this Section, the SSEP is, in a sense, the classical analogue of the quantum dynamics with $h=0$, while the TASEP of the dynamics with $h \neq 0$. It turns out, however, that the scaling functions describing the erosion of the corner (which occurs diffusively for SSEP and ballistically for TASEP), via the same mapping described in Sec.~\ref{sec:Lipschitz}, have a different functional form compared to $\Omega$ of Eq.~\eqref{eq:Omega} (see App.~\ref{app:sec:classical}). This fact highlights how the quantum and classical dynamics turn out to be quantitatively and qualitatively different in spite of their many similarities. In Sec.~\ref{sec:Young_diagrams} we will discuss how this difference emerges also in terms of \emph{concentration of probability measures}, showing that a simple entropic argument concerning the accessible configurations is not sufficient for explaining the limiting shapes of the interfaces, but that, as expected, also the classical or quantum nature of the underlying microscopic dynamics matters.

\subsection{Fluctuations of the interface}
\label{sec:fluct_interface}

\begin{figure}
    \centering
    \includegraphics[width=0.9\columnwidth]{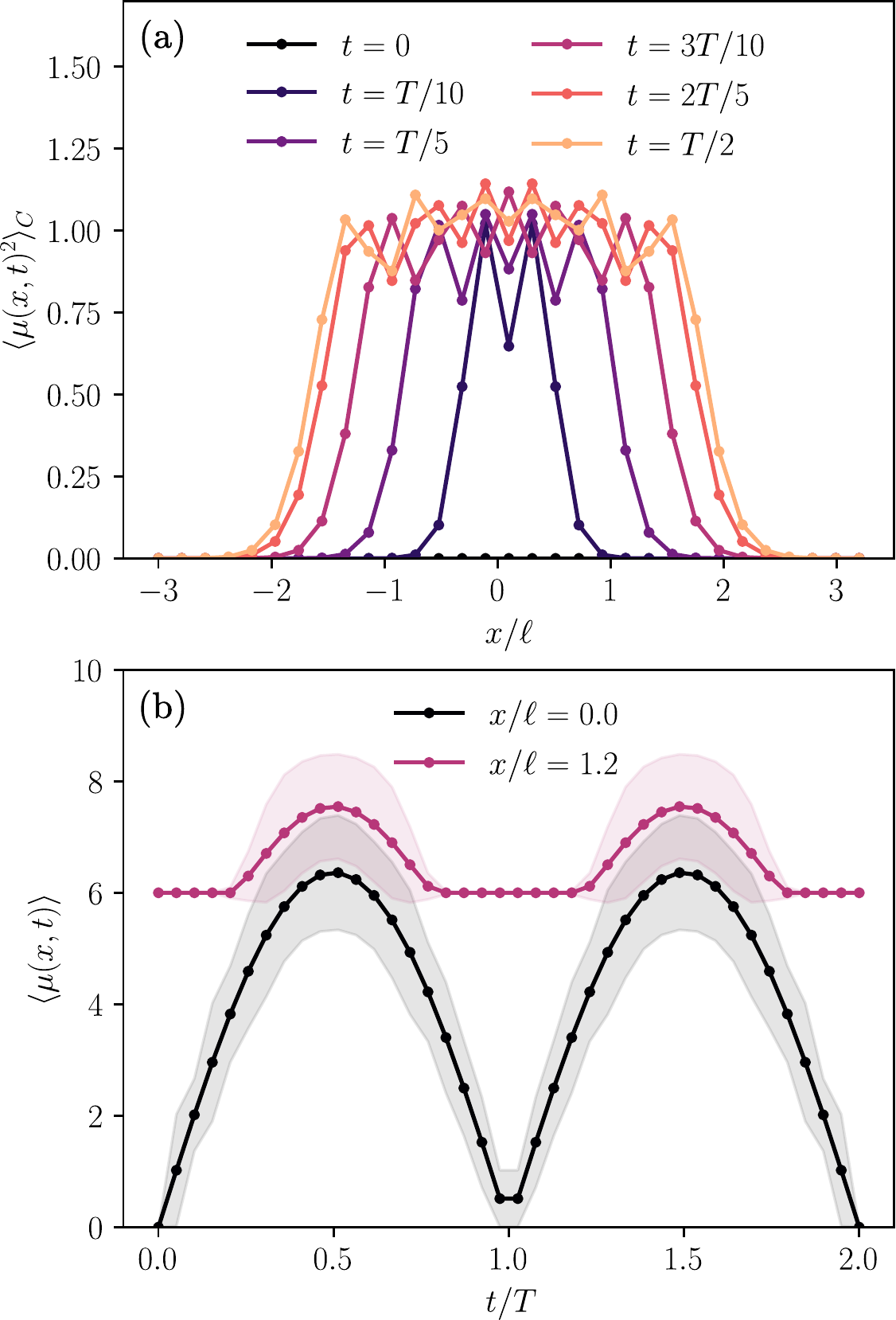}
    \caption{Fluctuations of the interface profile, quantified via $\av{\mu^2(x,t)}_C$, starting from the infinite corner at $t=0$.
    (a) Space dependence of $\av{\mu^2(x,t)}_C$ for various values of the time $t$ within a half period. Because of the presence of the linear external potential also the fluctuations of the interface periodically return to the original value.
    (b) Time dependence of the average interface position $\av{\mu(x,t)}$ (solid line) and of the corresponding fluctuations (shaded area), for two different values of the position.} 
    \label{fig:fluct_mu}
\end{figure}

The approach described in the previous Section allows one to determine not only the average position of the quantum-fluctuating interface, but also its fluctuations. While presenting the complete calculation in App.~\ref{app:2point_funct}, we report here the final result for the connected two-point function of the density:
\begin{equation}
\label{eq:corner_2pf}
    \av{n(x,t) n(y,t)}_C
    = \delta_{xy} \sum_{i>0}J_{i-x}^2(\omega_t) - \mathcal{B}^2(x,y; \omega_t),
\end{equation}
where we introduced the \emph{Bessel kernel}
\begin{equation}
\label{eq:Bessel_kernel}
    \mathcal{B}(x,y; \omega) := \omega \frac{J_{x-1}(\omega) J_{y}(\omega) - J_x(\omega) J_{y-1}(\omega)}{2(x-y)}.
\end{equation}
Note that, for $x=y$, Eq.~\eqref{eq:corner_2pf} straightforwardly reduces to
\begin{equation}
    \av{n^2(x,t)}_C = \av{n(x,t)} \big[ 1-\sav{n(x,t)} \big],
\end{equation}
which is actually expected for fermionic particles. Summing over $x$ and $y$ in Eq.~\eqref{eq:corner_2pf}---thus applying the prescription of Eq.~\eqref{eq:mu_from_n}---leads to the connected 2-point function of the interface profile $\av{\mu(x,t) \mu(y,t)}_C$ (Eq.~\eqref{app:eq:mu_fluct}). In Fig.~\ref{fig:fluct_mu} we show the fluctuations of the interface profile: in panel (a) we present the value of $\av{\mu^2(x,t)}_C$ as a function of position for different times, while in panel (b) we plot, for two values of $x$ along the chain, the average position of the interface with the corresponding fluctuations, over two periods of oscillation.

It is instructive to discuss the continuum limit also for the fluctuations of the shape $\mu$. As they involve the Bessel kernel in Eq.~\eqref{eq:Bessel_kernel}, they are related to the universal fluctuations found e.g.\ in Laguerre and Jacobi ensembles of random matrices~\cite{Tracy1994LevelBessel} (see also Ref.~\cite{Kuijlaars2011Universality}), and of random representations of the symmetric group~\cite{Borodin2000Asymptotics}. In particular, the presence of the Bessel functions implies a \emph{light-cone structure} for the correlations, see Fig.~\hyperref[fig:EntEntr]{\ref{fig:EntEntr}a}: if either $|x| \gg \omega_t$ or $|y| \gg \omega_t$, the correlations are exponentially suppressed (as follows from the large-index asymptotic behavior of the Bessel function discussed in Eq.~\eqref{app:eq:bessel_large_index}). 
If, instead, both $x,y \ll \omega_t$, then by virtue of the large-argument asymptotics of the Bessel functions presented in Eq.~\eqref{app:eq:bessel_large_argument}, the kernel reduces to the \emph{sine kernel}
\begin{equation}
    \mathcal{S}(x,y) = \frac{\sin \left(\pi (x-y)/2 \right)}{\pi (x-y)}.
    \label{eq:sine_kernel}
\end{equation}
The sine kernel is found in numerous contexts in physics and mathematics, among which gaussian ensembles of random matrices~\cite{Mehta2004Random}, and free fermionic chains \emph{without} a linear potential~\cite{Peschel2009Reduced}. Notice that, in passing from the Bessel kernel to the sine kernel, the explicit dependence on $\omega_t$ has been lost in the expression for the correlations, while it remains implicit in the maximum value attained by $x$ or $y$ (i.e.\ the border of the light cone), see Fig.~\hyperref[fig:EntEntr]{\ref{fig:EntEntr}b}. Finally, let us mention that a less trivial limit emerges in a region of order $\omega_t^{1/3}$ around the light cone, where by means of a uniform expansion (Eqs.~\eqref{app:eq:bessel_uniform_1}--\eqref{app:eq:bessel_uniform_2}) the Bessel kernel reduces to the celebrated Airy kernel~\cite{Tracy1993Level,Tracy1994LevelAiry}.

Despite all the connections mentioned above, we need to emphasize that in this quantum setting the fluctuations are given by the \emph{square} of the Bessel kernel, see Eq.~\eqref{eq:corner_2pf}: accordingly, they are quantitatively different from the cases mentioned above, which involve the kernels at their linear order.

\subsection{Entanglement dynamics}
\label{sec:entanglement}

\begin{figure}[t]
    \centering
    \includegraphics[width=0.9\columnwidth]{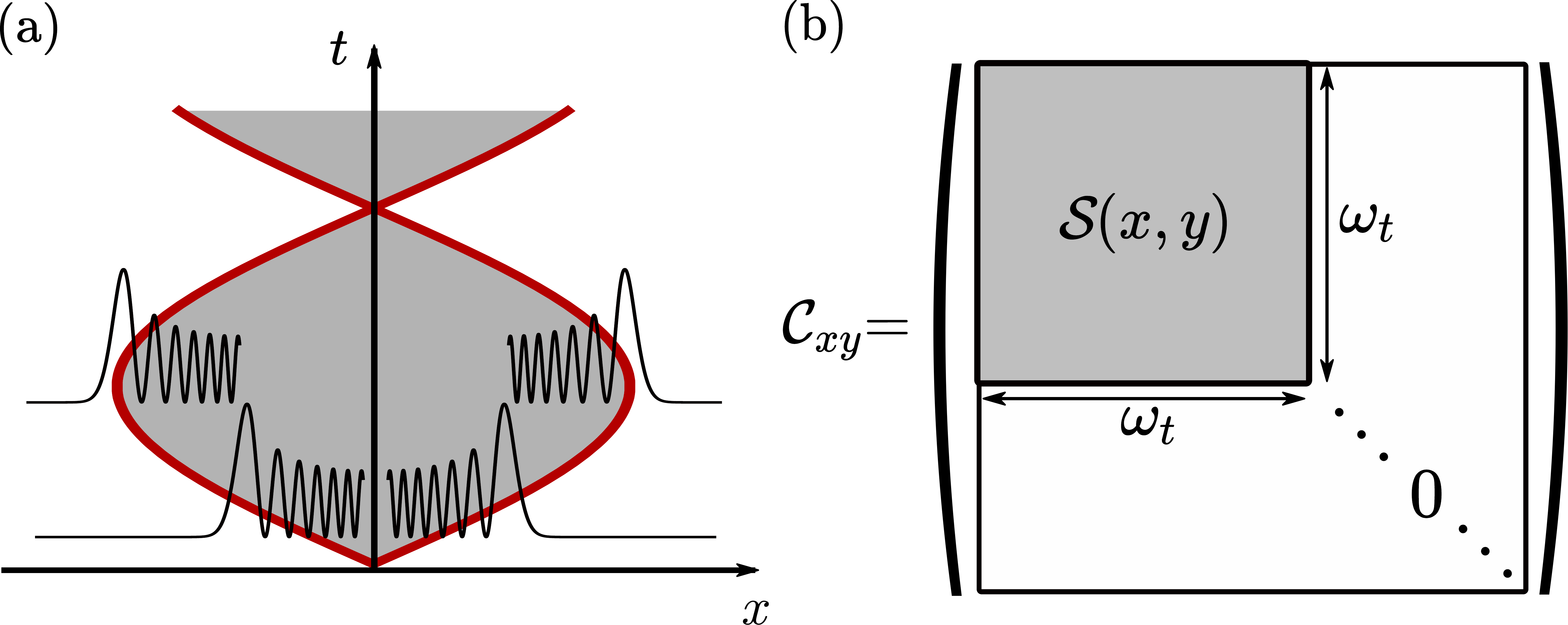}
    \caption{(a) Light-cone structure induced by the presence of the Bessel kernel, see Eqs.~\eqref{eq:corner_2pf} and \eqref{eq:Bessel_kernel}. The red line represents the position of the light cone $x=|\omega_t|$. Inside the light cone, the Bessel functions oscillate with a non-zero average value. It is within this region, in the continuum limit, that the Bessel kernel reduces to the sine kernel of Eq.~\eqref{eq:sine_kernel}. Outside the light cone, instead, the Bessel functions decay exponentially upon moving away from it, and in the continuum limit they approximately vanish. Therefore, if both $x$ and $y$ are inside the light cone, i.e. in the gray area, the Bessel kernel has a non-vanishing value. If at least one of the two points is outside the light-cone, the Bessel kernel approximately vanishes.
    (b) Correlation matrix (see Eq.~\eqref{eq:corrmatr}) in the continuum limit. As discussed in the text, in this regime $\mathcal{C}_{x,y}$ can be set to zero outside the light cone, while inside the light cone the Bessel kernel $\mathcal{B}(x,y;\omega_t)$ can be replaced by the sine kernel $\mathcal{S}(x,y)$. }
    \label{fig:EntEntr}
\end{figure}

The ``holographic'' description of the interface in terms of an integrable 1$d$ model (i.e.\ non-interacting fermions in a linear potential) allows one to extract much more information beyond averages and correlations, using the vast amount of analytical techniques developed in recent years~\cite{korepin1993quantum,giamarchi2003quantum,Doyon202Lecture,Eisler2020Entanglement}. For instance, one can compute the so-called \emph{full counting statistics}, i.e.\ the probability distribution of the fermions, with the techniques of Ref.~\cite{Eisler2013Full}. In fact, it turns out that the predictions of Ref.~\cite{Eisler2013Full} for the case $h=0$ carry over to our case $h \neq 0$ just by replacing $t \to \omega_t/(2g)$; this ``minimal'' substitution is motivated by the fact that in the analytical expressions discussed in Sec.~\ref{sec:Lipschitz} the time dependence occurred only via $\omega_t$ defined in Eq.~\eqref{eq:omega_t}, which encompasses both cases. Similarly, the growth of the entanglement across a bipartition of the lattice can be studied by using the results available for the $1d$ problem \cite{Eisler2009Entanglement}. In particular, one has to partition the $2d$ lattice in two halves by means of a ``vertical'' line (e.g. through the corner, corresponding to the time axis in Fig.~\hyperref[fig:EntEntr]{\ref{fig:EntEntr}a}), so that, on the chain, one has well defined subsystems. At this point, the entanglement of the $2d$ and the $1d$ problems are equal, as there is a one-to-one mapping linking all possible states in the two settings. The entanglement between the two subsystems can be computed as detailed in Ref.~\cite{Eisler2009Entanglement}: from the eigenvalues $\zeta_l(t)$ of the correlation matrix $\mathcal{C}_{xy}(t) := \av{\psi^{\dagger}_x(t) \psi_y(t)}$, the entanglement entropy is obtained as
\begin{equation}
    S_{\mathrm{ent}}(t) = -\sum_{l=0}^{\infty} \Big[ \zeta_l \ln \zeta_l + (1-\zeta_l) \ln (1 - \zeta_l)\Big].
    \label{eq:EE}
\end{equation}
The correlation matrix can be calculated explicitly, by using the properties of the Bessel functions which were used for calculating the average magnetization, with the result that
\begin{equation}
\label{eq:corrmatr}
    \mathcal{C}_{xy}(t) = e^{i\left( \frac{\pi}{2} + ht \right)(y-x)} \mathcal{B} (x,y; \omega_t),
\end{equation}
$\mathcal{B}$ being the Bessel kernel of Eq.~\eqref{eq:Bessel_kernel}. If one computes the entanglement entropy between two subsystems $A$ and $B$, separated by a vertical line in the $2d$ problem, the indices of the correlation matrix $\mathcal{C}_{xy}$ are such that $x,~y \in A$ (or $B$ equivalently). For a bipartition located in $0$, one has $x,~y >0$. Let us notice that the phase factor in the last equation does not affect the entanglement entropy; in fact, it can be removed via a unitary transformation. Accordingly, it is clear that the correlation matrix $\mathcal{C}_{xy}(t)$ (and thus $S_{\mathrm{ent}}(t)$) is a periodic function of time with period $|h|/\pi$, as its time dependence is only through $\omega_t$. Even if, to our knowledge, the eigenvalues of the correlation matrix of Eq.~\eqref{eq:corrmatr} cannot be obtained analytically, some analytical progress can be made in the continuum limit~\cite{Peschel2004OnReduced,Rottoli2022Entanglement}. Let us introduce the entanglement Hamiltonian $\mathcal{H}_A$ such that
\begin{equation}
\label{eq:ent_ham}
    \rho_A = \mathcal{K}_A e^{-\mathcal{H}_A},
\end{equation}
being $\rho_A$ the reduced density matrix of a subsystem $A$, and $\mathcal{K}_A$ a normalization constant. With this definition, one finds~\cite{Peschel2003Calculation,Peschel2004OnReduced,Cheong2004ManyBody}
\begin{equation}
\label{eq:enthamcorr}
    \mathcal{H}_A = \ln\left( \frac{1-\mathcal{C}_A}{\mathcal{C}_A} \right),
\end{equation}
where $\mathcal{C}_A$ is here the correlation matrix restricted to positions belonging to the considered subsystem $A$. This means that $\mathcal{H}_A$ and $\mathcal{C}_A$ are diagonal in the same basis, and the corresponding eigenvalues satisfy the relation in Eq.~\eqref{eq:enthamcorr}. 

\begin{figure}
    \centering
    \includegraphics[width=0.9\columnwidth]{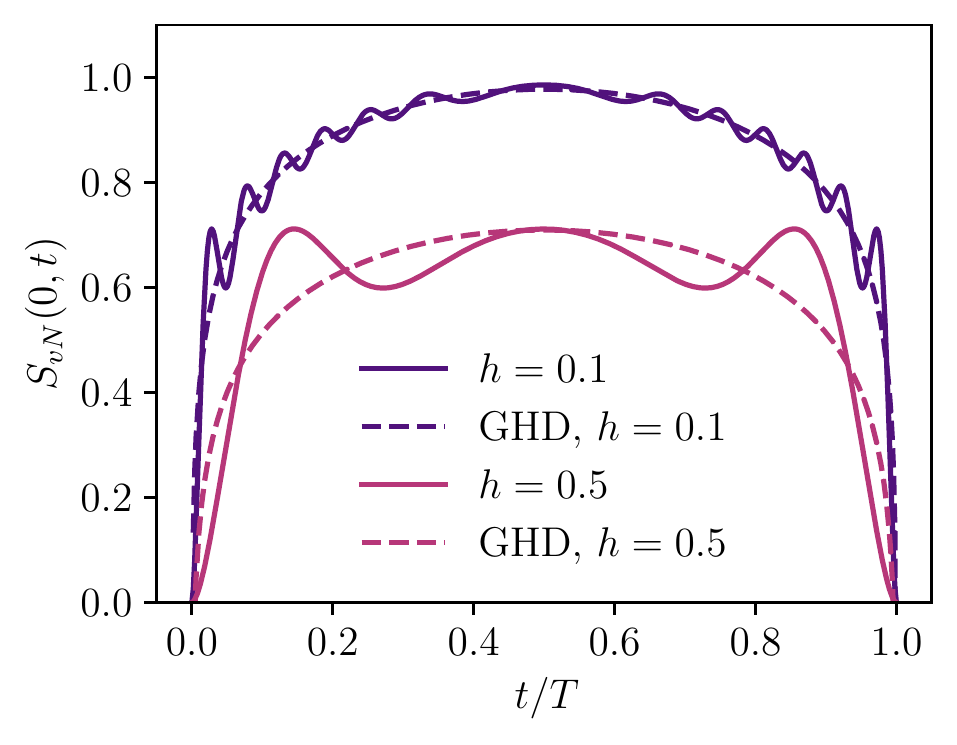}
    \caption{The plot shows the time dependence of the entanglement entropy for two different values of the external field $h$, over one period $T=\pi/h$, with $g=1$. The solid lines represent the entanglement entropy computed via diagonalization of the correlation matrix, according to Eq.~\eqref{eq:EE}. In both cases, the diagonalization has been performed on a chain of length $L=100 \gg \ell =g/h$. The dashed lines represent the prediction given by GHD, reported in Eq.~\eqref{eq:ent_hydro}. One can clearly see that, for smaller values of $h$, i.e. closer to the continuum limit, the agreement between numerical diagonalization and GHD improves.
    }
    \label{fig:ent_corrmatr}
\end{figure}

As discussed also in Sec.~\ref{sec:fluct_interface}, the Bessel kernel reduces to the sine kernel in the continuum limit inside the light cone. In this regime one can thus approximate the correlation matrix $\mathcal{C}_{xy}$ by setting to zero the entries with $x,~y \gtrsim \omega_t$, and therefore one is left with an effective matrix of size $\omega_t \times \omega_t$, as depicted in Fig.~\hyperref[fig:EntEntr]{\ref{fig:EntEntr}b}. Thanks to this approximation, one can obtain the eigenvalues $\epsilon_k$ of $\mathcal{H}_A$ as~\cite{Peschel2004OnReduced,Peschel2009Reduced}
\begin{equation}
    \epsilon_k(t) = \pm \frac{\pi^2}{2 \ln \omega_t} ( 2 k + 1 ),
    \label{eq:EE-eps}
\end{equation}
with $k = 0,\, 1,\, 2,\dots$. Denoting by $\zeta_k$ the eigenvalues of $\mathcal{C}_A$, one has from Eq.~\eqref{eq:enthamcorr}
\begin{equation}
    \zeta_k = \frac{1}{e^{\epsilon_k}+1}.
    \label{eq:zeta_k}
\end{equation}
Note, however, that the asymptotic value Eq.~\eqref{eq:EE-eps} needs a very large $\omega_t$ to be accurate. For smaller values of $\omega_t$, the eigenvalues vary as $1/(\ln{\omega_t} + b_k)$ rather than $1/\ln{\omega_t}$, where $b_k$ are constants depending on the specific eigenvalue~\cite{Peschel2004OnReduced}. The evolution of the entanglement entropy can now be determined by using Eqs.~\eqref{eq:zeta_k} and \eqref{eq:EE-eps} in Eq.~\eqref{eq:EE}. In Fig.~\ref{fig:ent_corrmatr} we show the numerical evaluation of the time evolution of the entanglement entropy, according to Eq.~\eqref{eq:EE}, for various values of $h$. The presence of a non-vanishing external field implies that also the entanglement entropy is periodic in time.

We have shown how the mapping of the original $2d$ problem onto a $1d$ chain can be used in order to calculate the half-system entanglement entropy. However, the computation was possible only because of a convenient choice of the bipartition of the $2d$ lattice (i.e.\ a vertical one in the rotated frame): more general bipartitions of the $2d$ lattice would instead map non-locally on the chain. It seems that computing the entanglement of the $2d$ system using the mapping into $1d$ is viable as long as the cut along which the entanglement is computed is parallel to the projection performed in the mapping itself.

As a final point of this Section, it is worth noticing that the above results, valid in general on the lattice for arbitrary values of the couplings $g$ and $h$, reduce, in the continuum limit $h \ll g$, to the predictions of conformal field theory in curved space~\cite{Dubail2017Conformal,Tonni2018Entanglement} or quantum generalized hydrodynamics (GHD) \cite{Scopa2022Exact}. The entanglement entropy is in fact given by
\begin{equation}
\label{eq:ent_hydro}
    S(x,t) = \frac{1}{6} \ln \left[ \omega_t \left(1 - \frac{x^2}{\omega_t^2}\right)^{3/2} \right] + c,
\end{equation}
where $c \simeq 0.475$ and $x$ is the position of the bipartition. Equation~\eqref{eq:ent_hydro} is clearly valid for $|x|\leq \omega_t$; otherwise, the entanglement entropy is zero because of the light-cone structure. In Fig.~\ref{fig:ent_corrmatr} we also compare the prediction given by Eq.~\eqref{eq:ent_hydro} with the results of the exact diagonalization on the lattice, showing that a good agreement is attained for small values of $h$, as expected. This relation was derived in Ref.~\cite{Scopa2022Exact} for $h=0$; the general case is obtained by means of the minimal substitution $2g |t| \to \omega_t$, coming from Eq.~\eqref{eq:omega_t}. In passing we mention that the GHD formalism allows one to predict the dynamics of one-dimensional integrable quantum systems directly in the continuum limit, even when the system is interacting; this is the reason why one needs the limit $h \ll g$ to match the GHD prediction.

\subsection{Connection with the asymptotics of the Plancherel measure}
\label{sec:Young_diagrams}

As pointed out at the beginning of Sec.~\ref{sec:corner}, the states in the Krylov sector connected with the infinite corner are in one-to-one correspondence with \emph{Young diagrams} (also known as \emph{Ferrers diagrams}). By definition, a Young diagram is a collection of boxes, arranged in a sequence of left-justified rows of non-increasing length~\cite{Fulton1996Young}. Young diagrams are a graphical tool commonly used to represent integer partitions, to compute dimensions of group representations, and for many other mathematical purposes~\cite{Fulton1996Young}.

In order to discuss the connection with the present work, let us recall here some basic facts concerning Young diagrams. A partition $\lambda = (\lambda_1 \geq \lambda_2 \geq \dots \geq \lambda_n \geq 0)$ of an integer $N$ indicates a possible decomposition of $N$ as a sum of $n$ positive integers, i.e.
\begin{equation}
\label{eq:partition}
    |\lambda| := \sum_{k=1}^n \lambda_k = N.
\end{equation}
Representing each integer $\lambda_k$ as a string of $\lambda_k$ adjacent boxes $\square \square \cdots \square$, one can easily see that a partition corresponds to a Young diagram, obtained by stacking all the $n$ strings, starting from the first. A theorem~\cite{Fulton1996Young} states that the irreducible representations of the symmetric group $S_N$ of degree $N$ are labelled by the possible partitions $\lambda$ of size $|\lambda| = N$. Moreover, the dimension of the representation corresponding to a certain $\lambda$ can be obtained via the \emph{hook length formula}
\begin{equation}
    \dim (\lambda) = \frac{|\lambda|!}{\prod_{\square \in \lambda} \mathfrak{h}(\square)},
\end{equation}
where $\mathfrak{h}(\square)$ is the so-called \emph{hook} of the square $\square$~\cite{Fulton1996Young}, an integer number determined as explained in Fig.~\hyperref[fig:mapping_lattice]{\ref{fig:mapping_lattice}b}.

For our purposes, the most interesting interpretation of $\dim(\lambda)$ resides in the fact that it gives the number of ways in which the diagram $\lambda$ can be constructed, starting from the empty diagram, by adding one square at a time in such a way that at each step one still has a partition~\cite{Okounkov2003Uses}. In the mathematical literature, it is common to define the \emph{Plancherel measure} on the set of partitions as~\cite{Logan1977Variational,Vershik1977AsymPlancherel,Vershik1985AsymMaximal,Okounkov2000Random,Borodin2000Asymptotics}
\begin{equation}
    \mu_{\mathrm{P}}(\lambda) := \frac{[\dim(\lambda)]^2}{|\lambda|!},
    \label{eq:mu-P}
\end{equation}
which is proved to be a normalized measure, i.e.\ a probability~\cite{Vershik2003TwoLectures}.

An important result of combinatorics is that the Plancherel measure $\mu_{\mathrm{P}}$ \emph{concentrates} at large $N$, i.e. it becomes a delta function on a particular set of diagrams~\cite{Logan1977Variational,Vershik1977AsymPlancherel,Vershik1985AsymMaximal,Okounkov2000Random,Borodin2000Asymptotics}. The diagrams belonging to this set have approximately the same shape; more precisely, after a $3\pi/4$ counterclockwise rotation of the diagrams (such that they are finally arranged as in Fig.~\hyperref[fig:mapping_lattice]{\ref{fig:mapping_lattice}a}) their shape is actually described by the function $\sqrt{N}\Omega(x/\sqrt{N})$ with $\Omega(v)$ given in Eq.~\eqref{eq:Omega}. It is thus quite surprising to find another, completely different growth process that leads to the same limiting shape as the one induced by the quantum dynamics of the $2d$ Ising model.

While we could not devise a mathematically rigorous proof, we heuristically understand the above correspondence as follows. Recalling that $\dim(\lambda)$ gives the number of paths that reach the diagram $\lambda$ from the empty one, always remaining within the set of Young diagrams, we notice that the Plancherel measure $\mu_{\mathrm{P}}$ weights each diagram with the \emph{square} of the number of paths. On the other hand, one can consider the Green's function 
\begin{equation}
    G(\lambda', \lambda; E) = \bra{\lambda'} \frac{1}{E-H_g} \ket{\lambda},
\end{equation}
where we denoted with $H_g$ the Hamiltonian $H_{\mathrm{PXP}}$ in Eq.~\eqref{eq:H_PXP_graphical}, making explicit the dependence on $g$, and $\lambda$ and $\lambda'$ are two Young diagrams. Performing the locator expansion of the resolvent~\cite{aizenman2015random,Pietracaprina2016Forward,Scardicchio2017Perturbation,Balducci2022Slow}
\begin{equation}
\label{eq:GreenFunct}
    G(\lambda, \lambda'; E) 
    = \frac{\delta_{\lambda \lambda'}}{E-E_{\lambda'}} + \frac{1}{E-E_{\lambda'}} \sum_{p \in \mathrm{P}(\lambda',\lambda)} \prod_{k=1}^{|p|} \frac{g}{E-E_{p_k}},
\end{equation}
where $\mathrm{P}(\lambda',\lambda)$ denotes the set of paths in configuration space from $\lambda'$ to $\lambda$, $|p|$ is the length of the path $p$ and we introduced the notation $H_{g=0}\ket{\lambda'} = E_{\lambda'} \ket{\lambda'}$, i.e. $E_{\lambda}$ denotes the energy of $\ket{\lambda}$ in the absence of hopping ($g=0$). In the spirit of the forward approximation~\cite{Pietracaprina2016Forward,Scardicchio2017Perturbation,Balducci2022Slow}, one can approximate the sum in Eq.~\eqref{eq:GreenFunct} by reducing $\mathrm{P}(\lambda',\lambda)$ to $\mathrm{SP}(\lambda',\lambda)$, i.e.\ the set of \emph{shortest} paths from $\lambda'$ to $\lambda$. This corresponds to work at the lowest order in the hopping $g$. Under this assumption, the argument of the sum does no longer depend on the specific path, but only on its length $d(\lambda',\lambda)$, because all the diagrams with a fixed number of blocks, viz. at the same distance form the empty diagram, have the same energy $E_\lambda = -h|\lambda|$, see Eq.~\eqref{eq:H_PXP_graphical}. This means that the sum gives the number of shortest paths from $\lambda'$ to $\lambda$ (with $\lambda \neq \lambda'$, otherwise it gives zero). Specializing Eq.~\eqref{eq:GreenFunct} to the case of the path from the empty diagram $\lambda' = 0$ (with $E_{\lambda' = 0} = 0$) to $\lambda$ ($\neq 0$), one finds
\begin{equation}
\label{eq:GreenFunct2}
\begin{split}
    G(\lambda, 0; E) &= \frac{\dim(\lambda)}{E} \prod_{k=1}^{d(0,\lambda)} \frac{g}{E + h k}\\
    &= \frac{\dim(\lambda)}{E} \left(\frac{g}{h}\right)^{|\lambda|}\frac{\Gamma\left(1 + E/h\right)}{\Gamma\left(1 + E/h + |\lambda|\right)},
\end{split}
\end{equation}
where, in the second line, we used the fact that $d(0,\lambda) = |\lambda|$. Taking the residue of this propagator at $E=0$, one finds the expression of the corresponding eigenfunction 
\begin{equation}
\label{eq:psi_FA}
    \psi_{E=0}(\lambda) = \frac{\dim(\lambda)}{|\lambda|!} \left(\frac{g}{h}\right)^{|\lambda|}.
\end{equation}
Accordingly, the probability $|\psi_{E=0}(\lambda)|^2$ of being in the state $\ket{\lambda}$ turns out to be proportional to $[\dim(\lambda)]^2$---i.e.\ to the square of the number of paths leading to it, according to the interpretation of $\dim(\lambda)$---and therefore to the Plancharel measure $\mu_P(\lambda)$ in Eq.~\eqref{eq:mu-P}. This motivates the connection between the quantum dynamics and the Plancherel measure concentration.

Before passing to the next Section, it is interesting to note that the forward approximation also gives the correct result for the decay of the eigenfunctions upon increasing $|\lambda|$. To see this, one must plug in Eq.~\eqref{eq:psi_FA} the value of $\dim(\lambda)$, which clearly depends on the specific form of the diagram associated with the state $\ket{\lambda}$. Referring for details to Ref.~\cite{Vershik1985AsymMaximal}, we just say here that it is possible to provide an upper (resp.\ lower) bound to the maximal (resp.\ typical) value of $\dim(\lambda)$: in both cases, the leading term scales as $\sqrt{|\lambda|!}$. Using Eq.~\eqref{eq:psi_FA}, one gets that the eigenfunctions approach zero faster than exponentially upon increasing $|\lambda|$, because of the overall factor $1/\sqrt{|\lambda|!}$. This estimate is in agreement with the exact result of Eq.~\eqref{eq:b_m}, since the Bessel functions decay as the inverse factorial of the (large) index, see Eq.~\eqref{app:eq:bessel_large_index}.

\section{Mechanisms of integrability breaking}
\label{sec:integrability_breaking}

In the previous Sections we showed that the Hilbert space of the $2d$ Ising model in the infinite-coupling limit $J\to\infty$ shatters in many disconnected Krylov sectors. Among these sectors, those corresponding to the wide class of interfaces discussed in Sec.~\ref{sec:Lipschitz} can be mapped onto a $1d$ model which turns out to be integrable. In this Section we discuss the dynamics of the interface beyond integrability and the robustness of the qualitative features of the exact solution, presenting in detail what was briefly anticipated in Ref.~\cite{Balducci2022Localization} by us. 

In Sec.~\ref{sec:bubble} we argue that the interfaces which do not satisfy the Lipschitz condition of Eq.~\eqref{eq:Lipschitz} may have a very different dynamical behaviour compared to the one described so far, because they can break into disconnected pieces. This is done by considering the case of an interface which is \emph{locally} Lipschitz, but which it is not the graph of a function $\mu$ at a larger scale. In Sec.~\ref{sec:1st_order_corrections} we consider, instead, another possible source of integrability breaking: the presence of a finite, albeit still large, coupling $J$. Specifically, we will discuss the $O(J^{-1})$ corrections to the infinite-coupling Hamiltonian~\eqref{eq:H_PXP} and address the ergodicity of the resulting model. In Sec.~\ref{sec:Stark_MBL} we discuss, using both analytical and numerical techniques, why the $O(J^{-1})$ corrections to the infinite-coupling Hamiltonian lead to a localization phenomenon, named Stark MBL, induced by the presence of the longitudinal field $h$. Finally, in Sec.~\ref{sec:false_vacuum} we compare our results for the time evolution of a domain on the lattice with the equivalent problem in the continuum, studied in the context of the false vacuum decay scenario, highlighting qualitative differences.

\subsection{Finite bubbles}
\label{sec:bubble}

Throughout Secs.~\ref{sec:Lipschitz} to \ref{sec:Young_diagrams} we assumed the presence of a single interface, separating the $2d$ lattice in two infinitely extended domains. It is then natural to investigate the extent to which the predictions derived therein carry over to finite domains. The easiest and natural case to be considered is that of a single, large bubble of ``down'' spins, surrounded by ``up'' spins (or vice-versa). Let us also introduce the notion of convexity on the lattice: we will say that a domain is \emph{convex} if any line \emph{parallel to the lattice axes} joining two points in the domain lies entirely within the domain itself. As already noted in Refs.~\cite{Hart2022Hilbert,Balducci2022Localization}, all convex bubbles are dynamically connected with the minimal rectangle (with sides parallel to the lattice axes) that contains them, i.e. they belong to the Krylov sector generated by this rectangle. Moreover, because of the perimeter constraint, the domain-wall dynamics is always confined within such a rectangle. Accordingly, we can directly assume that the shape of the bubble at the initial time $t=0$ is a rectangle, as all the other cases will follow from this one.

The early-stage dynamics of such a rectangular bubble can be predicted on the basis of the previous analysis. In fact, the sides of the bubble are immobile, since no spin can be flipped without modifying the perimeter, while the corners start to be eroded, as discussed in Sec.~\ref{sec:corner}. However, the evolution will deviate from that of an infinite and isolated corner as soon as two adjacent corners will start ``feeling'' the presence of the other. The timescale at which this happens can be be bounded from below by computing, in the fermionic language, the probability of finding two fermions, each coming from a different isolated corner, halfway along the flat portion of the interface which connects these two corners.

Let us denote by $L$ the length of the shortest side of the rectangular, finite bubble. There are now two possible cases. If the longitudinal field $h=0$ or, more generally, $h$ is small enough for the Bloch oscillations to have an amplitude $\ell \simeq |g/h|$ (Eq.~\eqref{eq:lloc}) larger than the distance $L/2$, then the excitations propagate ballistically on the chain with speed $2g$ (Eq.~\eqref{eq:mu_Omega}), and they meet at $L/2$ after a time
\begin{equation}
    T_{\text{corner}}(h=0) \sim \frac{L}{4g}. 
\end{equation}
If, instead, $h$ is nonzero and large enough to confine the dynamics in a region \emph{smaller} than $L/2$, one can estimate the probability $P(x,t)$ of having a fermion at a distance $x<0$ from the corner (equivalently, a hole at distance $x>0$) with $P(x,t) = 1 - \av{n(x,t)}$. On the maxima of the oscillations of the corresponding interface~\footnote{A very similar result is obtained if taking the average over a period, rather than the maximum of the oscillations.}, attained at times $t^*$ such that $\omega_{t^*}=2\gamma$ (Eqs.~\eqref{eq:omega_t} and~\eqref{eq:mu_Omega}), one finds $\av{n(x,t^*)} = \sum_{y<x} J^2_y (2 \gamma)$, cf.\ Eq.~\eqref{eq:corner_av_n}, and consequently
\begin{equation}
    P \left(L/2, t^* \right)
    = \sum_{y\geq L/2}J^2_y\left(2\gamma\right) .
\end{equation}
Recalling that the Bessel functions of large order decay exponentially fast to zero, one can approximate (see also Eq.~\eqref{app:eq:bessel_large_index})
\begin{equation}
    P\left(L/2, t^* \right)
    \approx J^2_{L/2}\left(2 \gamma \right)
    \approx \frac{1}{\pi L}\left(\frac{2 e g}{L h}\right)^{L}.
\end{equation}
With this result, the typical time after which two fermions, coming from different corners, interact can be estimated as $T_{\mathrm{corner}} (h \neq 0) \sim 1/P^2\left(L/2, t^* \right)$ or, more explicitly \footnote{This result can be obtained using Fermi Golden Rule. In particular, the interaction rate for two fermions coming from different corners is proportional to the probability of having both fermions at half chain. Being fermions of different species, they interact only at the scattering point and therefore such probability is the product of the single fermion probability of being at distance $L/2$ from the corner. Consequently, taking the inverse of the rate, one obtains $T_{\mathrm{corner}}$.},
\begin{equation}
\label{eq:T_corner}
    T_{\mathrm{corner}} (h \neq 0) \sim \frac{1}{g} \, e^{2L\ln L - 2L\ln (2e g/h)}.
\end{equation}
One can see that, in the case $h \neq 0$, a time which is more than exponentially large in the bubble size $L$ must pass, before integrability breaking starts to be manifest. 

It is natural to wonder what happens to the bubble after this timescale. Based on elementary reasoning, one can argue that two kinds of processes may take place: (a) the excitations coming from one corner may start to affect the dynamics of adjacent corners, transferring energy between corners and deteriorating the perfect coherence of the single-corner oscillations; (b) the interface may break because of the  detachment of isolated bubbles of flipped spins caused by the interface-splitting transitions  $\sket{\hhmove}\sbra{\vvmove} + \text{H.c.}$ of Eq.~\eqref{eq:H_PXP_graphical}. We note, however, that these detached parts can move away from the parent interface only via $g^2/J$ processes. A detailed study of this challenging problem is left for future investigations. 

We conclude by emphasising that the case of two adjacent corners we have considered here actually applies to any very large bubble, provided that its boundaries are ``smooth'' enough---i.e., that the Lipschitz condition is locally satisfied while the points responsible for its global violation are very dilute. If, instead, the initial interface is rather corrugated, i.e. it is not the graph of a function $\mu(x)$ even locally, then we expect a really complicate time evolution, during which all accessible configurations may be explored, and the single-interface description is no longer possible.

\subsection{Finite coupling}
\label{sec:1st_order_corrections}

We now relax the assumption that $J$ be strictly infinite, considering the effects of the corrections $\propto 1/J$, but still under the assumption that $J \gg |h|, g$. 

A large but finite $J$ still imposes an \emph{effective} dynamical constraint, valid up to a timescale which become exponentially long upon increasing $J$: this follows from the rigorous prethermalization bounds of Ref.~\cite{Abanin2017Rigorous}. Specifically, the perturbatively ``dressed'' version of the domain-wall length operator $D$ (defined in Eq.~\eqref{eq:D-dl}), arising from the Schrieffer-Wolff transformation, is accurately conserved for a long time that scales (at least) exponentially:
\begin{equation}
    \label{eq:T_preth}
    T_{\mathrm{preth}} \geq T_{\mathrm{preth}}^0 \equiv \frac{C}{g} \, \exp \left[ \frac{c J}{\max(g,|h|)} \right]
\end{equation}
(here $c$ and $C$ are numerical constants independent of $J$, $g$, and $h$). This is because the Schrieffer-Wolff effective Hamiltonian $H_{\mathrm{eff}}=H_{\mathrm{PXP}}+(1/J)(\cdots)+(1/J^2)(\cdots)+\cdots$, computed up to a suitable optimal perturbative order, commutes with $D$ in Eq.~\eqref{eq:D-dl} up to an exponentially small error~\cite{Abanin2017Rigorous}. In addition, the evolution of all local observables is well approximated by $H_{\mathrm{eff}}$ for $t \leq T_{\mathrm{preth}}$~\cite{Abanin2017Rigorous}. 

As anticipated above, the zeroth order of the Schrieffer-Wolff effective Hamiltonian $H_{\mathrm{eff}}$ was already determined in Sec.~\ref{sec:infinite_J_fragmentation} and is given by Eq.~\eqref{eq:H_PXP}. Computing higher-order corrections to $H_{\mathrm{eff}}$ becomes rapidly very complex, as the number of terms increases more than exponentially. In App.~\ref{app:sec:SW_PXP} we sketch the computation of the first-order corrections in $1/J$, while in App.~\ref{app:sec:SW_corner} we specialize it to the dynamical sector of a smooth interface, of the type defined in Sec.~\ref{sec:Lipschitz}. In this sector, the perturbative corrections takes a simpler form. The construction above can be translated in the fermionic representation. The Schrieffer-Wolff effective Hamiltonian
\begin{equation}
    \label{eq:H_fermions_1storder}
    H_{\mathrm{eff}} = H_F^{(0)} + H_F^{(1)} + O\left( J^{-2} \right),
\end{equation}
has the zeroth-order contribution $H_F^{(0)}$ given by Eq.~\eqref{eq:H_fermions}, while the first-order corrections turn out to be
\begin{align}
\label{eq:H_F_1}
    H_{\mathrm{F}}^{(1)} = &-\frac{g^2}{4J} \sum_x \left(\psi^\dagger_x \psi_{x+2}+ \mathrm{H.c.}\right) \nonumber \\
    &+ \frac{g^2}{4J}\sum_x\left(2\psi^\dagger_x \psi^\dagger_{x+1} \psi_{x+1}\psi_{x+2}+ \mathrm{H.c.} \right.  \nonumber \\
    & \qquad \qquad \qquad \left. -3\psi^\dagger_{x} \psi_{x}\psi^\dagger_{x+1} \psi_{x+1}\right).
\end{align}
One may recognize that they entail next-nearest-neighbour hoppings and density-density interactions. One can notice also that the density-density interactions, which are diagonal in the occupation number basis, do not depend on $h$. Consequently, the addition of the first-order corrections breaks the $h \to -h$ symmetry: changing the sign of $h$ modifies the expectation value of the energy. The terms in $H_{\mathrm{F}}^{(1)}$ are rather generic, and therefore one naturally expects them to break the integrability of the model, and make it thermalize rather quickly. However, as we anticipated in Ref.~\cite{Balducci2022Localization}, if $h$ is sufficiently large the perturbation is not able to restore ergodicity. In the next Section we describe this phenomenon in detail.

Let us briefly mention that, upon including the $O(J^{-1})$ corrections, an isolated flipped spin can spread in the $2d$ lattice with a hopping amplitude $\propto g^2/J$. This means that it is no longer possible to provide an effective $1d$ description even for initial configurations of the strip-like form, discussed in Sec.~\ref{sec:strip} for $J=\infty$.

Before continuing, it is important to emphasize a fundamental issue with the Schrieffer-Wolff transformation. For large but finite $J$, the dynamics of the initial product states considered so far in the form of the classical configurations will exhibit vacuum fluctuations even away from the existing domain walls. This is due to the perturbative dressing of the bare ferromagnetic state by virtual spin excitations. In practice, this arises from the application of the Schrieffer-Wolff unitary transformation $\exp(i S_1)$, c.f.\ Eq.~\eqref{app:eq:S1}, to the fully polarized initial product state. Accordingly, for such initial states one should think of the ferromagnetic vacua (i.e. those on the two sides of an infinite interface or the inner and outer regions of a bubble) as \emph{superpositions} of dilute spin flip excitations, of spatial density $\sim(g/J)^2$. Such excitations can be described as magnons, hopping on the $2d$ lattice with amplitude $\propto g^2/J$. In principle this dilute magnon gas contributes to the dynamics of the interface, but in the following we will ignore this fact, leaving its discussion to future investigations. This choice actually corresponds to taking as the initial state an interface in the \emph{the Schrieffer-Wolff transformed basis}, rather than in the classical one discussed so far.

\subsection{Arguments in favour of Stark many-body localization}
\label{sec:Stark_MBL}

The goal of this Section is to study the evolution induced by the Hamiltonian~\eqref{eq:H_fermions_1storder}. The first term in Eq.~\eqref{eq:H_fermions_1storder} is the Hamiltonian $H_F$ considered already in Secs.~\ref{sec:Lipschitz} and \ref{sec:corner}: it represents a chain of Stark-Wannier-localized, non-interacting fermions. The second term, i.e. $H_F^{(1)}$ of Eq.~\eqref{eq:H_F_1}, is a small perturbation containing both next-nearest-neighbour hoppings and two-body interactions. Accordingly, there is a competition between the localized nature of the dynamics induced by $H_F^{(0)}$ and the interactions in $H_F^{(1)}$, which are generally expected to drive the system towards a thermal phase. Previous works~\cite{Schulz2019Stark,Rafael2019Bloch,Morong2021Observation} have shown that, for interacting Hamiltonians very similar to Eq.~\eqref{eq:H_fermions_1storder}, an extended non-thermal phase is present for sufficiently strong field $h$, partly in analogy to what happens in the disorder-induced many-body localization (MBL). Indeed, the phenomenon has been dubbed \emph{Stark MBL}.

In order to quantify the competition between interactions and the linear potential responsible for the localization, we developed an analytical argument \`a la Basko-Aleiner-Altshuler (BAA)~\cite{Basko2006Metal} which goes as follows. Start from the integrable limit $J=+\infty$: the eigenfunctions are expressed in terms of the single-particle orbitals of Eq.~\eqref{eq:b_m} and are all spatially localized. Their localization length $\xi$ can be quantified by the participation ratio: using Eq.~\eqref{eq:b_m} and Neumann's addition theorem (see App.~\ref{app:sec:bessel}),
\begin{equation}
    \xi^{-1} = \sum_x J_x^4(\gamma) = \frac{1}{\pi} \int_0^{\pi} d\theta \, J_0^2 \left( \gamma \sqrt{2-2\cos{\theta}} \right),
    \label{eq:PR}
\end{equation}
where $\gamma$ is given after Eq.~\eqref{eq:b_m}. An approximate form of this relation is derived in App.~\ref{app:sec:participation_ratio}, where we perform the asymptotic expansion of the above integral for large values of $\gamma$ through the method of the Mellin transform, determining
\begin{equation}
    \xi^{-1} = \frac{\ln(\gamma C)}{\pi^2 \gamma} + O \left( \frac{\ln \gamma}{\gamma^3} \right),
    \label{eq:xi-1-expr}
\end{equation}
where $C=2^5\,e^{\gamma_E}\simeq 56$ and $\gamma_E = 0.5772\cdots$ is the Euler-Mascheroni constant. Such approximation is clearly accurate for $\gamma \gg 1$, i.e. for small $h$.

\begin{figure}[t]
    \centering
    \includegraphics[width=0.9\columnwidth]{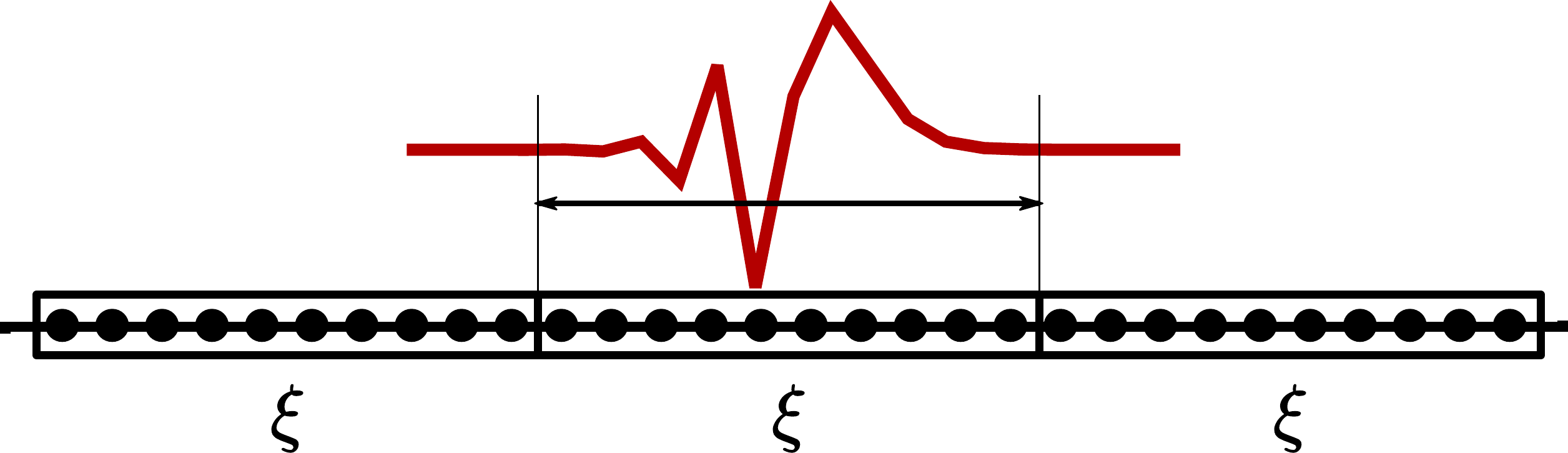}
    \caption{Graphical representation of adjacent regions (``quantum dots'') of size $\xi$ along the chain, each corresponding to one localization length, as represented by the eigenfunction (red, solid line). As described in the text, focusing on one of these intervals, one can derive an estimate for the critical value of $h$, above which the system is not ergodic even in the presence of a finite $J$. }
    \label{fig:quantumDot}
\end{figure}

We now assume that one can partition the system into boxes (``quantum dots'') of size $\xi$, as sketched in Fig.~\ref{fig:quantumDot}. Within each of them, the number of states is clearly $N_\xi = 2^{\xi}$ (there are $\xi$ sites that can be either empty or occupied), whereas the maximum energy difference between two many-particle states is $\Delta_{\max} \approx |h| \xi^2$. To understand this latter estimate, assume $h>0$: then, the minimum energy is attained when no particle is present ($E_{\min} = 0$), while the maximum when all sites are occupied (and thus $E_{\max} = \sum_{x=0}^\xi h x \approx h \xi^2$). With the same reasoning, with $h<0$ one gets $\Delta_{\max} \approx |h| \xi^2$, thus confirming the estimate.

\begin{figure}[t]
    \centering
    \includegraphics[width=\columnwidth]{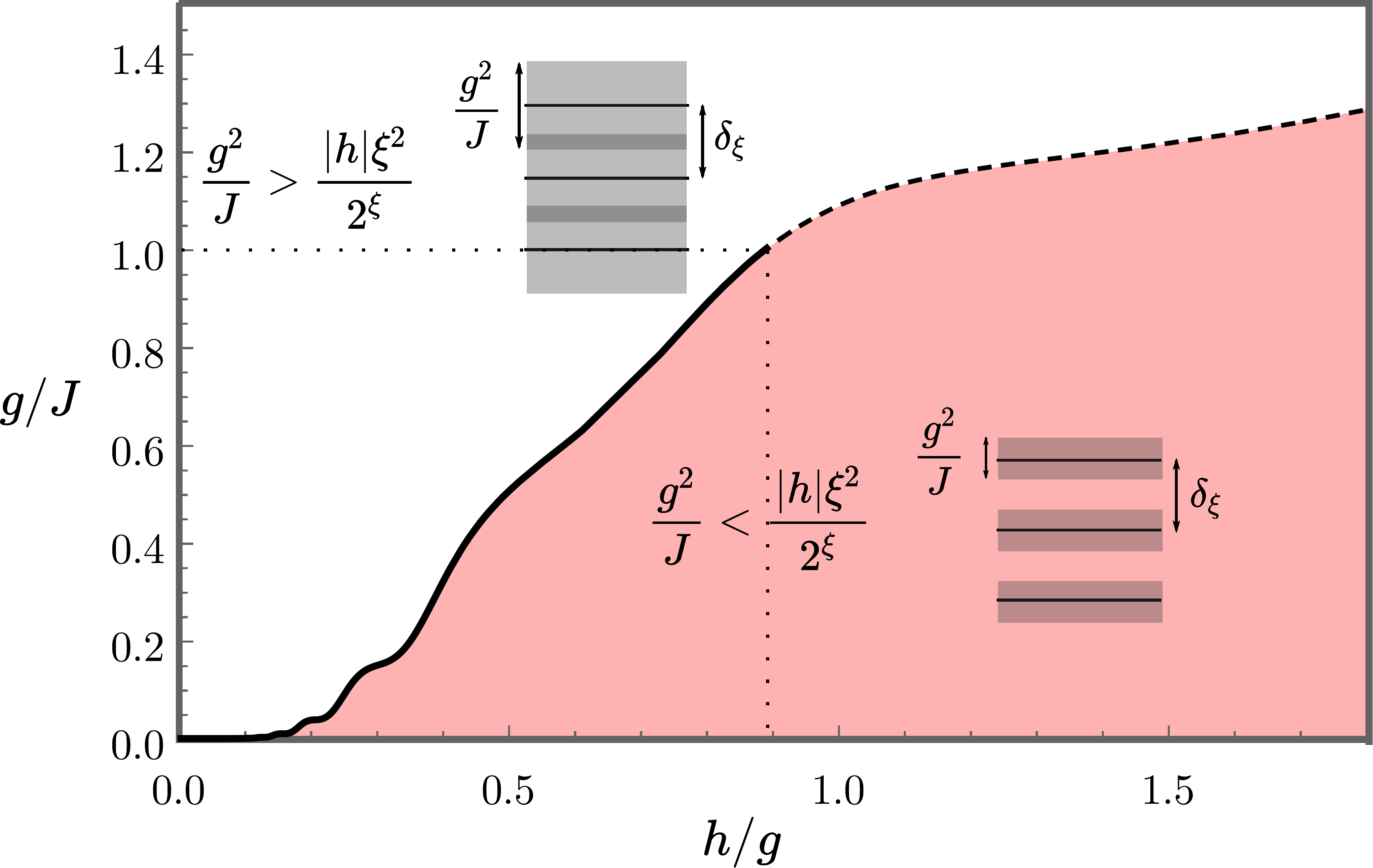}
    \caption{Region of the $(h/g,g/J)$-plane within which the condition of localization in Eq.~\eqref{eq:localiz_criterion} is valid (color), and region where delocalization is expected (white). For each region, a sketch is provided of the comparison between the unperturbed level spacings (solid black line) and the strength of the interactions (shaded gray area). We note that, for fixed $J$, the condition is satisfied for sufficiently large $h$. For $J/g \gtrsim 1$ (dashed line), the predicted boundary between the two regions is no longer reliable because the higher-order corrections neglected here become dominant. }
    \label{fig:phase_diag}
\end{figure}

Following BAA (and thus also building on Ref.~\cite{Altshuler1997Quasiparticle}), we say that interactions should not be able to restore ergodicity (at least perturbatively) if their strength $\lambda \sim g^2/J$ is smaller than the average local level spacing:
\begin{equation}
    \delta_{\xi} \approx \frac{\Delta_{\max}}{N_\xi} \approx \frac{|h| \xi^2}{2^\xi},
\end{equation}
i.e.\ when $\lambda < \delta_\xi$. This is equivalent to requiring
\begin{equation}
\label{eq:localiz_criterion}
    \frac{g^2}{J} < \frac{|h| \, \xi^2}{ 2^\xi},
\end{equation}
which is always satisfied for $0 \leq |\gamma| \lesssim 1$, i.e.\ for  sufficiently large $|h|$. It is interesting to note that the regime of validity of the heuristic criterion~\eqref{eq:localiz_criterion} depends only weakly on $J$. In Fig.~\ref{fig:phase_diag} we show, upon varying $J$ and $h$, the regions of validity of the inequality~\eqref{eq:localiz_criterion}, where $\xi = \xi(h/g)$ is given by Eq.~\eqref{eq:PR}. One can observe how, for fixed $J$, the criterion is satisfied for sufficiently large $h$; moreover, for $J/g\gtrsim 1$, the condition in Eq.~\eqref{eq:localiz_criterion} holds for $h/g \gtrsim 1$.

As a check for the above estimate, we performed numerical simulations, focusing in particular on the ``generalized imbalance'', a witness of ergodicity breaking. Given a generic initial state $\ket{\Psi_0}$, the time-evolved generalized imbalance for a system of length $L$ is
\begin{equation}
\label{eq:I-def}
    I_L(t) = \sum_{x=-L/2+1}^{L/2} \frac{1}{L} \bra{\Psi_0} m(x, t) m(x, 0) \ket{\Psi_0},
\end{equation}
where we defined $m (x, t) := 2 \, n(x, t) -1 $. If the initial state $\ket{\Psi_0}$ is a N\'eel state, then the generalized imbalance reduces to the standard imbalance between the occupation number of odd and even sites, used both in numerical simulations and cold-atom experiments. Taking the infinite-size limit and averaging over time, one obtains
\begin{equation}
    I = \lim_{L,T\to \infty} \frac{1}{T} \int_{0}^T dt \, I_L(t),
\end{equation}
which is zero in generic thermalizing systems. Accordingly, $I\neq 0$ is a sufficient condition for the system to be non-ergodic (even if it is not necessary). The infinite-time limit in the definition of $I$ can be obtained also by using the diagonal ensemble: assuming $\ket{\Psi_0}$ to be given as in Eq.~\eqref{eq:initial_state}, one finds
\begin{multline}
\label{eq:imbalance}
    I = \lim_{L \to \infty} \frac{1}{L} \sum_{x=-L/2+1}^{L/2} \av{m(x,0)} \\
    \times \sum_a \bra{E_a} m(x,0) \ket{E_a} |\braket{\Psi_0}{E_a}|^2,
\end{multline}
with the average $\av{\cdots}$ defined in Eq.~\eqref{eq:def_average}, and $\ket{E_a}$ the eigenbasis of the Hamiltonian Eq.~\eqref{eq:H_fermions}. 

The ergodicity test based on the valuer of $I$ should in principle be done for every initial configuration. However, there are states $\ket{\Psi_0}$ that will trivially give a non-ergodic result $I>0$. For example, states near the ground state will remain non-ergodic also in the presence of the $1/J$ corrections, just because they lie at the edges of the spectrum: we checked numerically that this is the case, for instance, for the domain-wall state of Eq.~\eqref{eq:domain_wall} (data not shown). A non-trivial test, instead, is provided by generic states which lie in the middle of the spectrum: for our purposes, the N\'eel state $\ket{\mathbb{Z}_2} = \prod_k \psi_{2k}^{\dagger} \ket{0}$, for which $\av{m(x,0)} = (-1)^x$, will suffice.

\begin{figure}
    \centering
    \includegraphics[width=\columnwidth]{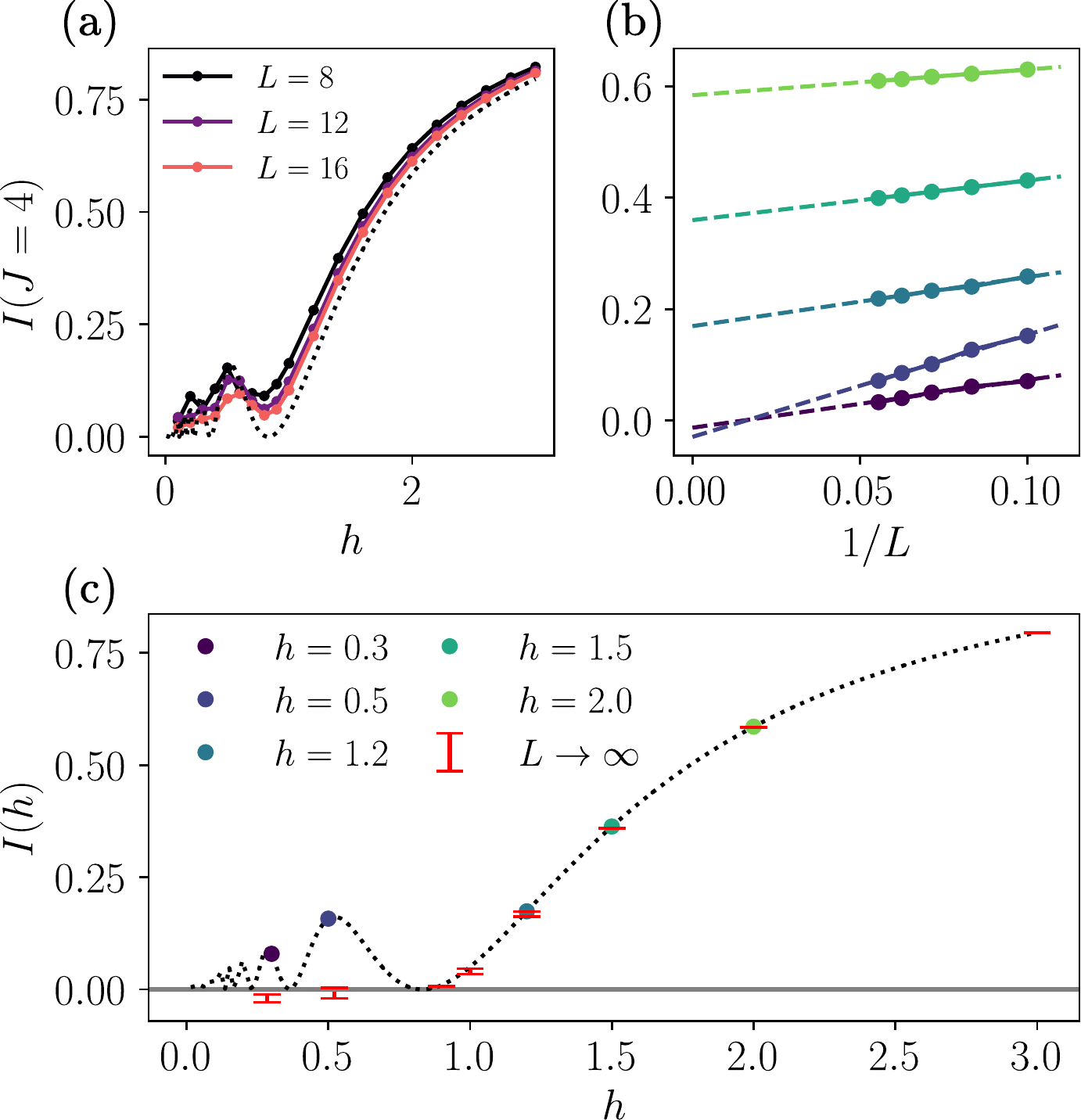}
    \caption{(a) Numerical values of the generalized imbalance $I$ obtained in the diagonal ensemble at $J=4$, $g=1$. In order to improve the readability, only the data for $L=8$,12,16 are reported. The dotted line represents the analytical prediction $\tilde I_\infty$ for $J=\infty$ and $L=\infty$ given in Eq.~\eqref{eq:imbalance_infty}.
    (b) Extrapolation to $L=\infty$ of the size-dependent generalized imbalance $I(L)$ obtained numerically, and reported in panel (a). The extrapolation is done using the ansatz $I(L) = I_{\infty} + A/L$. Different colors correspond to different values of $h$ (see panel (c) for the legend).
    (c) The colored dots correspond to the values of $h$ for which we reported the extrapolation in panel (b). The red error bars are the results of the extrapolation, with the error coming from the fit. The two points with the smaller values of $h$ are in correspondence of the local maxima of $I_{\infty}$ at $J=\infty$ and their extrapolations are compatible with zero within the error bars. At larger values of $h$, instead, the extrapolation provides values of the generalized imbalance which turn out to be compatible with those at $J=\infty$. }
    \label{fig:imbalance}
\end{figure}

In Fig.~\ref{fig:imbalance} we compare the numerical values of $I$ at finite $J$, with the analytical prediction $\tilde{I}$ at $J = + \infty$: using the definition in Eq.~\eqref{eq:I-def} and Eq.~\eqref{eq:sumBessel} one finds
\begin{equation}
\begin{split}
        \tilde{I}_\infty(t)
    &= \lim_{L\to \infty} \sum_{x=-L/2 + 1}^{L/2} \frac{1}{L} \bra{\mathbb{Z}_2} m(x,t) m(x,0) \ket{\mathbb{Z}_2} \\
    &= \lim_{L\to \infty} \frac{2}{L} \sum_{x=-L/2 + 1}^{L/2} \sum_{y=-\lceil L/4\rceil +1}^{\lceil L/4 \rceil} \!\!\!\!\! (-1)^x J_{2y-x}^2(2\gamma\sin(ht))\\ 
    &= J_0(4\gamma \sin(ht)).
\end{split}
\end{equation}
In the long-time limit, the temporal average $\tilde{I}_\infty$ of $\tilde{I}_\infty(t)$ is given by
\begin{equation}
\label{eq:imbalance_infty}
    \tilde{I}_\infty = \lim_{T\to \infty}\frac{1}{T}\int_0^T dt \, J_0(4\gamma \sin(ht)) = J_0^2(2\gamma),
\end{equation}
where the last step is a known property of the Bessel functions \cite{NIST:DLMF}, see Eq.~\eqref{app:eq:J0squared}.

The curve $\tilde{I}_\infty$ is represented, for $g=1$, by the dashed lines in Figs.~\hyperref[fig:imbalance]{\ref{fig:imbalance}a} and \hyperref[fig:imbalance]{\ref{fig:imbalance}c}. For finite values of $J$, instead, one is able to compute the generalized imbalance only numerically and for finite $L$. Accordingly, the estimate for $L \to \infty$ has to be obtained via extrapolation, which we perform in Fig.~\hyperref[fig:imbalance]{\ref{fig:imbalance}b}. The numerical values of the generalized imbalance show a linear dependence on $1/L$, allowing for a reliable extrapolation to $L = \infty$ (see the caption of Fig.~\ref{fig:imbalance} for more details). The final results are reported in Fig.~\hyperref[fig:imbalance]{\ref{fig:imbalance}c}: while for $h \lesssim 1$ the generalized imbalance is compatible with $0$, for $h>1$ the results at finite $J$ are perfectly compatible with the analytic prediction at $J = \infty$. These data provide numerical support to the argument \`a la BAA that we discussed above. In addition, we performed also numerical simulations for the time evolution of the generalized imbalance (which we do not report here), and we noticed that the relaxation time to the diagonal ensemble value depends on $J$ (the larger $J$, the longer the time needed), whereas the asymptotic value does not, again in agreement with the argument \'a la BAA.

\subsection{Implication for the dynamics of finite bubbles and the decay of false vacuum}
\label{sec:false_vacuum}

As mentioned in the Introduction, the problem addressed in the previous Sections is reminiscent of the false vacuum decay process, that received much attention in the field theory context, in particular starting from the works by Coleman~\cite{Coleman1977False,Callan1977Fate,coleman1988aspects}. In our work, we started directly from the situation in which a true vacuum bubble is already present in the false vacuum (or false vacuum in the true vacuum, which is equivalent in our setting). Therefore, we will not discuss here the timescale needed to create a bubble out of a uniform configuration (for the $1d$ quantum Ising model this issue has been addressed in Refs.~\cite{Rutkevich1999Decay,Lagnese2021False}). Here we will limit ourselves to compare the evolution of such domains on the lattice and in the continuum, the latter problem being solved in Ref.~\cite{Coleman1977False}.

Let us start by reminding the reader that a false vacuum, i.e.\ the state in which the spins are uniformly aligned in opposite direction to the longitudinal field, is a highly-excited state with finite energy density, which is expected to decay to configurations with equal total energy but larger entropy. Coleman identified and described this kind of decay process occurring in a field theory as the generation of a resonant true-vacuum bubble(s), the critical linear size $L_*$ of which is determined by the balance between the energetic cost for creating its interface ($\sim + 8 J L_*$ in our setup), and the bulk energy gain in having a bubble of true vacuum (in our setup, this comes from the the spin alignment, with gain $\sim - 2 h L_*^2$). The value of the critical linear size of the bubble is easily found: $L_*\sim J/h$.
  
In continuous space-time, the timescale associated with the formation of the bubble above can be calculated in the framework of relativistic quantum field theory using instantons. The total potential energy change $\Delta V\sim J L_*-2h L_*^2$, which vanishes at the moment of the formation of the bubble, starts {\it decreasing} to large negative values when the bubble increases its dimension and accelerates quickly to swallow the remaining false vacuum, transforming the gained potential energy in kinetic energy $\Delta T$ so that the total energy $E$ does not change, i.e., $\Delta E=\Delta T+\Delta V=0$. However, there are considerable differences between our lattice setting and what happens in a field theory on the continuum. First of all, in the limit $J\to\infty$ it is easily seen that $L_*\to\infty$ and therefore the bubble is not formed at all. But, for finite $J$, if one waits a time exponentially large in $J/h$ (using the result from Coleman's continuum calculations), the bubble will eventually form. The walls are now expected to accelerate, expanding the bubble indefinitely, accumulating excess potential energy in a way which conserves the total energy. This expansion and acceleration, however, cannot occur on the lattice: as it can be seen from the dual fermionic description, the kinetic energy of the domain wall is actually bounded on the lattice by the value of $g$, and this prevents an expansion to sizes larger than $g/h$. This is the reason why the bubble starts oscillating. We have proven in this paper that these oscillations survive the perturbative introduction of a finite $1/J$, because the mechanism of Stark MBL confines the holographic fermions. With finite $J$, $h$, and $g$, instead, Coleman's results for the expansion and the acceleration must proceed on timescales which are exponentially large (non-perturbative) in $J$. Our best attempt at calculating this rate is in Sec.~\ref{sec:integrability_breaking}. In the simpler case of $1d$ spin chains, in fact, it was found that Bloch oscillations also inhibit the expansion of the true vacuum bubble~\cite{Mazza2019Suppression,Lerose2020Quasilocalized,Pomponio2022Bloch}. Accordingly, the post-vacuum-decay scenario ought to be profoundly different from that described by Coleman.

\section{Conclusions}
\label{sec:conclusions}

In this work we have shown, expanding on the results of Ref.~\cite{Balducci2022Localization}, how to describe the dynamics of interfaces in the two-dimensional quantum Ising model with strong ferromagnetic coupling $J$. As a first step, we discussed the \emph{infinite}-coupling limit $J=\infty$, focusing both on the equilibrium properties of $1d$ linear spin domains embedded in the $2d$ lattice, and on the dynamics of infinite interfaces described by Lipschitz-continuous functions. In the first case, we have shown that the model reduces to a $1d$ PXP Hamiltonian, for which one can calculate exactly the equilibrium magnetization, assuming that the initial state has negligible overlap with quantum many-body scars. The interest in the second case, instead, has a twofold motivation: first, the corresponding configurations effectively describe smooth interfaces and, second, given the impossibility of breaking the domain wall (ensured by the Lipschitz condition which is conserved by the dynamics), an effective $1d$ description can be provided in terms of fermionic particles subject to a linear potential, which is amenable to an exact solution. We also discussed how to take the continuum limit of the dynamics, in order to predict the behaviour of the quantum-fluctuating interface at scales much larger than the lattice spacing. A semiclassical interpretation of the resulting formula naturally emerged. 

Then, we moved to the case of an interface shaped like an infinite corner. In particular, we discussed the properties of the average limiting shape, both on the lattice and in the continuum, and its relationship with classical corner growth models. We predicted the dynamics of the entanglement entropy between the two halves of the corner, and unveiled a deep connection between the quantum problem and the asymptotics of the Plancherel measure on random Young diagrams.

We finally relaxed the assumption of infinite strength of the Ising coupling $J$, making use of a Schrieffer-Wolff transformation to calculate the $O(1/J)$ corrections. The first-order corrections break the integrability of the model with $J=\infty$ but, remarkably, ergodicity is not restored. In fact, the presence of the longitudinal magnetic field in $2d$, which translates into a linear potential in $1d$, causes the emergence of Stark MBL, that we characterized both numerically, computing the generalized imbalance, and analytically, providing an argument for its validity. Even if a recent work provided analytical evidence against Stark MBL~\cite{Kloss2022Absence}, their results apply only to the infinite-time limit, where our perturbation theory in $1/J$ is no longer reliable. We expect therefore that, on the timescales considered in this work, the phenomenology of localization is quite robust.

In order to understand the temporal range of validity of our predictions, we investigated the relevant timescales controlling the dynamics of the system in generic conditions. In particular, we identified in $T_{\mathrm{preth}}$ the \emph{prethermal} timescale, after which the description in terms of Schrieffer-Wolff expansion is no longer valid: it turned out that $T_{\mathrm{preth}}$ becomes (at least) exponentially long upon increasing $J$. Moreover, the possibility of utilizing a $1d$ chain to describe interfaces in $2d$ is justified as long as the effects of possibly having a finite bubble size are negligible. Accordingly, we estimated the timescale $T_{\mathrm{corner}}$ below which this is a reliable assumption, and $T_{\mathrm{corner}}$ turned out to increase more than exponentially upon increasing the linear size of the domain. Both these timescales ensure that the results presented here, which were derived in the infinite-coupling or infinite-size limits, actually carry over to the case with finite but ``large'' coupling and sizes, up to very long times.

An intriguing question is about the dynamical effects arising at times longer than $T_{\mathrm{preth}}$ and $T_{\mathrm{corner}}$. While we leave this problem to future work, we can argue that the description given here is no longer valid, as the interface-splitting moves start playing a major role, and even the conservation of the interface length is no longer strictly guaranteed. As a consequence, the possibility of employing a $1d$ chain to describe the dynamics of a \emph{generic} $2d$ domain will likely be lost. However, for some initial configurations or at least in some regimes, we expect that it will still be possible to give a description in terms of a $1d$ effective problem, as we hope to address in a future work. 

Let us conclude by noting that, in the general case, the full $2d$ nature of the problem will emerge in the long-time limit, or for generic couplings. In these regimes no $1d$ description will be reliable and new techniques will be needed. Ultimately, it is natural to expect that a complete solution of the $2d$ quantum Ising model is at least as hard as the solution of the $3d$ classical Ising model.

\acknowledgements
F.B. and C.V. would like to thank P.\ Calabrese, L.\ Capizzi, G.\ Di Giulio, G.\ Giachetti,  A.\ Santini and S.\ Scopa for useful discussions.


\appendix
\onecolumngrid

\section{Magnetization in the linear strip}
\label{app:sec:strip}

We provide here a more detailed analysis of the dynamics of a linear strip of spins, treated in Sec.~\ref{sec:strip}.

Let us start from the computation of the number $C(L,l)$ of dynamically accessible configurations for a strip of spins of length $L$, in which $l\le L-2$ spins are flipped compared to the initial configuration: due to the perimeter constraint, $C(L,l)$ satisfies Eq.~\eqref{eq:recursion_strip}. This recursion relation is actually obtained by summing the number $C(L-1,l)$ of configurations in which the $L-1$-th spin along the chain is not flipped (we recall that the $L$-th spin cannot flip)  and the number $C(L-2,l-1)$ of configurations in which it is flipped. As there is only one configuration with no spin flip, the initial condition for Eq.~\eqref{eq:recursion_strip} is $C(L,0)=1$ and its solution is thus given by Eq.~\eqref{eq:numConfig}: this is a direct consequence of Pascal's property of the binomial coefficient~\cite{WolframBinomial}. In order to determine the total number of accessible configurations starting from the strip, one needs also to know the maximum number $l_{\rm max}$ of spins that can be flipped without violating the perimeter constraint. As anticipated above, the first and last spin cannot flip, and therefore one is left with $L-2$ potentially ``active'' spins. If $L-2$ is even, one can flip at most $(L-2)/2$ alternating spins, whereas if $L-2$ is odd, one can flip $(L-1)/2$ alternating spins. Accordingly, $l_{\rm max}$ is given by Eq.~\eqref{eq:n_max} in the main text.

Under the assumption of an ergodic dynamics (see the main text), the magnetization profile can be determined by using the following argument. Consider the $j$-th spin of a chain of length $L$. The number of configurations having the $j$-th spin ``up'' are given by the total number of allowed configurations for the two sub-chains (consisting of $j-1$ and $L-j$ spins, respectively) split by the $j$-th spin, that is $F_{j-1} \times F_{L-j}$. In the remaining $F_L-F_{L-j} F_{j-1}$ allowed configurations of the strip the $j$-th spin is down, 
and therefore the magnetization at site $j$ is simply given by
\begin{equation}
    \av{m_L(j)} = \frac{-(F_L -F_{L-j}F_{j-1})+F_{L-j}F_{j-1}}{F_L} 
    = 2\frac{F_{L-j}F_{j-1}}{F_L}-1,
\end{equation}
as reported in Eq.~\eqref{eq:magn1}, with the reflection symmetry $\av{m_L(j)} = \av{m_L(L-j+1)}$ and the boundary condition $\av{m_L(j)} = -1$. Using the explicit expression of the Fibonacci numbers in Eq.~\eqref{eq:FN}, one gets, for $L \to \infty$
\begin{equation}
    \av{ m_{\infty}(j)} =\frac{2}{(2\phi-1)\phi} - 1 + \frac{2}{(\phi-1)(2\phi-1)} \left(\frac{1}{\phi}-1\right)^{j},
    \label{eq:Msemi-inf}
\end{equation}
$\phi$ being the golden ratio. In the limit $j \to \infty$, one finds the magnetization at center of the strip $\av{m_{\infty,{\rm bulk}}} = \lim_{j\to\infty}\av{m_{\infty}(j)}$ to be given by Eq.~\eqref{eq:m-bulk}, after using the expression of the golden ratio provided before Eq.~\eqref{eq:FN}. The dependence on $j$ of Eq.~\eqref{eq:Msemi-inf} also implies that the approach to this asymptotic value is exponential with the typical length $\xi_b$ indicated after Eq.~\eqref{eq:m-bulk}.

\section{Useful properties of the Bessel functions}
\label{app:sec:bessel}

In this Appendix we collect a number of properties of Bessel functions which are useful and widely used to derive the results presented in the main text, and we provide also a sketch of their proofs. Many of these properties can actually be found in Refs.~\cite{NIST:DLMF,WolframBessel,Abramowitz1964Handbook}.

One of the equivalent definitions of the Bessel function of the first kind $J_n$ is in terms of the integral:
\begin{equation}
    J_n(\gamma) =  \int_{-\pi}^{\pi} \frac{d\tau}{2\pi}  ~e^{i(n\tau - \gamma\sin{\tau})}.
\end{equation}
From this definition it follows immediately that
\begin{equation}
    \sum_{n=-\infty}^{\infty} J_n(x) = 1,
\end{equation}
$J_n(-\gamma) = J_{-n}(\gamma)$, and, for $n \in \mathbb{Z}$,
\begin{equation}
    J_{-n}(\gamma) = (-1)^n J_n(\gamma).
    \label{eq:sym-Jn-1}
\end{equation}
Using again the definition, one can compute the following relation, useful in the calculation of the average of the number operator in Eq.~\eqref{eq:av_density}, Sec.~\ref{sec:Lipschitz}
\begin{align}\label{eq:sumBessel}
    \sum_{n=-\infty}^{\infty} J_{x-n}(\gamma)J_{y-n}(\gamma)e^{-2 i t h n} &= \sum_{n=-\infty}^{\infty} \int_{-\pi}^{\pi} \frac{d\tau}{2\pi}\frac{d\tau'}{2\pi} ~ e^{i((x-n)\tau - \gamma \sin{\tau})}e^{i((y-n)\tau' - \gamma \sin{\tau'})} e^{-2 i t h n}\\
    &= e^{-2iyht} \int_{-\pi}^{\pi} \frac{d\tau}{2\pi} ~ e^{i[(x-y)\tau - \gamma(\sin{\tau}-\sin{(\tau + 2ht)})]}\\
    &= e^{-i(x+y)ht} \int_{-\pi}^{\pi} \frac{d\tau}{2\pi} ~ e^{i[(x-y)\tau + 2\gamma \sin{ht}\cos{\tau}]}\\
    &= e^{-i(x+y)ht} ~ i^{x-y} ~ J_{x-y}(2\gamma\sin(ht)),
\end{align}
where we used $\sum_{n=-\infty}^{\infty} e^{inx} = 2\pi \delta(x + 2 k \pi)$. Setting $t=0$ we obtain the completeness relation
\begin{equation}
\label{eq:completeness_Bessel}
    \sum_{n=-\infty}^\infty J_{n-m}(\gamma)J_{n-k}(\gamma)=\delta_{mk},
\end{equation}
that also leads immediately to
\begin{equation}
    \sum_{n=-\infty}^\infty J_{n}^2(\gamma) = 1.
\end{equation}

If the sums of the previous equation is restricted to positive integer values, using telescopic sums one obtains
\begin{equation}\label{eq:prodSum}
\sum_{j=1}^{\infty}J_{j+m}(\gamma)J_{j+n}(\gamma) 
=\frac{\gamma [J_{m}(\gamma)J_{n+1}(\gamma)-J_{m+1}(\gamma)J_{n}(\gamma)]}{2(m-n)},
\end{equation}
that reduces to
\begin{equation}
    \sum_{j=0}^{\infty}J^2_{j+n}(\gamma) 
    =\frac{\gamma}{2}\left[J_n(\gamma)\partial_n J_{n-1}(\gamma)-J_{n-1}(\gamma)\partial_n J_n(\gamma)\right]
\end{equation}
when the limit $m \to n$ is taken. This relations allows us to compute explicitly the fluctuations of the number operator in Eq.~\eqref{eq:fluct}, App.~\ref{app:2point_funct}.
Using the same procedure as in Eq.~\eqref{eq:sumBessel} we can compute also
\begin{equation}
\label{app:eq:bessel_sum_multiples}
   \sum_{k \in \mathbb{Z}} J_{mk-x}^2(\gamma) = \frac{1}{m} \sum_{0 \leq n < m} e^{2 i x n \pi / m} J_{0}\left(2 \gamma \sin \frac{n \pi}{m}\right),
\end{equation}
which is used in Eq.~\eqref{eq:average_number_zigzag}, in Sec.~\ref{sec:Lipschitz}.

Another very useful tool is the asymptotic expansion of the Bessel functions for large order and argument (see Ref.~\cite{NIST:DLMF}), which are useful for discussing the continuum limit of the dynamics in Sec.~\ref{sec:Lipschitz-cont}. For fixed $\gamma$ and $x \to \infty$ one finds
\begin{equation}
\label{app:eq:bessel_large_index}
    J_{x}(\gamma)\sim \frac{1}{\sqrt{2\pi x}}\left(\frac{e \gamma}{2 x} \right)^x,
\end{equation}
i.e. the Bessel functions vanish faster than exponentially upon increasing $x \gg \gamma$. In the limit $\gamma \to \infty$ with fixed $x$, instead,  we have, at the leading order
\begin{equation}
\label{app:eq:bessel_large_argument}
    J_{x}(\gamma) \sim \sqrt{\frac{2}{\pi \gamma}} \cos{\left( \gamma -\frac{\pi}{2} x - \frac{\pi}{4} \right)}.
\end{equation}
If both the order and the argument of the Bessel function diverge, the asymptotic expansion is different if the argument is larger than the order or vice versa: at the leading order for $x\to+\infty$, one has
\begin{gather}
    \label{app:eq:bessel_uniform_1}
    J_x(x \sech \alpha) \sim \frac{e^{x(\tanh \alpha - \alpha)}}{\sqrt{2 \pi x \tanh \alpha}},
    \quad \mbox{for} \quad \sech \alpha < 1,\, \alpha>0 \\
    \label{app:eq:bessel_uniform_2}
    J_x(x\sec \beta) \sim \sqrt{\frac{2}{\pi x \tan \beta}} \cos \left(x (\tan \beta - \beta) - \frac{\pi}{4} \right), \quad \mbox{for} \quad  \sec \beta > 1, \, \beta \in \left(0, \frac{\pi}{2}\right)
\end{gather}
from which we notice the strong similarity between Eqs.~\eqref{app:eq:bessel_large_index}--\eqref{app:eq:bessel_uniform_1} and Eqs.~\eqref{app:eq:bessel_large_argument}--\eqref{app:eq:bessel_uniform_2}. This means that the asymptotic expansions of the Bessel function remain unaltered even when both the argument and the order scale linearly, but with different powers. Indeed, to see the transition between the regimes described by Eqs.~\eqref{app:eq:bessel_uniform_1} and~\eqref{app:eq:bessel_uniform_2} one has to consider $J_x(x + a x^{1/3})$ for $x \to \infty$ and fixed $a$, which is not important for our discussions. For the purpose of taking the continuum limit as explained in Sec.~\ref{sec:Lipschitz-cont} and Sec.~\ref{sec:av_interface} we note that Eqs.~\eqref{app:eq:bessel_uniform_1} and \eqref{app:eq:bessel_uniform_2} imply that, in the limit $x,\ y \to+\infty$ with fixed $y/x$, one has
\begin{equation}
\label{eq:J2-continuum}
J^2_x(y) \sim \theta (y-x)\frac{1}{\pi} \frac{1}{y\sqrt{1-(x/y)^2}},
\end{equation}
where $\theta(x)$ is the unit step function which equals 1 for $x\ge 0$ and vanishes otherwise. This equality is valid after integration with a smooth function, i.e.\ in the sense of distributions. In fact, the rapidly oscillating $\cos^2$ term deriving from Eq.~\eqref{app:eq:bessel_uniform_2} has been replaced with its average value $1/2$, while the rapidly decaying exponential in Eq.~\eqref{app:eq:bessel_uniform_1} has been set to zero, as indicated by $\theta(y-x)$ which appears in the expression above.

Another useful formula is
\begin{equation}
    \label{app:eq:neumann_bessel}
    \sum_{k=-\infty}^{\infty} J_k^4 (\gamma) = \frac{1}{\pi} \int_{0}^{\pi} d\theta \, J_{0}^2 \left(\sqrt{2 \gamma^2 - 2\gamma^2 \cos{\theta}}\right),
\end{equation}
which is used in Eq.~\eqref{eq:PR}. In order to prove it, one can use a modified version of Neumann's addition theorem, i.e.\  the Graf's and Gegenbauer's addition theorem~\cite{NIST:DLMF}
\begin{equation}
    J_0 \left( \sqrt{x^2+y^2-2xy \cos{\theta}}\right) = J_0(x) J_0(y) + 2 \sum_{k=1}^{\infty} J_k(x)J_k(x) \cos(k\theta).
\end{equation}
By setting $x=y=\gamma$, taking the square of both sides, and taking the angular average for $\theta \in [0,\pi]$, one gets
\begin{equation}
    \frac{1}{\pi} \int_{0}^{\pi} d\theta \, J_{0}^2 \left(\sqrt{2 \gamma^2 - 2\gamma^2 \cos{\theta}}\right) = J_{0}^4 (\gamma) + 2 \sum_{k=1}^{\infty} J_k^4(\gamma) = \sum_{k=-\infty}^{\infty} J_k^4 (\gamma),
\end{equation}
thus proving the identity.

We conclude this Section by reporting from Ref.~\cite{NIST:DLMF} the relation
\begin{equation}
    J_{\nu}(z) J_{\nu}(\zeta) = \frac{2}{\pi} \int_{0}^{\pi/2} d\theta \, J_{2\nu} \left( 2\sqrt{z \zeta} \sin{\theta} \right) \cos\left((z-\zeta)\cos{\theta}\right), 
\end{equation}
that, setting $z=\zeta$ and $\nu=0$ reduces to
\begin{equation}
    \label{app:eq:J0squared}
    J^2_0(z) = \frac{2}{\pi} \int_{0}^{\pi/2} d\theta \, J_0\left( 2 z \sin{\theta} \right).
\end{equation}
This expression is used to derive Eq.~\eqref{eq:imbalance_infty}.

\section{Two-point functions}
\label{app:2point_funct}

In order to obtain the fluctuations of the limiting shape $\mu$ of the Young's diagrams, one needs the 2-point function of the number operator $n$, see also Eq.~\eqref{eq:mu_from_n}. For simplicity, we report here the computation done at equal times, but the same procedure can be extended also for different times. Let us start by computing
\begin{equation}
    \bra{\Psi_0}n(x,t)n(y,t)\ket{\Psi_0}
    =\bra{\Psi_0}\psi_x^{\dagger}(t)\psi_x(t)\psi_y^{\dagger}(t)\psi_y(t)\ket{\Psi_0}.
\end{equation}
Also in this case, one can expand the initial state and use the time evolution of the fermionic operators. The expectation value one gets, using Wick contractions, is
\begin{equation}
    \bra{0}\psi_{\infty}\dots \psi_1~\psi_j^{\dagger}\psi_i \psi_l^{\dagger}\psi_k~\psi_1^{\dagger}\dots \psi_{\infty}^{\dagger}\ket{0} 
    = -\delta^+_{jk}\delta^+_{il} + \delta_{jk}\delta^+_{il} + \delta^+_{ij}\delta^+_{kl},
\end{equation}
being, by definition,
\begin{equation}
    \delta_{ab}^+ :=
    \begin{cases}
    1 &\text{if } a=b>0\\
    0 &\text{otherwise}.
    \end{cases}
\end{equation}
After some straightforward steps one arrives at
\begin{align}
\label{eq:2point}
    \bra{\Psi_0}n(x,t)n(y,t)\ket{\Psi_0}
    = & \left(\sum_{i>0}J_{i-x}^2(\omega_t)\right) \left(\sum_{i>0}J_{i-y}^2(\omega_t)\right)  - \left(\sum_{i>0}J_{i-x}(\omega_t)J_{i-y}(\omega_t)\right)^2  +\delta_{x,y}\left(\sum_{i>0} J_{i-x}(\omega_t)J_{i-y}(\omega_t)\right),
\end{align}
being $\omega_t = 2|\gamma \sin(ht)|$, as in the main text. Therefore, the connected 2-point function is
\begin{equation}
\label{eq:fluct}
\bra{\Psi_0}n(x,t)n(y,t)\ket{\Psi_0}_C =\delta_{xy}\left(\sum_{i>0}J_{i-x}^2(\omega_t)\right)-\left(\sum_{i>0}J_{i-x}(\omega_t)J_{i-y}(\omega_t)\right)^2.
\end{equation}
Using Eq.~\eqref{eq:prodSum}, one arrives at (see also Ref.~\cite{Antal2008Logarithmic})
\begin{equation}
    \bra{\Psi_0}n(x,t)n(y,t)\ket{\Psi_0}_C 
    =\delta_{xy}\left(\sum_{i>0}J_{i-x}^2(\omega_t)\right) - \left(\frac{\omega_t [J_{x}(\omega_t)J_{y-1}(\omega_t)-J_{x-1}(\omega_t)J_{y}(\omega_t)]}{2(y-x)}\right)^2.
\end{equation}
The fluctuations of the number operator of the fermions along the chain is readily obtained from Eq.~\eqref{eq:2point} by setting $x=y$:
\begin{equation}
    \delta n(x,t) =\bra{\Psi_0}n(x,t)^2\ket{\Psi_0}_C = \sum_{i>0}J_{i-x}^2(\omega_t)\left(1-\sum_{i>0}J_{i-x}^2(\omega_t)\right) .
\end{equation}

At this point, summing over space as it was done before, one arrives at the correlation function for the shape operator
\begin{align}
\label{app:eq:mu_fluct}
    \bra{\Psi_0}\mu(x',t)\mu(y',t)\ket{\Psi_0}_C = &-\sum_{x\leq x'}\sum_{y\leq y'}\left( \frac{\omega_t\left[ J_x(\omega_t)J_{y-1}(\omega_t)-J_y(\omega_t)J_{x-1}(\omega_t) \right]}{x-y} \right)^2 + 4 \sum_{x\leq x'}\sum_{y\leq y'}\delta_{xy}\left(\sum_{i>0}J_{i-x}^2(\omega_t)\right)\\
    = & \, 4 \sum_{x\leq x'}\sum_{y\leq y'} \left[ \delta_{xy}\left(\sum_{i>0}J_{i-x}^2(\omega_t)\right) - \mathcal{B}(x,y;\omega_t)^2 \right].
\end{align}

With the same procedure one can compute the expectation value of the current operator, defined as:
\begin{equation}
    j(x,t)\equiv i(\psi_x^{\dagger}(t) \psi_{x+1}(t) - \psi_{x+1}^{\dagger}(t) \psi_{x}(t)).
\end{equation}
Using this definition, one obtains
\begin{equation}
    \bra{\Psi_0}j(x,t)\ket{\Psi_0} = \gamma \sin{2ht}\left[ J_x^2\left( 2\gamma \sin{ht} \right) - J_{x+1}\left( 2\gamma \sin{ht} \right) J_{x-1}\left( 2\gamma \sin{ht} \right) \right].
\end{equation}
At $x=0$ it reduces to
\begin{equation}
    \bra{\Psi_0}j(0,t)\ket{\Psi_0} = \gamma \sin{2ht}\left[ J_0^2\left( 2\gamma \sin{ht} \right) + J_{1}^2\left( 2\gamma \sin{ht} \right) \right].
\end{equation}
Also the current-current correlator can be computed with the same tools: we report here the result for $x=0$, which is given by
\begin{equation}
    \bra{\Psi_0}j(0,t)j(0,0)\ket{\Psi_0} = J_{1}^2\left( 2\gamma \sin{ht} \right) - J_{0}^2\left( 2\gamma \sin{ht} \right).
\end{equation}

\section{Comparison with classical simple exclusion processes}
\label{app:sec:classical}

In this Appendix we compare the predictions presented in Sec.~\ref{sec:av_interface} for the dynamics of the corner-shaped interface in the quantum Ising model, with those obtained for the classical simple exclusion processes (SEP) which, as discussed in the main text, represents the classical counterpart of the quantum Hamiltonian~\eqref{eq:H_fermions}.
In particular, we focus on the totally asymmetric simple exclusion process (TASEP) and the symmetric exclusion process (SSEP), discussing them in the appropriate continuum limits, which  makes the comparison with Eqs.~\eqref{eq:num_density} and~\eqref{eq:mu_Omega} immediate. For a discussion of these processes on the lattice, instead, we refer to the vast literature on the topic, e.g.\ Refs.~\cite{Derrida1993Exact,Derrida1998Exactly,Schutz2001Exactly,Derrida2009Current,Derrida2002Large,Krapivsky2021Stochastic}.

First, we note that, while the time evolution of the TASEP is ballistic~\cite{Corwin2010Limit}, the one of SSEP is characterized by a diffusive scaling~\cite{DeMasi2015Symmetric,Derrida2002Large}.
Denoting respectively by $n(x,t)$ and $\mu(x,t)$ the density of particles and the interface height (as in Eq.~\eqref{eq:mu_from_n}), it turns out \cite{Corwin2010Limit} that the dynamics of the rescaled density
\begin{equation}
\label{app:eq:limit_n_rho}
n_T(\xi,\tau) := \lim_{N \to \infty} \rho_T(\xi N, \tau N)    
\end{equation}
of the TASEP (the subscript $T$ stands for TASEP) obeys the Burgers equation
\begin{equation}
    \label{app:eq:Burgers}
    \frac{\partial}{\partial \tau}n_T = \frac{\partial}{\partial \xi} [n_T (1 - n_T)].
\end{equation}
We have denoted with $\rho_T$ the average number of particles on the lattice, while $n_T$ denotes the average number in the continuum, which is obtained by taking the limit in Eq.~\eqref{app:eq:limit_n_rho}, where $N$ is the inverse of the lattice spacing. With the step initial condition $n_T(\xi,0)=\theta(\xi)$---corresponding to the corner---one obtains the solution $n_T(x,t) \equiv n_T(x/t)$, with the scaling function
\begin{equation}
\label{app:eq:nT}
    n_T(v) = 
    \begin{cases}
        0      & \mbox{for} \quad v \leq - 1, \\[1mm]
    \displaystyle \frac{1+v}{2}
        & \mbox{for} \quad |v| < 1, \\[1mm]
        1       & \mbox{for} \quad  v \ge 1.
    \end{cases}
\end{equation}
The corresponding interface height $\mu_T(x,t)$, determined according to Eq.~\eqref{eq:mu_from_n} on the continuum, turns out to be given by $\mu_T(x,t)=t \, \Omega_T(x/t)$, with the scaling function
\begin{equation}
\label{app:eq:Omega_T}
    \Omega_T(v) = 
    \begin{cases}
        |v|      & \mbox{for} \quad |v| \geq 1, \\[1mm]
    \displaystyle \frac{1 + v^2}{2}
        & \mbox{for} \quad |v| < 1.
    \end{cases}
\end{equation}

\begin{figure}
    \centering
    \includegraphics[width=\textwidth]{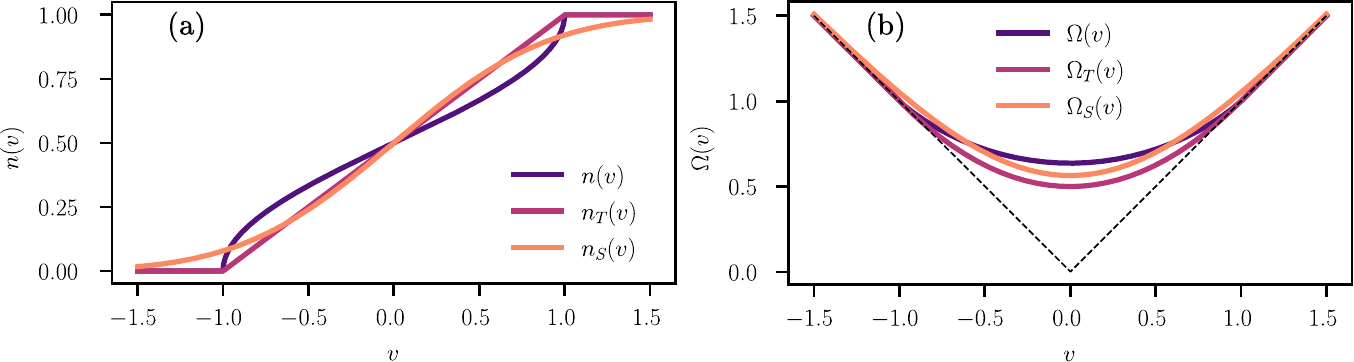}
    \caption{(a) Comparison between the average number density $n(v)$ in the quantum case, and the density profiles $n_T(v)$ and $n_S(v)$ of TASEP and SSEP, respectively. In particular, $n(v=x/\omega_t) \equiv \av{n(x,t)}$ is given by Eq.~\eqref{eq:num_density}, expressed in terms of the ratio $v=x/\omega_t$, while $n_T(v=x/t)$ and $n_S(v= x/\sqrt{2t})$ are given by Eqs.~\eqref{app:eq:nT} and~\eqref{app:eq:nS}, respectively.
    (b) Comparison between the interface limit shape $\mu(v)$ of the quantum problem and the corresponding quantities $\mu_T(v)$ and $\mu_S(v)$ for TASEP and SSEP, repectovely. The scaling function $\Omega(v=x/\omega_t)$ is given by Eq.~\eqref{eq:Omega} and is the limit shape of the quantum system. $\Omega_T(v=x/t)$ and $\Omega_S(v=x/\sqrt{2t})$, instead, are given by Eqs.~\eqref{app:eq:Omega_T} and~\eqref{app:eq:Omega_S}, and correspond to the evolution of the shape of an initial corner (dashed line).}
    \label{app:fig:mu_n_classical}
\end{figure}

In the case of the SSEP, instead, the dynamics is diffusive and a different scaling of space and time has to be taken in order to obtain a non-trivial continuum limit.
Specifically, the rescaled density $n_S(\xi,\tau) = \lim_{N \to \infty} \rho_S(\xi \sqrt{N}, \tau N)$, (the subscript $S$ stands for SSEP) 
satisfies the heat equation~\cite{DeMasi2015Symmetric}
\begin{equation}
\label{app:eq:heat}
    \frac{\partial }{\partial \tau} n_S =\frac{1}{2} \frac{\partial^2}{\partial \xi^2} n_S.
\end{equation}
The same initial condition as before, i.e. $n_S(\xi,0)=\theta(\xi)$, leads to the solution $n_S(x,t) \equiv n_S (x/\sqrt{2t})$ with the scaling function
\begin{equation}
\label{app:eq:nS}
    n_S(v) = \frac{1 + \text{erf}\left( v \right)}{2},
\end{equation}
where we introduced the error function $\text{erf}(x)=2/\sqrt{\pi}\int_0^x e^{-t^2}dt$. According to the continuum version of Eq.~\eqref{eq:mu_from_n}, the interface profile is given by $\mu_S(x,t)=\sqrt{2t} \, \Omega_S(x / \sqrt{2t})$, with the scaling function
\begin{equation}
\label{app:eq:Omega_S}
    \Omega_S(v) = \frac{e^{-v^2}}{\sqrt{\pi}} + v \, \text{erf}(v).
\end{equation}
In Fig.~\ref{app:fig:mu_n_classical} we compare the scaling forms obtained above for TASEP and SSEP with the one of the quantum model. In doing this comparison one should remember that the very scaling variables differ in the various cases with the sole exception of TASEP and the quantum Ising model with $h=0$: in spite of the fact that they both show a ballistic scaling, the corresponding scaling functions are still different.

\section{Second order Schrieffer-Wolff and integrability breaking}
\label{app:sec:SW}

In this Section, we perform a Schrieffer-Wolff transformation \cite{Schrieffer1966Relation} to get a renormalized Hamiltonian, describing the effective degrees of freedom in each sector $\mathcal{H}_l$ when $g,h \ll J < + \infty$. We remind that the Schrieffer-Wolff transformation consists in a renormalization procedure that progressively eliminates, order by order in perturbation theory, all the block-off-diagonal Hamiltonian matrix elements, i.e.\ the ones coupling different sectors $\mathcal{H}_l$ and $\mathcal{H}_{l'}$ with $l \neq l'$. Mathematically, it is a unitary rotation $U = e^S$, with $S = S_1 + S_2 + \cdots$, that gives
\begin{equation}
    e^S H e^{-S} = H_0 + H_1 + H_2 + \cdots,
\end{equation}
where $S_n$ and $H_n$ are of order $n$ in the perturbative coupling. Moreover, performing the expansion up to a finite $n$ yields a rotated Hamiltonian in which the block-off-diagonal terms are of order $n+1$ or higher.

We will follow a recent derivation of the transformation, given in Refs.~\cite{Abanin2017Rigorous,Lerose2020Quasilocalized}, that gives directly the correct result at any desired order.

\subsection{First-order corrections: PXP Hamiltonian}
\label{app:sec:SW_PXP}

Let us start by separating the original $2d$ Ising Hamiltonian in Eq.~\eqref{eq:Ising2d_ham} as follows:
\begin{equation}
    H = H_{\mathrm{Is}} = \bigg( -J\sum_{\langle i,j\rangle} \sigma_i^z \sigma_j^z \bigg) + \bigg( - g \sum_i \sigma_i^x - h \sum_i \sigma_i^z \bigg) \equiv H_0 + V_1 .
\end{equation}
Setting, for the time being, $h=0$, the Schrieffer-Wolff transformation amounts to the following iterative algorithm (starting from $n=1$):
\begin{enumerate}
    \item Split $V_n \equiv H_n + R_n$, where $H_n$ contains only the block-diagonal terms and $R_n$ only the block-off-diagonal ones.
    
    \item Determine $S_n$ from the equation
    \begin{equation}
        \label{eq:eqS_SW}
        \big[ S_n, H_0 \big] + R_n = 0.
    \end{equation}
    
    \item Set
    \begin{equation}
        \label{eq:expression_V_SW}
        V_{n+1} = \sum_{(k_1,\dots,k_p)\in [n+1]'} \frac{1}{p!} [S_{k_1}, [S_{k_2}, \dots, [S_{k_p}, H_0]\dots]] + \sum_{(k_1,\dots,k_p) \in [n]} \frac{1}{p!} [S_{k_1}, [S_{k_2}, \dots, [S_{k_p},V] \dots]],
\end{equation}
    where the summations run over the set $[m]$ of the ordered partitions $(k_1,\dots,k_p)$ of an integer $m (= k_1 + k_2 + \cdots +k_p) $, and $[m]'$ excludes the partition $(k_1=m)$ with $p=1$.
\end{enumerate} \medskip

Let us apply the algorithm described above to our case, up to order $n=2$. First of all, we identify in $V_1$ the block-diagonal terms:
\begin{equation}
\begin{aligned}
    H_1 = - g\sum_i \big(
    & P_{\mathrm{L}i}^\uparrow P_{\mathrm{D}i}^\uparrow \sigma_i^x P_{\mathrm{R}i}^\downarrow P_{\mathrm{U}i}^\downarrow
    + P_{\mathrm{L}i}^\uparrow P_{\mathrm{D}i}^\downarrow \sigma_i^x P_{\mathrm{R}i}^\downarrow P_{\mathrm{U}i}^\uparrow
    + P_{\mathrm{L}i}^\downarrow P_{\mathrm{D}i}^\downarrow \sigma_i^x P_{\mathrm{R}i}^\uparrow P_{\mathrm{U}i}^\uparrow \\
    &+ P_{\mathrm{L}i}^\downarrow P_{\mathrm{D}i}^\uparrow \sigma_i^x P_{\mathrm{R}i}^\uparrow P_{\mathrm{U}i}^\downarrow
    + P_{\mathrm{L}i}^\uparrow P_{\mathrm{D}i}^\downarrow \sigma_i^x P_{\mathrm{R}i}^\uparrow P_{\mathrm{U}i}^\downarrow
    + P_{\mathrm{L}i}^\downarrow P_{\mathrm{D}i}^\uparrow \sigma_i^x P_{\mathrm{R}i}^\downarrow P_{\mathrm{U}i}^\uparrow\big).
\end{aligned}
\end{equation}
and the block-off-diagonal terms:
\begin{multline}
    R_1 = - g \sum_{i} \big(
    P_{\mathrm{L}i}^\downarrow P_{\mathrm{D}i}^\downarrow \sigma_i^x P_{\mathrm{R}i}^\downarrow P_{\mathrm{U}i}^\downarrow
    + P_{\mathrm{L}i}^\downarrow P_{\mathrm{D}i}^\downarrow \sigma_i^x P_{\mathrm{R}i}^\downarrow P_{\mathrm{U}i}^\uparrow
    + P_{\mathrm{L}i}^\downarrow P_{\mathrm{D}i}^\downarrow \sigma_i^x P_{\mathrm{R}i}^\uparrow P_{\mathrm{U}i}^\downarrow
    + P_{\mathrm{L}i}^\downarrow P_{\mathrm{D}i}^\uparrow \sigma_i^x P_{\mathrm{R}i}^\downarrow P_{\mathrm{U}i}^\downarrow
    + P_{\mathrm{L}i}^\downarrow P_{\mathrm{D}i}^\uparrow \sigma_i^x P_{\mathrm{R}i}^\uparrow P_{\mathrm{U}i}^\uparrow \\
    + P_{\mathrm{L}i}^\uparrow P_{\mathrm{D}i}^\downarrow \sigma_i^x P_{\mathrm{R}i}^\downarrow P_{\mathrm{U}i}^\downarrow 
    + P_{\mathrm{L}i}^\uparrow P_{\mathrm{D}i}^\downarrow \sigma_i^x P_{\mathrm{R}i}^\uparrow P_{\mathrm{U}i}^\uparrow
    + P_{\mathrm{L}i}^\uparrow P_{\mathrm{D}i}^\uparrow \sigma_i^x P_{\mathrm{R}i}^\downarrow P_{\mathrm{U}i}^\uparrow
    + P_{\mathrm{L}i}^\uparrow P_{\mathrm{D}i}^\uparrow \sigma_i^x P_{\mathrm{R}i}^\uparrow P_{\mathrm{U}i}^\downarrow
    + P_{\mathrm{L}i}^\uparrow P_{\mathrm{D}i}^\uparrow \sigma_i^x P_{\mathrm{R}i}^\uparrow P_{\mathrm{U}i}^\uparrow \big).
\end{multline}
In the previous equations, the projectors $P^{\uparrow,\downarrow}_i$ are those given in Eq.~\eqref{eq:projectors}, while $\mathrm{L}i/\mathrm{R}i/\mathrm{U}i/\mathrm{D}i$ stands for the left/right/above/below neighbour of the site $i$, as in the main text. One easily gets convinced that the terms in $H_1$ couple states within each $\mathcal{H}_l$, since they conserve the number of domain walls; contrarily, each term in $R_1$ changes their number. 

Then, we need to solve Eq.~\eqref{eq:eqS_SW}, specified for $S_1$:
\begin{equation}
    \big[ S_1, H_0 \big] + R_1 = 0.
\end{equation}
A bit of reasoning leads to the conclusion that one can compensate each term in $R_1$, of the form $P_{\mathrm{L}i} P_{\mathrm{D}i} \sigma_i^x P_{\mathrm{R}i} P_{\mathrm{U}i}$, with a term in $S_1$ of the form $P_{\mathrm{L}i} P_{\mathrm{D}i} \sigma_i^y P_{\mathrm{R}i} P_{\mathrm{U}i}$. Fixing the correct signs, one finds
\begin{multline}
\label{app:eq:S1}
    S_1 = - \frac{ig}{4J} \sum_{i} \bigg(
    \frac{1}{2} P_{\mathrm{L}i}^\downarrow P_{\mathrm{D}i}^\downarrow \sigma_i^y P_{\mathrm{R}i}^\downarrow P_{\mathrm{U}i}^\downarrow
    + P_{\mathrm{L}i}^\downarrow P_{\mathrm{D}i}^\downarrow \sigma_i^y P_{\mathrm{R}i}^\downarrow P_{\mathrm{U}i}^\uparrow
    + P_{\mathrm{L}i}^\downarrow P_{\mathrm{D}i}^\downarrow \sigma_i^y P_{\mathrm{R}i}^\uparrow P_{\mathrm{U}i}^\downarrow
    + P_{\mathrm{L}i}^\downarrow P_{\mathrm{D}i}^\uparrow \sigma_i^y P_{\mathrm{R}i}^\downarrow P_{\mathrm{U}i}^\downarrow
    - P_{\mathrm{L}i}^\downarrow P_{\mathrm{D}i}^\uparrow \sigma_i^y P_{\mathrm{R}i}^\uparrow P_{\mathrm{U}i}^\uparrow  \\
    + P_{\mathrm{L}i}^\uparrow P_{\mathrm{D}i}^\downarrow \sigma_i^y P_{\mathrm{R}i}^\downarrow P_{\mathrm{U}i}^\downarrow 
    - P_{\mathrm{L}i}^\uparrow P_{\mathrm{D}i}^\downarrow \sigma_i^y P_{\mathrm{R}i}^\uparrow P_{\mathrm{U}i}^\uparrow
    - P_{\mathrm{L}i}^\uparrow P_{\mathrm{D}i}^\uparrow \sigma_i^y P_{\mathrm{R}i}^\downarrow P_{\mathrm{U}i}^\uparrow
    - P_{\mathrm{L}i}^\uparrow P_{\mathrm{D}i}^\uparrow \sigma_i^y P_{\mathrm{R}i}^\uparrow P_{\mathrm{U}i}^\downarrow
    - \frac{1}{2} P_{\mathrm{L}i}^\uparrow P_{\mathrm{D}i}^\uparrow \sigma_i^y P_{\mathrm{R}i}^\uparrow P_{\mathrm{U}i}^\uparrow \bigg).
\end{multline}
Finally, applying Eq.~\eqref{eq:expression_V_SW} for $n=2$ yields
\begin{equation}
\label{eq:expression_V2}
    V_2 = \frac{1}{2} [S_1,[S_1,H_0]] + [S_1,V_1] = -\frac{1}{2} [S_1,R_1] + [S_1,V_1].
\end{equation}
The expression above generates a plethora of terms; however, we are interested \emph{only in the block-diagonal part} of $V_2$, namely $H_2$: indeed, the block-off-diagonal part $R_2$ can be removed by going to the next order in the perturbative construction. For now, we will compute only the terms in Eq.~\eqref{eq:expression_V2} that are \emph{diagonal} in $\sigma^z$ (thus leaving out terms involving $\sigma^x$ and $\sigma^y$ that are still block-diagonal). It is easy to identify them, since they come from commuting $\sigma^x_i$ in $R_1$ and $V_1$ with $\sigma_i^y$ in $S_1$, while leaving the projectors untouched (and therefore the 4 projectors around $i$ have to be the same both in $R_1$, $V_1$ and $S_1$). With a bit of patience, one may work out all the details, finding
\begin{multline}
    \label{eq:H2_SW}
    \big[ H_2 \big]_{\mathrm{diag}} = \frac{g^2}{4J} \sum_{i} \bigg(
    \frac{1}{2} P_{\mathrm{L}i}^\downarrow P_{\mathrm{D}i}^\downarrow \sigma_i^z P_{\mathrm{R}i}^\downarrow P_{\mathrm{U}i}^\downarrow
    + P_{\mathrm{L}i}^\downarrow P_{\mathrm{D}i}^\downarrow \sigma_i^z P_{\mathrm{R}i}^\downarrow P_{\mathrm{U}i}^\uparrow
    + P_{\mathrm{L}i}^\downarrow P_{\mathrm{D}i}^\downarrow \sigma_i^z P_{\mathrm{R}i}^\uparrow P_{\mathrm{U}i}^\downarrow\\
    + P_{\mathrm{L}i}^\downarrow P_{\mathrm{D}i}^\uparrow \sigma_i^z P_{\mathrm{R}i}^\downarrow P_{\mathrm{U}i}^\downarrow
    - P_{\mathrm{L}i}^\downarrow P_{\mathrm{D}i}^\uparrow \sigma_i^z P_{\mathrm{R}i}^\uparrow P_{\mathrm{U}i}^\uparrow 
    + P_{\mathrm{L}i}^\uparrow P_{\mathrm{D}i}^\downarrow \sigma_i^z P_{\mathrm{R}i}^\downarrow P_{\mathrm{U}i}^\downarrow 
    - P_{\mathrm{L}i}^\uparrow P_{\mathrm{D}i}^\downarrow \sigma_i^z P_{\mathrm{R}i}^\uparrow P_{\mathrm{U}i}^\uparrow\\
    - P_{\mathrm{L}i}^\uparrow P_{\mathrm{D}i}^\uparrow \sigma_i^z P_{\mathrm{R}i}^\downarrow P_{\mathrm{U}i}^\uparrow
    - P_{\mathrm{L}i}^\uparrow P_{\mathrm{D}i}^\uparrow \sigma_i^z P_{\mathrm{R}i}^\uparrow P_{\mathrm{U}i}^\downarrow
    - \frac{1}{2} P_{\mathrm{L}i}^\uparrow P_{\mathrm{D}i}^\uparrow \sigma_i^z P_{\mathrm{R}i}^\uparrow P_{\mathrm{U}i}^\uparrow \bigg).
\end{multline}

\subsection{First-order corrections: corner Hamiltonian}
\label{app:sec:SW_corner}

\begin{figure}[t]
    \centering
    \includegraphics[width=0.9\textwidth]{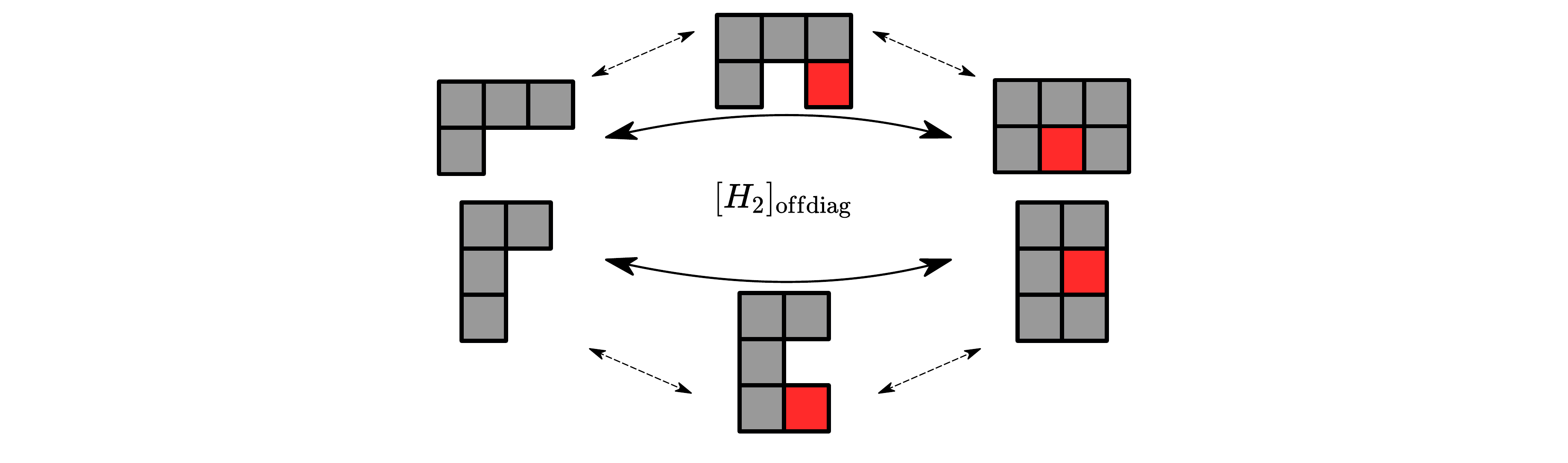}
    \caption{Graphical representation of the off-diagonal part of $H_{2,Y}$, corresponding to next-nearest-neighbor hoppings (see Eq.~\eqref{eq:H2_SW}), constrained to Young diagrams configurations.}
    \label{fig:H2SW}
\end{figure}

Now we specify the expression derived in the previous Section to the sector within $\mathcal{H}_l$ which is dynamically connected to the corner considered in the main text, i.e.\ we restrict our attention to the Young diagrams subspace $\mathcal{H}_{\mathrm{Y}}$. In the previous Section we have already determined the diagonal part of the second-order correction $H_2$, see Eq.~\eqref{eq:H2_SW}. We just need to determine the off-diagonal (but block-diagonal) part. With a bit of reasoning, one may get convinced that the only allowed moves at the second-order perturbation theory, which bring a state out of $\mathcal{H}_{\mathrm{Y}}$ and then back in, are those represented in Fig.~\ref{fig:H2SW}. 
Correspondingly, the Schrieffer-Wolff Hamiltonian reads
\begin{multline}\label{eq:H2SW}
    H_{2,\mathrm{Y}} = \big[ H_2 \big]_{\mathrm{diag}} - \frac{g^2}{4J} \sum_i \Big[ 
    P_{\mathrm{L}i}^\uparrow P_{\mathrm{LU}i}^\uparrow P_{\mathrm{UU}i}^\uparrow 
    \big(\sigma_i^+\sigma_{\mathrm{U}i}^+ + \sigma_i^-\sigma_{\mathrm{U}i}^-\big) 
    P_{\mathrm{R}i}^\downarrow P_{\mathrm{RU}i}^\downarrow P_{\mathrm{D}i}^\downarrow 
    \\
    + P_{\mathrm{L}i}^\uparrow P_{\mathrm{U}i}^\uparrow P_{\mathrm{RU}i}^\uparrow 
    \big(\sigma_i^+\sigma_{\mathrm{R}i}^+ + \sigma_i^-\sigma_{\mathrm{R}i}^- \big)
    P_{\mathrm{D}i}^\downarrow P_{\mathrm{RD}i}^\downarrow P_{\mathrm{RR}i}^\downarrow \Big].
\end{multline}
The factor in front of the sum is fixed by a careful use of Eq.~\eqref{eq:expression_V2}. Now that we have the Hamiltonian in $2d$, we can express it in the $1d$ language of fermions. Before, however, it is convenient to expand all the projectors $P^{\uparrow,\downarrow}$ in terms of $\sigma^z$: one finds
\begin{multline}
    \big[ H_2 \big]_{\mathrm{diag}} = - \frac{5g^2}{64J} \sum_i \big( \sigma_{Li}^z \sigma_i^z + \sigma_i^z \sigma_{Ri}^z + \sigma_i^z \sigma_{Ui}^z + \sigma_{Di}^z \sigma_i^z \big) \\
    + \frac{3g^2}{64J} \sum_i \big( \sigma_{Li}^z \sigma_{Di}^z \sigma_i^z \sigma_{Ri}^z + \sigma_{Li}^z \sigma_{Di}^z \sigma_i^z \sigma_{Ui}^z + \sigma_{Di}^z \sigma_i^z \sigma_{Ri}^z \sigma_{Ui}^z + \sigma_{Li}^z \sigma_i^z \sigma_{Ri}^z \sigma_{Ui}^z \big).
\end{multline}
The term with only two Pauli matrices gives a constant contribution on the Young diagram states, since it counts the number of horizontal and vertical frustrated bonds (it is a constant energy shift in the whole sector $\mathcal{H}_l$). The term with four spins, instead, can be represented, up to a constant term in the subspace $\mathcal{H}_{\mathrm{Y}}$, by an operator which counts the number of corners in each diagram. Accordingly, in the fermion language, one finds the Hamiltonian
\begin{align}
    H_{2,\mathrm{F}} &= -\frac{g^2}{4J} \sum_x \left( \psi^\dagger_x e^{-i\pi n_{x+1}} \psi_{x+2} + \mathrm{H.c.} + 3 n_x n_{x+1} \right) \\
    &= -\frac{g^2}{4J} \sum_x \left(\psi^\dagger_x \psi_{x+2}+ \mathrm{H.c.}\right) + \frac{g^2}{4J}\sum_x\left(2\psi^\dagger_x \psi^\dagger_{x+1} \psi_{x+1}\psi_{x+2}+ \mathrm{H.c.}-3\psi^\dagger_{x} \psi_{x}\psi^\dagger_{x+1} \psi_{x+1}\right),
\end{align}
where the first term is a correction to the kinetic energy and the second a four-fermions interaction.

\section{Participation ratio and localization length}
\label{app:sec:participation_ratio}

In this Section we compute the (inverse) participation ratio, from which one can easily derive the localization length of the eigenfunctions. By definition
\begin{equation}
    \mathrm{IPR} = \sum_{k} J_k^4(\gamma) = \frac{1}{\pi}\int_0^{\pi} d\theta J_0^2(\gamma \sqrt{2 - 2 \cos{\theta}}),
\end{equation}
where we used both Neumann's addition theorem to write the sum as an integral and the explicit form of the eigenfunctions. With a change of variables, the integral can be cast in the form
\begin{equation}
    \mathrm{IPR} = \frac{2}{\pi}\int_0^{1} dx \frac{J_0^2(2 \gamma x)}{\sqrt{1 - x^2}} = {}_2 F_3 \left( \frac{1}{2}, \frac{1}{2}; 1, 1, 1; -4\gamma^2 \right),
    \label{eq:app-IPR-def}
\end{equation}
where ${}_2 F_3$ is the generalized hypergeometric function.  
For large $\gamma$, one can take the asymptotic expansion of the latter and the (non-oscillating part of the) $\mathrm{IPR}$ turns out to be given by
\begin{equation}
\label{eq:IPR_hyper}
    \mathrm{IPR} = \frac{2 \gamma_E + 5\ln{4} + \ln{\gamma^2}}{2\pi^2 \gamma} + \frac{3-\gamma_E - \ln(32 \gamma)}{64 \pi \gamma^3} + O\left(\frac{1}{\gamma^5}\right),
\end{equation}
being $\gamma_E$ the Euler constant. Since the localization length of the eigenfunctions is roughly $\xi \approx 1/\mathrm{IPR}$, one finds
\begin{equation}
    \xi \sim \frac{2\pi^2 \gamma}{2 \gamma_E + 5\ln{4} + \ln{\gamma^2}},
\end{equation}
which gives Eq.~\eqref{eq:xi-1-expr}. 

Alternatively, one can also determine the asymptotic expansion for the IPR directly from the integral, using the Mellin transform. In particular, for two functions $f_{1,2}(x)$ and their Mellin transforms $\tilde{f}_{1,2}(s)$, it holds
\begin{equation}
    \int_{0}^{\infty}\!dx\, f_1(x) f_2(x)  = \frac{1}{2\pi i} \int_{c-i\infty}^{c+i\infty}\!ds\, \tilde{f}_1(1-s) \tilde{f}_2(s)  ,
\end{equation}
being, in our case, $f_1(x)=\frac{1}{\sqrt{1-x^2}} \theta(1-|x|)$ and $f_2(x)=J_0^2(2\gamma x)$. One gets then
\begin{equation}
    \tilde{f}_{1}(s)=\frac{\sqrt{\pi}}{2}\frac{\Gamma\left(\frac{s}{2}\right)}{\Gamma\left(\frac{1+s}{2}\right)},\qquad \tilde{f}_{2}(s)=\frac{1}{(2\gamma)^s}\frac{\Gamma\left(\frac{1-s}{2}\right)\Gamma\left(\frac{s}{2}\right)}{2\sqrt{\pi} \Gamma^2\left(1-\frac{s}{2}\right)}.
\end{equation}
Accordingly, the first equality in Eq.~\eqref{eq:app-IPR-def} can be alternatively be written as
\begin{equation}
\label{eq:IPR_residue}
    \mathrm{IPR} = \frac{1}{2\pi i} \int_{c-i\infty}^{c+i\infty} ds \frac{1}{2\pi}\frac{1}{(2\gamma)^s} \frac{\Gamma^2\left(\frac{1-s}{2}\right) \Gamma\left(\frac{s}{2}\right)}{\Gamma^3\left(1-\frac{s}{2}\right)} \equiv  \frac{1}{2\pi i} \int_{c-i\infty}^{c+i\infty} ds \, \mathcal{F}(s) .
\end{equation}
The poles structure of the $\Gamma$-function sets $c\in (0,1)$. To compute the integral, one can move the Bromwich path towards increasing values of $\Re(s)$; this way one has to go around the poles of the integrand, which are double poles located on the odd integer numbers, and use the residue theorem to compute their contribution to the integral. As an example, we report the residue at $s=1$, for other values of $s$ the computation is analogous. By definition one has
\begin{equation}
    \text{Res}\left[ \mathcal{F}(s), s=1 \right] = \frac{d}{ds} \left[ (s-1)^2 \frac{1}{2\pi}\frac{1}{(2\gamma)^s} \frac{\Gamma\left(\frac{1-s}{2}\right)^2 \Gamma\left(\frac{s}{2}\right)}{\Gamma\left(1-\frac{s}{2}\right)^3} \right] \Biggr\rvert_{s=1}.
\end{equation}
Expanding around $s=1$ one has $(s-1)^2 ~ \Gamma\left(\frac{1-s}{2}\right) = 4 (1 + \gamma_E (s-1)) + O(s-1)^2$, from which it follows
\begin{equation}
    \text{Res}\left[ \mathcal{F}(s), s=1 \right] = - \frac{2 \gamma_E + 5\ln{4} + \ln{\gamma^2}}{2\pi^2 \gamma}.
\end{equation}
Applying the residue theorem, one obtains a $2 \pi i$ factor that cancels the one in front of Eq.~\eqref{eq:IPR_residue} and a minus sign given by the index of the contour, which is clockwise, obtaining the first term of Eq.~\eqref{eq:IPR_hyper}. The other terms are obtained with the residues of the other poles. In general, from the dependence on $1/\gamma^s$ in the integral, one can see that the residue of the pole at $s=2n+1$ gives the order $1/\gamma^{2n+1}$ of the asymptotic expansion. In this way, one obtains the same result as in Eq.~\eqref{eq:IPR_hyper} from the asymptotic expansion of the hypergeometric function. Notice that the residues of the poles give only the power series contribution to the whole integral. There is a bounded oscillating term missing, that comes from the remaining part of integral on the Bromwich path.

\twocolumngrid

\bibliography{references}
\end{document}